\documentclass[preprint,11pt, aps,superscriptaddress,groupedaddress,nofootinbib]{revtex4-2}   

\RequirePackage{rotating}

\RequirePackage[T1]{fontenc}
\RequirePackage{graphicx}
\RequirePackage{mathptmx}      
\RequirePackage{flushend}
\RequirePackage[colorlinks,citecolor=blue,urlcolor=blue,linkcolor=blue]{hyperref}

\usepackage{verbatim}
\usepackage{multirow}
\usepackage{dcolumn}   
\usepackage{bm}        
\usepackage{amssymb}   
\usepackage{amsmath}
\usepackage{amsfonts}
\usepackage{slashed}
\usepackage{graphicx}
\usepackage[ngerman, english]{babel}
\usepackage{float}
\usepackage{subfigure}
\usepackage[utf8]{inputenc}
\usepackage{mathrsfs}
\usepackage{hhline}
\usepackage{braket}
\usepackage{bbold}
\usepackage{nicefrac}
\usepackage{xcolor}
\usepackage{hyperref}
\usepackage{titleref}
\usepackage{nameref}
\usepackage{comment}
\usepackage{pifont}
\usepackage[section]{placeins}
\usepackage{rotating}
\usepackage{nameref}
\usepackage{placeins}
\usepackage{multirow}
\usepackage{makecell}
\usepackage{xpatch}

\newcommand{\xmark}{\textcolor{red}{\ding{55}}}

\definecolor{darkblue}{cmyk}{1, 1, 0, .45} 
\definecolor{darkred}{cmyk}{0, 1, 1, .45} 
\newcommand \supsetsim{\mathrel{\substack{
  \textstyle\supset\\[-0.2ex]\textstyle\sim}}}

\begin{document}

\title{Fermionic Singlet Dark Matter in One-Loop Solutions to the $R_K$ Anomaly: \\ A Systematic Study}

 \preprint{DO-TH 21/11}
 \preprint{TUM-HEP-1322/21}
 
 \author{Mathias Becker}
 \email[]{mathias.becker@tum.de}
 
 \affiliation{Fakult\"at f\"ur Physik,
	 Technische Universit\"at Dortmund, 44221 Dortmund,
 Germany} 
 
 \affiliation{Physik Department T70, Technische Universit\"at M\"unchen, 85748 Garching, Germany}
 
 \author{Dominik D\"oring}
 \email[]{dominik.doering@tu-dortmund.de}

 \affiliation{Fakult\"at f\"ur Physik,
	 Technische Universit\"at Dortmund, 44221 Dortmund,
 Germany}

\author{Siddhartha Karmakar}
\email[]{karmakars@iitb.ac.in}

\affiliation{Discipline of Physics, Indian Institute of Technology Indore, Simrol, Indore - 453 552, India}
\affiliation{Department of Physics, Indian Institute of Technology Bombay, Powai, Mumbai- 400 076, India}

 \author{Heinrich P\"as}
 \email[]{heinrich.paes@tu-dortmund.de}

\affiliation{Fakult\"at f\"ur Physik,
	 Technische Universit\"at Dortmund, 44221 Dortmund,
 Germany} 
 
\begin{abstract}
We study the dark matter phenomenology of Standard Model extensions addressing the reported anomaly in the $R_K$ observable at one-loop. The article covers the case of fermionic singlet DM coupling leptophilically, quarkphilically or amphiphilically to the SM. The setup utilizes a large coupling of the new particle content to the second lepton generation to explain the $R_K$ anomaly, which in return tends to diminish the dark matter relic density. Further, dark matter direct detection experiments provide stringent bounds even in cases where the dark matter candidate only contributes a small fraction of the observed dark matter energy density. In fact, direct detection rules out all considered models as an explanation for the $R_K$ anomaly in the case of Dirac dark matter. Conversely, for Majorana dark matter, the $R_K$ anomaly can be addressed in agreement with direct detection in coannihilation scenarios. For leptophilic dark matter this region only exists for $M_\text{DM} \lesssim 1000 \, \mathrm{GeV}$ and dark matter is underabundant. Quarkphilic and amphiphilic scenarios even provide narrow regions of parameter space where the observed relic density can be reproduced while offering an explanation to $R_K$ in agreement with direct detection experiments.    
\end{abstract}

\maketitle

\section{Introduction}

The Standard Model (SM) of particle physics provides an excellent description of the electroweak and strong interactions. However, one of its most pressing shortcomings is the lack of a suitable dark matter (DM) candidate. Various astrophysical observations, such as the cosmic microwave background (CMB)\cite{Aghanim:2018eyx}, big bang nucleosynthesis (BBN) \cite{Schramm:1997vs} or gravitational lensing effects \cite{Clowe:2006eq}, provide strong evidence for the existence of (particle) DM. The energy density of DM is precisely determined by the Planck satellite, $\Omega_{DM} h^2 = 0.120 \pm 0.001$ \cite{Aghanim:2018eyx}. Considering particle DM, this part of the universe energy content must consist of massive, electrically neutral or millicharged particles, which if thermally produced, must be cold, i.e. non-relativistic, in order not to spoil the structure formation in the early universe.

If DM is produced thermally via interactions with the SM either directly or via a portal, these interactions are expected to induce signals in DM direct detection experiments, such as XENON \cite{Aprile:2018dbl,Aprile:2019dbj}. However, so far no direct detection experiment observed a positive signal, which in return provides tight constraints on the interaction of DM with the SM. \\
While the LHC mostly confirms the predictions of the SM and did not observe any direct production of a beyond the SM particle, the LHCb experiment observed tensions within the flavor sector. More precisely, it observes a $2.6 \sigma$ discrepancy \cite{Aaij:2019wad} with the SM in the theoretically clean observable $R_K = \frac{Br \left( B \rightarrow K \mu^+ \mu^- \right)}{Br \left( B \rightarrow K e^+ e^- \right)}$ \cite{Hiller:2003js}. This anomaly was addressed extensively in the past including very different approaches, for instance the tree-level exchange of leptoquarks \cite{Hiller:2016kry,Fajfer:2015ycq,Cornella:2019hct,Pas:2015hca}, possibly emerging from a grand unified theory, or via the introduction of a new $U\left(1\right)$ gauge group \cite{Bian:2017rpg}. 

Our work builds on a class of models introduced in \cite{Arnan:2016cpy} to address the $R_K$ anomaly. The authors analyze a scenario where contributions to $b \rightarrow s \mu \mu$ are realized at one-loop level via three new particles charged under the SM gauge groups.
Such new physics scenarios can also provide sizable contribution to the muon magnetic moment ($(g-2)_{\mu}$). Allowing for representations up to the adjoint, they find and analyze 48 different models. Several of those models include an electrically neutral color singlet, which, if stabilized by a symmetry, can be a valid DM candidate. In this work, we enlarge the analysis of this class of models and address the question if both the $R_K$ anomaly and the observed DM relic density can be explained within this framework for models including a fermionic singlet. The connection of this class of models and DM has been addressed previously, for instance in \cite{Kawamura:2017ecz,Cline:2017qqu,Vicente:2018xbv,Barman:2018jhz,Cerdeno:2019vpd}. All of these specify on one out of the 48 available models. References \cite{Kawamura:2017ecz,Vicente:2018xbv} discuss a scalar singlet DM realization (bIA model), while \cite{Cline:2017qqu} discusses DM in the context of a leptophilic fermionic singlet setup (bIIA) for Majorana DM. The article \cite{Barman:2018jhz} discusses the bIIA setup in the context of scalar doublet DM. Reference \cite{Cerdeno:2019vpd} addresses fermionic Majorana DM in the aIA model specification. The connection of $B$-anomalies and DM has also been studied in different one-loop setups \cite{Belanger:2015nma} and tree-level realizations \cite{Guadagnoli:2020tlx,Carvunis:2020exc}.

Typically, this setup requires a relatively large coupling of the new particles to the second lepton generation to explain the $R_K$ anomaly, since, in comparison to tree-level realizations, the one-loop suppression has to be compensated and, additionally, the coupling to quarks is constrained by $B_0$-$\overline{B}_0$ mixing. This in turn can lead to an underpopulated dark sector, as the strong coupling to the second lepton generation tends to delay the DM freeze-out. Moreover, this coupling can generate a large vector current coupling of DM to the $Z$-boson at one-loop level that is tightly constrained from the XENON experiment  \cite{Aprile:2018dbl}. However, in for instance models with a fermionic singlet DM candidate, DM can be either a Dirac or Majorana fermion, where the latter has a vanishing vector coupling to the $Z$-boson and weakens the constraints from $B_0$-$\overline{B}_0$ mixing. Thus, these models seem especially promising and are the main subject of our studies. 

This article is structured as follows: In Section \ref{sec:ModelandAnalysis}, we review the model and constraints presented in \cite{Arnan:2016cpy}. Subsequently, we discuss the implications of the models on the relic density and direct detection experiments in Section \ref{sec:DMPheno}, where we also provide a summary of our analysis strategy for our numerical analysis. The results of this analysis for each model are presented in Section \ref{sec:Results}. In Section \ref{sec:conclusion}, we conclude.


\section{Model Classification and Coupling Constraints} \label{sec:ModelandAnalysis}
\subsection{Setup and Classification}
The model classes proposed by \cite{Arnan:2016cpy} as a one-loop solution to the $b \to s \mu^- \mu^+$ anomalies are distinct in their particle content. In either of the two model realizations, three additional particles are added to the SM. In realization a, two heavy scalars, $\phi_Q$ and $\phi_L$, and a vector-like fermion $\psi$ are present, whereas in realization b there exist two vector-like fermions, $\psi_Q$ and $\psi_L$, and a heavy scalar $\phi$. The indices $Q$ and $L$ denote their coupling to quarks/leptons respectively. The corresponding Lagrangians for each realization are  
\begin{align}
	\label{eq: ModelLagrangiana}
	\mathcal{L}_{int}^{a} &= \Gamma_{Q_i} \bar{Q}_i P_R \psi \phi_Q + \Gamma_{L_i} \bar{L}_i P_R \psi \phi_L + \text{h.c.} \\
	\label{eq: ModelLagrangianb}
	 \mathcal{L}_{int}^{b} &= \Gamma_{Q_i} \bar{Q}_i P_R \psi_Q \phi + \Gamma_{L_i} \bar{L}_i P_R \psi_L \phi + \text{h.c.} \, .
\end{align}
where $Q_i$ and $L_i$ denote the left-handed quark/lepton doublets, while $i$ is a flavor index.
The Eqns.\@\eqref{eq: ModelLagrangiana} and \eqref{eq: ModelLagrangianb} reveal that the additional heavy particles couple only to left-handed SM fermions. This is a phenomenologically driven feature, implemented to ensure that $\mathcal{C}_9=-\mathcal{C}_{10}$ holds, which is the preferred scenario to fit the \textit{B} anomalies, where $\mathcal{C}_9$ and $\mathcal{C}_{10}$ are Wilson coefficients of
\begin{align}
\mathcal{O}_9 &= (\bar{s}\gamma^{\nu}P_{L} b)(\bar{\mu}\gamma_{\nu}\mu), \,\,\,\,\,
\mathcal{O}_{10} = (\bar{s}\gamma^{\nu}P_{L} b)(\bar{\mu}\gamma_{\nu}\gamma^5\mu) \, . 
\end{align}

The model classification up to the adjoint representation can be extracted from Table \ref{tbl:SUclassification}, where the representations of the new particles under the SM gauge group are presented. The roman numerals I-VI classifies the representation of the new fields under $SU(2)_L$ and capital latin letters A-B specifies the $SU(3)_C$ charges.     

\begin{table}[h]
 \centering
 \begin{tabular}{c|ccc}
  $SU(2)_L$ & $\phi_Q,\psi_Q$ & $\phi_L,\psi_L$ & $\psi,\phi$ \\ [1pt]  \hline 
  I & \textbf{2} & \textbf{2} & \textbf{1} \\ [1pt] 
  II & \textbf{1} & \textbf{1} & \textbf{2} \\ [1pt] 
  III & \textbf{3} & \textbf{3} & \textbf{2} \\ [1pt] 
  IV & \textbf{2} & \textbf{2} & \textbf{3}  \\ [1pt] 
  V & \textbf{3} & \textbf{1} & \textbf{2} \\ [1pt] 
  VI & \textbf{1} & \textbf{3} & \textbf{2} \\ [1pt]  \hline \hline 
  $SU(3)_C$ & & &  \\ [1pt]  \hline 
  A & \textbf{3} & \textbf{1} & \textbf{1} \\[1pt] 
  B & \textbf{1} & $\overline{\textbf{3}}$ & \textbf{3} \\[1pt] 
  C & \textbf{3} & \textbf{8} & \textbf{8} \\[1pt] 
  D & \textbf{8} & $\overline{\textbf{3}}$ & \textbf{3} \\ [1pt]  \hline \hline  
  $Y$ & & &  \\ [1pt]  \hline 
  & $\nicefrac{1}{6} \mp X$ & $-\nicefrac{1}{2} \mp X$ & $\pm X$
 \end{tabular}
    \caption{All possible choices for the combinations of representation of the new particles such that they allow for an one-loop contribution to $b \rightarrow s \mu^+ \mu^-$. The upper sign of $\pm$ belongs to a-type models, the lower to b-type models.}
        \label{tbl:SUclassification}
\end{table}

In this context, $X$ is defined as the hypercharge of $\psi$ in model class a, while it is defined as the negative hypercharge of $\phi$ in model realization b. The parameter $X$ can be freely chosen in units of $\nicefrac{1}{6}$ in the interval $X\in(-1,1)$. 

In this work, we want to analyze whether the model classes presented above contain a viable DM candidate, while still being able to explain the \textit{B} anomalies. A dark matter candidate is constrained to be a colorless, electrically neutral, massive particle \footnote{In fact, DM could be colored and exist in form of eventually colorless bound states. We, however, only consider single particle DM. Please note also, that in principle DM could possess a small electric charge, such as in scenarios of millicharged DM \cite{Berlin:2018sjs,Munoz:2018pzp}.}. This statement alone eliminates one half of the 48 possible model configurations, since in the categories $C$ and $D$ there is not a single colorless particle. The remaining 24 models can be classified in terms of the properties of their DM candidates.
In this article, we limit ourselves to models with a fermionic singlet dark matter candidate, which amounts to five models where the DM can be either Majorana or Dirac fermion. All singlet DM models are categorized in Table \ref{tbl:DMclassification}. Note that in order to stabilize the DM candidate against a decay, we assume all BSM particles to carry an odd charge under a $\mathtt{Z}_2$ symmetry, while all SM particles are evenly charged.
\begin{table}[H]
 \centering
 \begin{tabular}{c|c}
  fermionic singlet & scalar singlet  \\ [1pt] \hline
  \textcolor{darkred}{aIA}  & aIIA \\ [1pt]
  \textcolor{darkred}{bIIA} & aIIB \\ [1pt]
  \textcolor{darkred}{bIIB} & aVA \\ [1pt]
  \textcolor{darkred}{bVA}  & aVIB \\ [1pt]
  \textcolor{darkred}{bVIB} & bIA
 \end{tabular}
    \caption{Models containing a singlet DM candidate. Only the models containing a fermionic singlet DM candidate, highlighted in red, are considered in this work. Note that the value for $X$ is fixed within each model by the condition that there is a singlet DM candidate.}
        \label{tbl:DMclassification}
\end{table}

\subsection{Constraints on new Yukawa Couplings} \label{sec:YukConstraints}
Following \cite{Arnan:2016cpy}, we obtain constraints on the couplings $\Gamma_\mu$ from constraints on the Wilson coefficients $\mathcal{C}_9$ and $\mathcal{C}_{B\bar{B}}$, 
which are obtained from global fits of LFUV observables and $B$-$\bar{B}$-mixing respectively, which is generated by the effective operator
\begin{align}
 \mathcal{O}_{B\bar{B}}= (\bar{s}_\alpha \gamma^\mu P_L b_\alpha)(\bar{s}_\beta \gamma_\mu P_L b_\beta) \, .
\end{align}

The Wilson coefficients read
\begin{align}
 \mathcal{C}_9^\text{box, a}=-\mathcal{C}_{10}^\text{box, a}&= \frac{\sqrt{2}}{4G_F V_{tb}V_{ts}^*} \frac{\Gamma_s\Gamma_b^* |\Gamma_\mu|^2}{32 \pi \alpha_\text{em} M_\psi^2}(\chi \eta F(x_Q,  x_L)+2\chi^M \eta^M G(x_Q, x_L)) \, , \\
 \mathcal{C}_9^\text{box, b}=-\mathcal{C}_{10}^\text{box, a}&= -\frac{\sqrt{2}}{4G_F V_{tb}V_{ts}^*} \frac{\Gamma_s\Gamma_b^* |\Gamma_\mu|^2}{32 \pi \alpha_\text{em} M_\phi^2}(\chi \eta-\chi^M \eta^M) F(y_L, y_L) \, , \\
 \mathcal{C}_{B\bar{B}}^\text{a}&= \frac{(\Gamma_s \Gamma_b^*)^2}{128 \pi^2 M_\psi^2} \left( \chi_{B\bar{B}} \eta_{B\bar{B}} F(x_Q,x_Q) + 2 \chi_{B\bar{B}}^M \eta_{B\bar{B}}^M G(x_Q,x_Q) \right) \, , \\
 \mathcal{C}_{B\bar{B}}^\text{b}&= \frac{(\Gamma_s \Gamma_b^*)^2}{128 \pi^2 M_\phi^2} \left( \chi_{B\bar{B}} \eta_{B\bar{B}} -\chi_{B\bar{B}}^M \eta_{B\bar{B}}^M \right) F(y_Q, y_Q) \, ,
\end{align}
where $x_{\nicefrac{Q}{L}}=\nicefrac{M^2_{\phi_{\nicefrac{Q}{L}}}}{M^2_\psi}$ and 
$y_{\nicefrac{Q}{L}}=\nicefrac{M^2_{\psi_{\nicefrac{Q}{L}}}}{M^2_\phi}$.
$F$ and $G$ are the dimensionless loop-functions
\begin{align}
 F(x,y)&=\frac{1}{(1-x)(1-y)}+\frac{x^2 \ln{x}}{(1-x)^2(1-y)}+\frac{y^2 \ln{y}}{(1-x)(1-y)^2} \\
 G(x,y)&=\frac{1}{(1-x)(1-y)}+\frac{x \ln{x}}{(1-x)^2(1-y)}+\frac{y \ln{y}}{(1-x)(1-y)^2} \, .
\end{align}

The $SU(2/3)$-factors $\eta^{(M)}_{(B\bar{B})}, \chi^{(M)}_{(B\bar{B})}$ can be extracted from Table \ref{tbl:SUfactors}. 
\begin{table}[H]
 \centering
 \begin{tabular}{c|c|c|c|c|c|c}
  $SU(2)_L$ & $\eta$ & $\eta^M$ & $\eta_{B\bar{B}}$ & $\eta^M_{B\bar{B}}$ & $\eta_{a_\mu}$ & $\tilde{\eta}_{a_\mu}$ \\ [1pt] \hline
  I & $1$ & $1$ & $1$ & $1$ & $-1 \mp X$ & $ \pm X$  \\ [1pt]
  II & $1$ & $0$ & $1$ & $0$ & $-\frac{1}{2} \mp X $ & $-\frac{1}{2} \pm X $  \\ [1pt]
  III & $\frac{5}{16}$ & $0$ &  $\frac{5}{16}$ &  $0$ & $-\frac{7}{8} \mp \frac{3}{4}X $ & $ \frac{1}{8} \pm \frac{3}{4}X$  \\ [1pt]
  IV & $\frac{5}{16}$ & $\frac{1}{16}$ &  $\frac{5}{16}$ &  $\frac{1}{16}$ & $-\frac{1}{4} \mp \frac{3}{4}X$ & $-\frac{1}{2} \pm \frac{3}{4}X$  \\ [1pt]
  V & $\frac{1}{4}$ & $0$ &  $\frac{5}{16}$ & $0$ & $-\frac{1}{2} \mp X$ & $-\frac{1}{2}\pm X$  \\ [1pt]
  VI & $\frac{1}{4}$ & $0$ & $1$ & $0$ & $-\frac{7}{8} \mp \frac{3}{4}X$ & $\frac{1}{8} \pm \frac{3}{4}X$  \\ [1pt] \hline \hline
  $SU(3)_C$ & $\chi$ & $\chi^M$ & $\chi_{B\bar{B}}$ & $\chi^M_{B\bar{B}}$ & $\chi_{a_\mu}$ &  \\ [1pt] \hline
  A & $1$ & $1$ & $1$ & $1$ & $1$ \\ [1pt]
  B & $1$ & $0$ & $1$ & $0$ & $3$ \\ [1pt]
 \end{tabular}
    \caption{$SU(2)$ and $SU(3)$ factors entering Wilson coefficients $\mathcal{C}_9^\text{box}$ and $\mathcal{C}_{B\bar{B}}$.}
        \label{tbl:SUfactors}
\end{table}

As in \cite{Arnan:2016cpy}, we neglect the influence of photon penguin diagram contributions to $\mathcal{C}_9$ and therefore assume $\mathcal{C}_9 \approx \mathcal{C}_9^\text{box}$. The $2\sigma$ bounds on $\mathcal{C}_9=-\mathcal{C}_{10}$ and $\mathcal{C}_{B\bar{B}}$ are \cite{Alguero:2021anc,FermilabLattice:2016ipl}
\begin{align}
 \mathcal{C}_9=-\mathcal{C}_{10} &\in [-0.46,-0.29] \\
 \mathcal{C}_{B\bar{B}} &\in [-2.1,0.6] \cdot 10^{-5} \,\text{TeV}^{-2} \,.
\end{align}
Starting from these premises, we can construct an upper bound on $\Gamma_s\Gamma_b^*$ from $B$-$\bar{B}$ mixing and use this to construct a lower bound on $\Gamma_\mu$ by taking into account the bounds on $\mathcal{C}_9$.
\\
We assume mass degeneracy between the new non-dark matter particles $\phi_Q$ and $\phi_L$ in a-type models and $\psi_Q/\psi_L$ and $\phi$ in b-type models respectively.\footnote{We comment on the effects of lifting this assumption for the DM phenomenology at the beginning of section \ref{sec:Results}} For convenience, we further introduce the dimensionless parameter $\kappa$, which we define as 

\begin{align}
	\kappa = \left\{\begin{array}{ll}
	 			\nicefrac{M_{\phi_Q}}{M_{\psi}}   = \nicefrac{M_{\phi_L}}{M_{\psi}}, & \text{in a-type models} \\
				\nicefrac{M_{\psi_Q}}{M_{\psi_L}} = \nicefrac{M_{\phi}}{M_{\psi_L}}, & \text{in $\psi_L$-DM b-type models} \\
				\nicefrac{M_{\psi_L}}{M_{\psi_Q}} = \nicefrac{M_{\phi}}{M_{\psi_Q}}, & \text{in $\psi_Q$-DM b-type models}\, ,
	 		 \end{array}\right. 
\label{kappadefs}	 		 
\end{align}
quantifying the mass gap between DM and the non-DM exotic particles in the model. \\
Additionally, we restrict our analysis to single-component DM scenarios. This translates into a lower bound on the decay rate of the heavier dark sector particles to ensure their decay proceeds sufficiently fast. For the mass splitting, this in turn translates to $(\kappa -1) m_{\text{DM}} > m_{\pi}$ \cite{Cirelli:2005uq}.
Within this model it is possible to construct a solution to the anomalous magnetic moment of the muon, commonly dubbed $(g-2)_\mu$. In these models we have a contribution to $a_\mu = \nicefrac{(g-2)_\mu}{2}$ as
\begin{align}
 \Delta a_\mu^{a} = \frac{m^2_\mu \left|\Gamma_\mu \right|^2}{8 \pi^2 M_\psi^2} \chi_{a_\mu} \left[\eta_{a_\mu} F_7(x_L) - \tilde{\eta}_{a_\mu} \tilde{F}_7(x_L) \right]\, , \\
 \Delta a_\mu^{b} = \frac{m^2_\mu \left|\Gamma_\mu \right|^2}{8 \pi^2 M_\phi^2} \chi_{a_\mu} \left[\tilde{\eta}_{a_\mu} \tilde{F}_7(y_L) - \eta_{a_\mu} F_7(y_L) \right]\, ,
\end{align}
where the group factors $\tilde{\eta}_{a_\mu}/\eta_{a_\mu}$ and $\chi_{a_\mu}$ can be extracted from Table \ref{tbl:SUfactors} and the functions $F_7(x)$ and $\tilde{F}_7$ are characterized as
\begin{align}
 F_7(x)= \frac{x^3 - 6x^2 + 6x \log{(x)} +3x +2}{12(x-1)^4}, \, \tilde{F}_7(x)=\frac{F_7(\frac{1}{x})}{x} \, .
\end{align}
Recently, the g-2 collaboration updated the earlier results of \cite{Bennett:2006fi} and reported on a $4.2\,\sigma$ deviation from the SM, where the difference $\Delta a_\mu \equiv \Delta a_\mu^{\text{Exp}} - \Delta a_\mu^{\text{SM}}$ amounts to \cite{Muong-2:2021ojo}
\begin{align}
 \Delta a_\mu = (251\pm 59)\cdot 10^{-11} \, .
\end{align}
This translates to upper and lower bound on the leptonic Yukawa coupling $\Gamma_\mu$ summarized in Table \ref{tbl:g-2}.\\

The constraints on the new Yukawa couplings $\Gamma_\mu$,$\Gamma_b$ and $\Gamma_s$ are summarized in Table \ref{tbl:couplingConstraintsYuk}. In general, Dirac and Majorana versions of the models can lead to different constraints on the Yukawa couplings, since additional contributions to the $B$-$\bar{B}$-mixing and $b \to  s \, l^+ l^-$ transitions can arise. These contributions, however, only occur in the case of model aIA \footnote{Note that this is generally a feature of models, where $SU(2)\in[I,IV], SU(3)=[A,C]$ and $X=0$. The model aIA happens to be the only model of those with a fermionic singlet DM candidate.} because of the constellations of $SU(2)$ representations present in this model. This feature is thus ultimately incorporated in the Wilson coefficients $\mathcal{C}_{B\bar{B}}$ and $\mathcal{C}^{\text{box}}_{9}$, together with Table \ref{tbl:SUfactors}, where the product $\chi^M_{(B\bar{B})} \eta^M_{(B\bar{B})}$ is non-zero only in the case aIA. In general, the the constraints of the new Yukawa couplings read
\begin{align}
\Gamma_s \Gamma_b^* &\leq \mathcal{B}^\text{model}_{bs} \left( \kappa \right) \frac{M_\text{DM}}{\text{GeV}} \, , \label{eq:GammaBSBound} \\
\Gamma_\mu &\geq \mathcal{B}^\text{model}_\mu \left( \kappa \right) \sqrt{\frac{M_\text{DM}}{\text{GeV}}} , . \label{eq:GammaMuBound}
\end{align}
The coefficient functions $\mathcal{B}^\text{model}_i \left( \kappa \right)$ are model dependent and summarized in Table \ref{tbl:couplingConstraintsYuk}. 


\subsection{Constraints on Parameters from the Scalar Potential}\label{sec:ScalarPotConstraints}
The extended scalar sector of the models leads to additional terms in the scalar potential. These parameters are mass parameters $\mu_i$ as well as quartic couplings $\lambda$. In this work, we assume that no other scalar than the SM $SU(2)$-doublet scalar field $H$ acquires a vacuum expectation value since a non-zero vacuum expectation value in the dark sector would break the stabilizing $\mathtt{Z}_2$ symmetry. The resulting condition on the model parameters is given specifically below. \\
Further, we demand perturbativity from all allowed quartic couplings. This yields 
\begin{align}
	|\lambda| \leq (4 \pi)^2 \, .
\end{align}
We also want to constrain ourselves to scenarios, where the vacuum is stable at tree-level. Since the particle content in the scalar sector differs between model realizations a or b, the implications for the scalar quartic couplings differ as well.

\subsubsection{a-type Models}
There exist two other scalar doublets $\phi_Q$ and $\phi_L$ in a-type models in addition to the SM $SU(2)$-doublet $H$. The most general scalar potential, which is invariant under the DM stabilizing symmetry in this scenario is consequently
\begin{align}
\begin{split}
	V_\text{scalar}=& \mu_H^2 H^\dagger H + \mu_{\phi_{Q}}^2 \phi_Q^\dagger \phi_Q + \mu_{\phi_{L}}^2 \phi_L^\dagger \phi_L \\
	 &+ \lambda_H \left(H^\dagger H \right)^2 + \lambda_{\phi_{Q}} \left( \phi_Q^\dagger \phi_Q \right)^2 + \lambda_{\phi_L} \left( \phi_L^\dagger \phi_L \right)^2 \\
	 &+ \lambda_{\phi_{Q},H,1} \left( \phi_Q^\dagger \phi_Q \right)\left( H^\dagger H \right)+  \lambda_{\phi_{L},H,1} \left( \phi_L^\dagger \phi_L \right)\left( H^\dagger H \right) \\
	 &+ \lambda_{\phi_{Q},H,2} \left( \phi_Q^\dagger H \right)\left( H^\dagger \phi_Q \right) +  \lambda_{\phi_{L},H,2} \left( \phi_L^\dagger H \right)\left( H^\dagger \phi_L \right) \\
	 &+ \lambda_{\phi_L,\phi_Q,1} \left( \phi_L^\dagger \phi_L \right)\left( \phi_Q^\dagger \phi_Q \right) + \lambda_{\phi_L,\phi_Q,2} \left( \phi_L^\dagger \phi_Q \right)\left( \phi_Q^\dagger \phi_L \right) \\ 
	  &+ \left[ \lambda_{\phi_{Q},H,3} \left( \phi_Q^\dagger H \right)^2 + \text{h.c.}\right] + \left[ \lambda_{\phi_{L},H,3} \left( \phi_L^\dagger H \right)^2 + \text{h.c.}\right] \\
	  &+\left[ \lambda_{\phi_{L},\phi_Q,3} \left( \phi_Q^\dagger \phi_L \right)^2 +\text{h.c.} \right]
	  \, .
\end{split}
\end{align}
Adopting the limits from \cite{Keus:2014isa}, the vacuum stability bounds at tree-level for this kind of potential read
\begin{align}
 \lambda_H, \lambda_{\phi_Q}, \lambda_{\phi_L} &> 0 \\
 \lambda_{\phi_L, \phi_Q, 1} + \lambda_{\phi_L, \phi_Q, 2} &> -2 \sqrt{\lambda_{\phi_Q} \lambda_{\phi_L}} \nonumber \\
 \lambda_{\phi_L, H, 1} + \lambda_{\phi_L, H, 2} &> -2 \sqrt{\lambda_{H} \lambda_{\phi_L}} \nonumber \\
 \lambda_{\phi_Q, H, 1} + \lambda_{\phi_Q, H, 2} &> -2 \sqrt{\lambda_{\phi_Q} \lambda_{H}} \nonumber \\
 \left|\lambda_{\phi_{L},\phi_Q,3} \right| &< \left|\lambda_{\alpha} \right|, \left|\lambda_{\alpha, \beta, i} \right| \, 
\end{align}
where $\alpha \in [\phi_Q, \phi_L, H]$  and $i \in[1,2,3]$.
Since we assume mass degeneracy within the non-DM dark sector, the Higgs portal couplings $\lambda_{\phi_L, H, 2}$ and $\lambda_{\phi_Q, H, 2}$ vanish. Another important feature of the model is the DM stabilizing symmetry $\mathtt{Z}_2$, under which all dark sector particles are oddly charged. For this symmetry to be intact, we require the other Higgs portal couplings to satisfy
\begin{align}
 \lambda_{\phi_L, H, 1}, \lambda_{\phi_Q, H, 1} < \frac{2}{v^2} \kappa^2 M^2_{\text{DM}}
\end{align}

\subsubsection{b-type Models}
In addition to $H$, there exists another scalar doublet $\phi$ in b-type models. The most general form (again respecting the DM stabilizing symmetry) of the scalar potential in this scenario is therefore
\begin{align}
\begin{split}
	V_\text{scalar}=& \mu_H^2 H^\dagger H + \mu_\phi^2 \phi^\dagger \phi + \lambda_H \left(H^\dagger H \right)^2 + \lambda_\phi \left( \phi^\dagger \phi \right)^2 \\
	 &+ \lambda_{\phi,H,1} \left( \phi^\dagger \phi \right)\left( H^\dagger H \right) + \lambda_{\phi,H,2} \left( \phi^\dagger H \right)\left( H^\dagger \phi \right) + \left[ \lambda_{\phi,H,3} \left( \phi^\dagger H \right)^2 + \text{h.c.}\right] \, .
\end{split}
\end{align}

For this kind of models we adopt the limits of \cite{Branco:2011iw,Lindner:2016kqk}, leading to
\begin{align}
	\lambda_{\phi}, \lambda_{H}  &> 0, \nonumber \\
	\lambda_{\phi,H,1} &> -2\sqrt{\lambda_\phi \lambda_H},  \nonumber \\
	\lambda_{\phi,H,1} + \lambda_{\phi,H,2} - |2\lambda_{\phi,H,3}| &> -2\sqrt{\lambda_\phi \lambda_H} \, .
\end{align}

As in a-type models, we can infer constraints for the Higgs portal couplings from mass degeneracy of the non-DM exotic particles and the requirement that the new scalar does not acquire a non-zero vev, yielding
\begin{align}
 \lambda_{\phi, H, 1} &< \frac{2}{v^2} \kappa^2 M^2_{DM} \\
 \lambda_{\phi, H, 2} &= 0
\end{align}

\section{Dark Matter Phenomenology} \label{sec:DMPheno}

In this section, we analyze whether the aforementioned models can simultaneously create the correct DM relic density $\Omega_{DM} h^2 = 0.120 \pm 0.001$ (at $68\%$ CL) observed by Planck \cite{Aghanim:2018eyx} and withstand the bounds set by DM direct detection (DD) experiments like XENON \cite{Aprile:2019dbj}.
To carry out our phenomenological analysis, we use \textsc{FeynRules} \cite{Alloul:2013bka,Christensen:2009jx} and its interface with \textsc{FeynArts} \cite{Hahn:2000kx} and \textsc{FormCalc} \cite{Hahn:1998yk} and/or \textsc{FeynCalc} to calculate contributions to direct detection cross sections up to one-loop level.  
We also use another \textsc{FeynRules} interface with \textsc{CalcHEP} \cite{Belyaev:2012qa} to compute direct detection cross sections with \textsc{micrOMEGAs}, which uses said \textsc{CalcHEP} output. We additionally use \textsc{micrOMEGAs} to numerically solve Boltzmann's equations \cite{Belanger:2013oya,Belanger:2018ccd} and obtain the relic density. Since the Yukawa coupling required to generate the observed $R_K$ is sizable, the DM production mechanism present in this model can only be the standard \textit{freeze-out} mechanism.    

\subsection{Relic Density}\label{sec:RD}
\begin{figure}
    \centering
    \includegraphics[width=8cm]{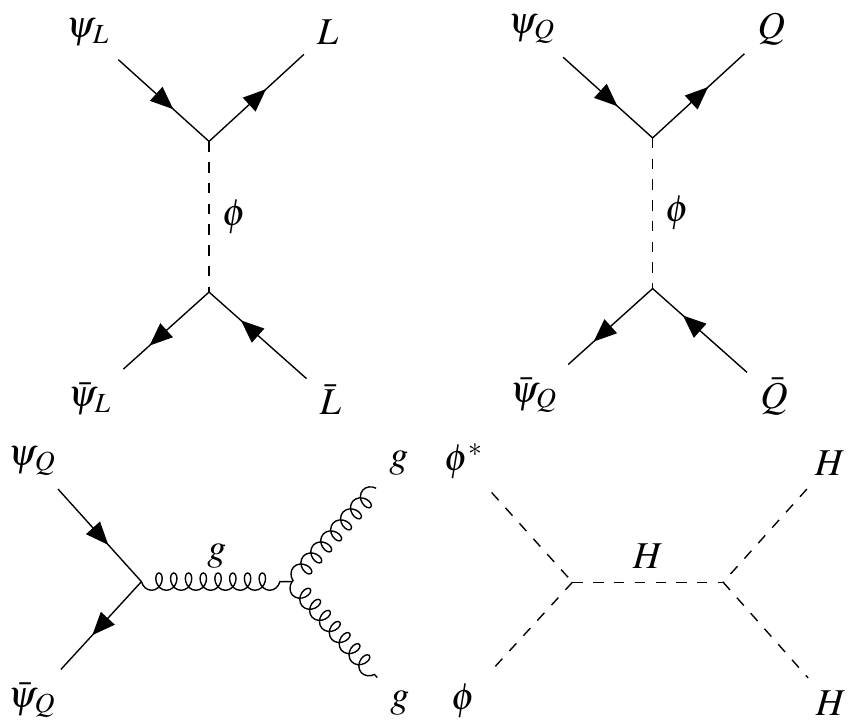}
    \caption{Examples of dark sector annihilations via Yukawa couplings (upper row), Higgs portal couplings and gauge couplings in leptophilic DM models.}
    \label{fig:annihilation}
\end{figure}

The relic density can be estimated by the effective annihilation cross section $\sigma_\text{eff}$ of DM. Depending on the parameter $\kappa$, which parametrizes the mass ratio of the various particles in the dark sector according to Eq.~\eqref{kappadefs}, $\sigma_\text{eff}$ is dominated solely by the direct annihilation of DM anti-DM pairs or can receive sizable contributions from annihilations of heavier dark sector particles, commonly referred to as coannihilations. The latter case can only arise if the masses in the dark sector are comparable, which is given for $\kappa \lesssim 1.2$. For larger values of $\kappa$, coannihilations are typically absent. Assuming that conversions among the different dark sector particles are efficient during freeze-out and that the heavier dark sector particles decay sufficiently fast, the coupled system of Boltzmann equations describing the time evolution of all dark sector particles can be reduced to a single effective Boltzmann equation for DM relic density \footnote{Note that \textsc{micrOMEGAs 5.0} does not validate that conversions between the dark sector particles are efficient but assumes them to be efficient enough to allow for the treatment described above. We numerically verified the efficiency of $2 \leftrightarrow 2$ conversion processes and $1 \leftrightarrow 2$ (inverse) decays between dark sector particles for coannihilating scenarios $\kappa \lesssim 1.2$ during freeze-out, i.e. $\Gamma_\text{conversion} > H$ for $T \gtrsim M_\text{DM}/30$. Reference \cite{Garny:2017rxs} discusses the efficiency of conversions in a slightly simpler but similar setup and find that effects of out-of equilibrium conversions only arise for couplings $\Gamma_i \lesssim 10^{-6}$, which is far below the couplings considered in this work.} \cite{Griest:1990kh,Edsjo:1997bg}
\begin{align}
\frac{\mathrm{d} \, \tilde{Y}_\text{DM}}{\mathrm{d} \, x} = -\sqrt{\frac{\pi}{45}} \frac{M_\text{Pl} M_\text{DM}g_{*,\text{eff}}^\frac{1}{2}}{x^2} \left\langle \sigma_\text{eff} v \right\rangle \left( \tilde{Y}^2 - \tilde{Y}_\text{eq}^2 \right) \, . \label{eq:effectiveBoltzmann}
\end{align}  
Here, we use $x=m_\text{DM} / T$ and $M_\text{Pl} = 1.2 \cdot 10^{19} \, \mathrm{GeV}$ is the Planck mass. The yield $\tilde{Y}$ is defined as the sum over all coannihilating particles
\begin{align}
\tilde{Y}^\text{eq} = \left\{\begin{array}{ll} \sum \limits_{i=\phi_L,\phi_Q,\psi} Y_i^\text{eq}, & \text{for a-type models} \\
         \sum \limits_{i=\psi_L,\psi_Q,\phi} Y_i^\text{eq}, & \text{for b-type models}\end{array}\right. ,
\end{align}
while the Yield $Y_i$ of the particle species $i$ is related to its number density $n_i$ via $Y_i = n_i / s$, with the entropy density $s$ of
\begin{align}
s = \frac{2 \pi^2}{45} g_{*s}T^3 \, .
\end{align}
The quantity $g_{*,\text{eff}}$ combines the entropy degrees of freedom $g_{*s}$ and the energy degrees of freedom $g_*$ via
\begin{align}
g_{*,\text{eff}}^\frac{1}{2} = \frac{g_{*s}}{\sqrt{g_*}} \left( 1 + \frac{T}{3 g_{*s}} \frac{\mathrm{d} \, g_{*s}}{\mathrm{d}\, T} \right) \, .
\end{align}
Since we assume mass-degenerate unstable dark sector particles with a mass of $ \kappa M_\text{DM}$ the yields can be expressed as
\begin{align}
\label{eqn:equilibrium}
Y_\text{DM}^\text{eq} &= \frac{90}{\left( 2 \pi \right)^\frac{7}{2}} \frac{g_\text{DM}}{g_{*s}} x^\frac{3}{2} \exp \left( -x \right) \, , \\
Y_\text{non-DM}^\text{eq} &= \frac{90}{\left( 2 \pi \right)^\frac{7}{2}} \frac{g_\text{non-DM}}{g_{*s}} \left( \kappa x \right)^\frac{3}{2} \exp \left( - \kappa x \right) \, .
\end{align}
Here, $g_\text{DM}$ and $g_\text{non-DM}$ refer to the degrees of freedom of the DM and the non-DM dark sector particles, respectively. For instance, in a b-type model with $\psi_L$ DM we have $g_\text{DM} = g_{\psi_L}$ and $g_\text{non-DM}=g_{\psi_Q} + g_\phi$. The effective annihilation cross section is given by a weighted sum of the thermally averaged cross sections of the various annihilation channels
\begin{align}
\left\langle \sigma_\text{eff} v \right\rangle = \sum_{i,j} \left\langle \sigma_{ij} v_{ij} \right\rangle \frac{Y_i^\text{eq}}{\tilde{Y}^\text{eq}} \frac{Y_j^\text{eq}}{\tilde{Y}^\text{eq}} \, , \label{eqn:sigmaveff}
\end{align} 
where $i$ and $j$ run over all dark sector particles and $\left\langle \sigma_{ij} v_{ij} \right\rangle$ is the thermally averaged  cross section for for the annihilation of the dark sector particles $i$ and $j$. \\
Due to the singlet nature of DM, the direct DM annihilation proceeds via the Yukawa couplings $\Gamma_{i}$ connecting DM to the SM leptons and/or quarks\footnote{For fermionic DM in an a-type model DM directly couples to both quarks and leptons, while $\psi_{L}$ ($\psi_Q$) DM couples to leptons (quarks) exclusively.}, while the heavier dark sector particles may be charged under the SM gauge groups and therefore can additionally annihilate via gauge interactions. Various annihilation channels are depicted in Figure \ref{fig:annihilation}. 
The relic density of a particle species\footnote{In case of Dirac DM the total relic density is given as the sum of the particle and antiparticle contribution.} can be estimated from the leading order velocity contribution of the annihilation cross section $\left\langle \sigma_\text{eff} v \right\rangle \sim \sigma_\text{eff,0} T^{n}$ \cite{Kolb:1990vq}
\begin{align}
\Omega_\text{DM} h^2 = \frac{\sqrt{g_*}}{g_{*s}} \frac{3.79 \left( n+1 \right) x_f^{n+1}}{M_\text{Pl} m_B \sigma_\text{eff,0}} \frac{\Omega_B}{Y_B} \, , \label{eq:DMEstimate}
\end{align}
where $x_f \sim 25$ parametrizes the freeze-out temperature, $m_B$ is the typical baryon mass and $\Omega_B$ and $Y_B$ describe the energy density and the yield of baryons today, respectively. Assuming that the annihilation cross section is dominated by the annihilations of dark matter into SM leptons and quarks of negligible mass\footnote{This scenario arises frequently in models with $\psi_L$, $\psi$-DM or for $\psi_Q$-DM significantly heavier than the top.} via the new Yukawa couplings, the leading order contribution of the thermally averaged cross section results in
\begin{align}
\label{eqn:CrossSectionAnnihi}
\sigma_\text{eff,0} &= \frac{1}{4 \pi M_\text{DM}^2}\frac{1}{\left( 1+ \kappa^2 \right)^2} \left[ C_l^m \Gamma_\mu^4 + 3 C_q^m \left( \Gamma_b^2 +\Gamma_s^2 \right)^2 \right] \times \nonumber \\
&\times \left\{\begin{array}{ll} \frac{1}{4}, & \text{for Dirac DM} \\
         \left( 1 + \kappa^4 \right) \left( 1+ \kappa^2 \right)^{-2}, & \text{for Majorana DM}\end{array}\right. \, .
\end{align}
Further, we find $n=0$($n=1$), corresponding to s-wave(p-wave) annihilation, for Dirac(Majorana) DM. The coefficients $C_{l/q}^m$ are model dependent and result in $C_l^\text{bIIB}=C_l^\text{bVIB}=C_q^\text{bIIA}=C_q^\text{bVA}=0$ and the remaining six coefficients are equal to $1$. \\  
Given the $\sigma_{\text{eff,}0}^{-1}$ scaling of the relic density in Eq.\@ \ref{eq:DMEstimate}, we expect to generate the observed relic density in such a scenario neglecting the logarithmic mass dependence of $x_f$ for 
\begin{align}
\left[ C_l^m \Gamma_\mu^4 + 3 C_q^m \left( \Gamma_b^2 +\Gamma_s^2 \right)^2 \right] &\approx 1.65 \cdot 10^{-7} \left( \frac{M_\text{DM}}{\text{GeV}} \right)^2 \nonumber \\ &\times \left\{\begin{array}{ll} \left( 1+ \kappa^2 \right)^{2}, & \text{for Dirac DM} \\
         5.6 \left( 1 + \kappa^4 \right)^{-1} \left( 1+ \kappa^2 \right)^{4}, & \text{for Majorana DM}\end{array}\right. \, . \label{eqn:correctRD}
\end{align}
In the following we derive some implications for limiting behaviors of this condition for the various models.  
\begin{enumerate}
\item Leptophilic DM (bIIA,bVA) \\ In this scenario Eq.\@ \eqref{eqn:correctRD} results in a scaling behavior $\Gamma_\mu = \mathcal{B}_{RD}^m \left( \kappa \right) \sqrt{\nicefrac{M_\text{DM}}{\text{GeV}}}$, with $\mathcal{B}_{RD}^m \left( \kappa \right)$ a model-dependent coefficient function. Comparing this result with the lower bound on $\Gamma_\mu$ for a successful explanation of the $R_K$ anomaly, given in Eq. \eqref{eq:GammaMuBound}, we find a lower bound on $\kappa>\kappa_0^m$ for a possible simultaneous explanation of $R_K$ and DM of
\begin{align}
\label{eqn:RDleptophilic}
\kappa_0^\text{bIIA} \approx 11.8 \, , \quad \kappa_0^\text{bIIA,Maj} \approx 4.7  \, , \quad \kappa_0^\text{bVA} \approx 26.5 \, , \quad \kappa_0^\text{bVA,Maj} \approx 10.9 
\end{align}  

\item Quarkphilic DM (bIIB,bVIB) \\ In this scenario the observed relic density is reproduced for $\left( \Gamma_s^2 + \Gamma_b^2  \right) \sim \mathcal{B}_{RD}^\text{m} \left( \kappa \right) \nicefrac{M_\text{DM}}{\text{GeV}}$. The product of $\Gamma_s$ and $\Gamma_b$ is constrained from above by $B-\bar{B}$-mixing, given in Eq.\@ \eqref{eq:GammaBSBound}. The annihilation cross section is given by the sum $\Gamma_s^2 + \Gamma_b^2$. Thus, it cannot be constrained from $B$-$\bar{B}$ mixing since either the second or third generation coupling could be arbitrarily small. However, we obtain an upper bound on the annihilation cross section by setting $\Gamma_s \sim 0$ and $\Gamma_b = 4 \pi$ at its perturbative limit \footnote{Note that the second generation coupling $\Gamma_s$ cannot be close to its perturbative limit, as it is constrained from $D$-$\bar{D}$ mixing.}. We find $\nicefrac{M_\text{DM}}{\text{GeV}} \leq \left[ \mathcal{B}_\text{RD}^\text{m} \left( \kappa \right) \right]^{-1} 16 \pi^2$. For the four different configurations we find
\begin{align}
M_\text{DM,max}^\text{bIIB}=M_\text{DM,max}^\text{bVIB} \approx \frac{674 \, \mathrm{TeV}}{1+ \kappa^2} \, , \quad M_\text{DM,max}^\text{bIIB,Maj} = M_\text{DM,max}^\text{bVIB,Maj} \approx \frac{254 \, \mathrm{TeV} \sqrt{1+\kappa^4}}{\left( 1+ \kappa^2 \right)^2}
\end{align}

\item Amphiphilic DM (aIA) \\ Since both, couplings to leptons an quarks, are present in this scenario, we can apply the limits derived in the two scenarios above if $\Gamma_\mu \gg \Gamma_q$ or $\Gamma_q \gg \Gamma_\mu$.   
\end{enumerate} 
\FloatBarrier
\subsection{Direct Detection} \label{sec:DD}
In this section we discuss the direct detection limits arising for the exemplary case of a fermionic singlet DM particle $\chi$ interacting with the SM quarks via a vector~($V_{\mu}$) or scalar~($A$) mediator.  
Dark matter direct detection (DD) is a means of experimental measurement of DM properties via the attempted observation of recoil of nuclei from scattering with DM particles. This scattering is assumed to occur at low momentum transfer, as the typical relative velocity of the Earth and DM lies at $v \sim 10^{-3}$.
As experiments like for instance XENON1T, LUX, PICO and also IceCube put bounds on the DM-nucleon cross section, we shortly discuss the estimates of the spin-independent (SI) as well as spin-dependent (SD) parts of this quantity for the different diagram structures present in this work. For this discussion, we mainly use the results of \cite{Berlin:2014tja},\cite{Agrawal:2010fh} and \cite{Mohan:2019zrk}. We put the main focus on dependence of the cross sections on the DM mass and the couplings.

An important note is that the bounds presented by the aforementioned experiments are given under the assumption that the total DM relic density is constituted of the particle in question. In the case where a model underproduces the relic density of a DM candidate, the DD bounds have to be rescaled such that only the produced fraction of the observed relic density $\Omega_{\text{DM}}h^2$ is regarded.
Thus, the relation 
\begin{align}
 \label{eqn:BoundRescaling}
 \sigma_{\text{DD}}  \leq \frac{\Omega_{\text{DM}}}{\Omega_\chi} \text{Bound}(m_\chi) 
\end{align}
must hold for a model to be in accordance with DD bounds. Bounds for underproduced DM are relaxed, as the allowed DM-nucleon cross section rises by a factor $\nicefrac{\Omega_{\text{DM}}}{\Omega_\chi}>1$.
If the RD is in turn overprocuded, the bounds are tightened in our setup. Note however, that this part of the parameter space is already excluded by the RD and therefore DD results in these regions are of no special interest.

\begin{figure}
 \centering
 \includegraphics[width=\textwidth]{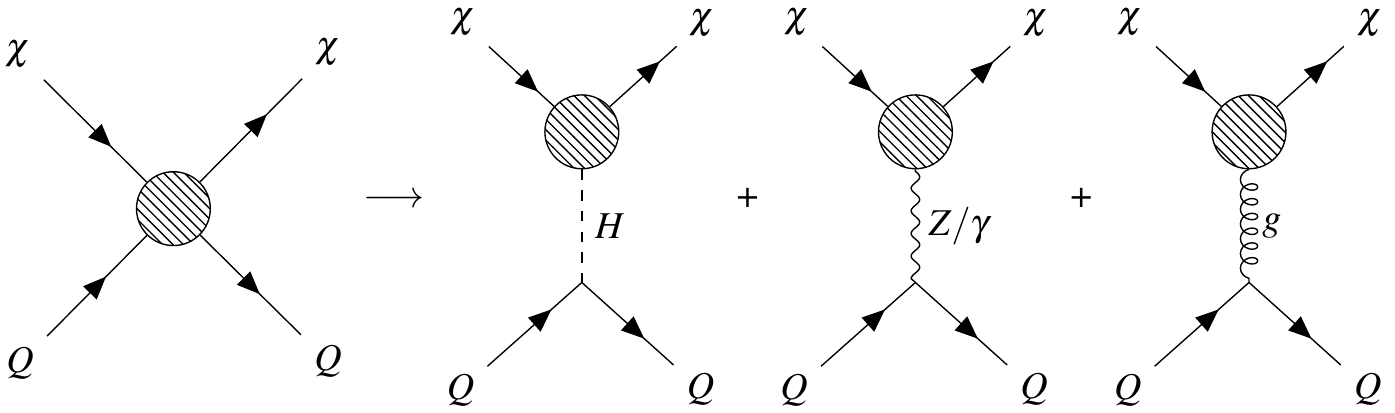}
 \caption{Several $t$-channel diagrams contributing to the DM-nucleon cross section relevant for DD.}
 \label{fig:tChannelDD}
\end{figure}

We estimate the results for $t$-channel DD interactions\footnote{Note that $t$-channel DD typically implies $s$-channel annihilation into quarks and vice versa.} as shown in Figure \ref{fig:tChannelDD}, where an EFT reduction is made, which is eventually matched to the operator structure in Eq.\@ \eqref{eq:Effectivetchannel}. If the DM-quark operator structure differs from this arrangement, as for instance in an $s$-channel diagram, this operator structure must be translated to the $t$-channel structure via Fierz transformations.
This is done because for the $t$-channel operators, DM and SM particle content strictly factorizes. \\
In this work, we evaluate the interaction within the blobs in Fig. \ref{fig:tChannelDD} up to one-loop level with \textsc{FeynArts}. The effective $\bar{\chi} \chi H/Z/\gamma/g$ - vertices are then implemented in the \textsc{CalcHEP}-files, which are used by \textsc{micrOMEGAs} to calculate the DM-nucleon cross section.

Typical $t$-channel operators with bosonic mediators contributing to DD are structured as 
\begin{align}
\mathcal{L}_{\text{DD}} &\supset \mathcal{O}_{\text{DM}}\cdot \Pi_{\text{med}} \cdot  \mathcal{O}_{q} \label{eq:Effectivetchannel} \\
&\stackrel{q^2 \ll m_{\text{med}}^2}{\to} \left(\frac{1}{m_{\text{med}}^2} \mathcal{O}_{\text{DM}}\right)  \mathcal{O}_{N}  \, ,
\end{align}
where the $ \Pi_{\text{med}}$ is the propagator of the mediator and $ \mathcal{O}_{N}$ denotes the nucleon-level operator. As we are interested in the nucleon rather than the partons, a summation over the quark content is necessary. This procedure, even up to the level of the whole nucleus, is explained in detail in the appendix of \cite{Berlin:2014tja} and references therein.

\subsubsection{Fermionic DM $\chi$ and Scalar Mediator $A$} \label{sec:DDscalar}
In the fermionic DM and scalar mediator case, the Lagrangian yields
\begin{align}
 \mathcal{L} \supset \left[ \left(\frac{1}{2}\right) \bar{\chi} (\lambda_{\chi s} + \lambda_{\chi p} \mathrm{i}\gamma^5) \chi + \bar{q} (\lambda_{q s} + \lambda_{q p} \mathrm{i}\gamma^5) q \right] A \, ,
\end{align}
where the factor of $\frac{1}{2}$ enters in the Majorana case. In this type of interaction, the combinations scalar-scalar ($s,s$),scalar-pseudoscalar ($s,p$), pseudoscalar-scalar ($p,s$) and pseudoscalar-pseudoscalar ($p,p$) between the DM and the nucleon operators $\mathcal{O}_{\text{DM}}$ and $\mathcal{O}_{N}$ arise. The DM-nucleon cross sections for each combination scale like 
\begin{align}
\label{eqn:sigmaScaMed}
	\begin{split}
 \sigma^{\text{SI}}_{s,s} &\sim \frac{\mu_{\chi N'}^2 \lambda_{\chi s}^2}{m^4_A}  \\
 \sigma^{\text{SI}}_{p,s} &\sim \frac{\mu_{\chi N'}^2 v^2}{m^2_\chi} \frac{\mu_{\chi N'}^2 \lambda_{\chi p}^2}{m^4_A} \\
 \sigma^{\text{SD},p/n}_{s,p} &\sim \frac{\mu_{\chi N'}^2 v^2}{m^2_{N'}} \frac{\mu_{\chi N'}^2 \lambda_{\chi s}^2}{m^4_A} \\
 \sigma^{\text{SD}, p/n}_{p,p} &\sim \left( \frac{\mu_{\chi N'}^2 v^2}{m_\chi m_{N'}} \right)^2 \frac{\mu_{\chi N'}^2 \lambda_{\chi p}^2}{m^4_A} \, ,
 \end{split}
\end{align}
where $\mu_{\chi N'}= \nicefrac{m_\chi m_{N'}}{m_\chi + m_{N'}}$ denotes the reduced mass of DM $\chi$ and the nucleus $N'$.
Interactions that are scalar on the nucleon operator contribute to the SI cross section, whereas SD contributions arise from interactions that are pseudoscalar on the nucleon operator side.
Note here especially that the ($s,s$) interaction is the only interaction unsuppressed by the relative velocity $v$. The mixed terms ($s,p$) and ($p,s$) are suppressed by two powers of $v$, while the ($p,p$) interaction features a suppression of $v^4$.
In this work, those interactions are mainly SM-Higgs-exchange induced.

\subsubsection{Fermionic DM $\chi$ and Vector Mediator $V_\mu$} \label{sec:DDvector}
In the fermionic DM and vector-boson mediator case, the Lagrangian reads 
\begin{align}
 \mathcal{L} \supset \left[ \left(\frac{1}{2}\right) \bar{\chi} \gamma^\mu(g_{\chi v} + g_{\chi a} \gamma^5) \chi + \bar{q} \gamma^\mu (g_{q v} + g_{q a} \gamma^5) q \right] V_\mu \, ,
\end{align}
where the factor $\frac{1}{2}$ enters only in the Majorana case. A special caveat here is that $g_{\chi v}$ \textit{vanishes} in the Majorana case, since terms of the form $\bar{\chi} \gamma^\mu \chi$ are forbidden.
For the possible interactions, vector-vector ($v,v$), vector-axial vector ($v,a$), axial vector-vector ($a,v$) and axial vector-axial vector ($a,a$), we obtain 
\begin{align}
\label{eqn:sigmaVecMed}
\begin{split}
 \sigma^{\text{SI}}_{v,v} &\sim \frac{\mu_{\chi N'}^2 g_{\chi v}^2}{m^4_V} \\
 \sigma^{\text{SI}}_{a,v} &\sim \frac{\mu^2_{\chi N'} v^2}{m_\chi^2} \frac{\mu_{\chi N'}^2}{\mu_{\chi N}^2} \frac{\mu_{\chi N'}^2 g_{\chi a}^2}{m^4_V} \\
 \sigma^{\text{SD},p/n}_{v,a} &\sim \frac{\mu^2_{\chi N'}}{\mu^2_{\chi N}} \frac{g^2_{\chi v} v^2}{m_V^4}  \\
 \sigma^{\text{SD},p/n}_{a,a} &\sim \frac{\mu_{\chi N'}^2 g^2_{\chi a}}{ m_V^4} \, .
 \end{split}
\end{align}

Analogously to the scalar mediator case, the nucleon current containing $\gamma^5$, the axial vector current, leads to SD contributions, while the SI contribution stems from the vector current.
The velocity suppression in these configurations is, however, not quite analogous. The ($v,v$) \textit{and} ($a,a$) interactions are unsuppressed, while the mixed terms ($a,v$) and ($v,a$) are suppressed by two powers of $v$. Therefore, there does exists a contribution to the SD cross section unsuppressed by velocity.

\subsubsection{$s$-channel Lagrangians and Fierz Identities} \label{sec:Fierz}
In the case of $s$-channel-type direct detection, it is possible to relate the Lagrangian to the $t$-channel interactions whose cross sections we presented in Section \ref{sec:DDscalar} and \ref{sec:DDvector} via Fierz identities. The $s$-channel type DD interactions that are typically present in quarkphilic models are mediated by a scalar. Starting from the structure of the Lagrangian 
\begin{align}
  \mathcal{L} \supset - \bar{\chi} (\lambda_{s}- \lambda_{p} \gamma^5) q \, A - \bar{q} (\lambda^*_{s}+ \lambda^*_{p}\gamma^5) \chi A^\dagger \, ,
\end{align}
we obtain the effective Lagrangian \cite{Agrawal:2010fh}
\begin{align}
  \mathcal{L}_{\text{eff}} \supsetsim \frac{1}{m_A^2} \left[ |\lambda_{s}|^2 (\bar{\chi} q)(\bar{q}\chi ) -|\lambda_p|^2 (\bar{\chi}\gamma^5 q)(\bar{q}\gamma^5\chi) + \lambda_s \lambda_p^* (\bar{\chi}q)(\bar{q}\gamma^5 \chi) - \lambda_s^* \lambda_p (\bar{\chi}\gamma^5 q)(\bar{q}\chi) \right]  
\end{align}
by integrating out the mediator. Further manipulation with Fierz transformations yields
\begin{align}
\label{eqn:LeffFierz}
\begin{split}
  \mathcal{L}_{\text{eff}} &\supsetsim \frac{1}{4m_A^2} \left[ (|\lambda_s|^2 - |\lambda_p|^2) (\bar{q}q\bar{\chi}\chi + \frac{1}{2}\bar{q}\sigma^{\mu \nu}q \bar{\chi}\sigma_{\mu \nu}\chi) \right. \\ &\left. + (|\lambda_s|^2 + |\lambda_p|^2) (\bar{q}\gamma^\mu q \bar{\chi}\gamma_\mu \chi - \bar{q}\gamma^\mu \gamma^5 q \bar{\chi}\gamma_\mu \gamma^5 \chi ) \vphantom{\frac{1}{2}} \right] \, ,
  \end{split}
\end{align}
where velocity suppressed combinations are neglected. 
Eq. \eqref{eqn:LeffFierz} suggests that in the limit $|\lambda_s|=|\lambda_p|$, the scalar-scalar and tensor-tensor operators vanish completely.
The remaining quadrilinears can be related to the scaling of the cross sections presented in Sections \ref{sec:DDvector} and \ref{sec:DDscalar}.
Note that both vector-vector and tensor-tensor contributions vanish in the Majorana case.

\subsubsection{Twist-$2$ Contributions to the SI DM-Nucleon Cross Section} \label{sec:twist2}
\begin{figure}
	\centering
	\includegraphics[width=0.7\textwidth]{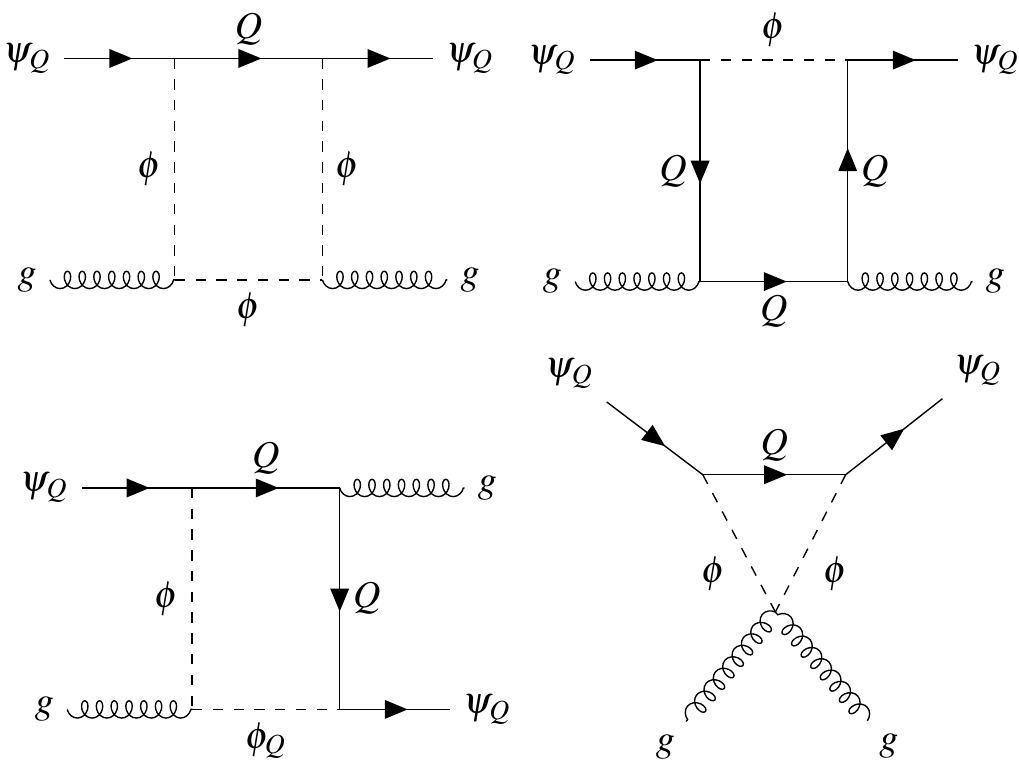}
	\caption{Leading order diagrams for DM-gluon scattering for quarkphilic DM.}
	\label{fig:DMGluonScattering}
\end{figure}

In some Majorana versions of models presented in this work, there is no unsuppressed contribution to the SI cross section from quark operators at leading order. In this case, twist-2 operators become important. While the twist-2 quark operators come from a higher order propagator expansion of the tree-level quark diagrams, also gluonic twist-2 operators generated by the box-diagrams shown in Fig. \ref{fig:DMGluonScattering} have to be considered. These diagrams are taken into account by \textsc{micrOMEGAs} automatically. 
In this case, the SI cross sections receive additional contributions from these diagrams, which scale like \cite{Mohan:2019zrk}

\begin{align}
\label{eqn:sigmaTwist2}
 \sigma^{\text{SI}} &\sim \left( \frac{m_\chi m_N}{m_{\text{med}} + m_N} \right)^2 \left| -\frac{8\pi}{9\alpha_s} 0.8 f_G + \frac{3}{4} 0.416 \left( g_G^{(1)} + g_G^{(2)} \right) \right|^2 \, , \nonumber \\
 f_G &\approx \frac{\alpha_s \Gamma_{s/b}^2 m_\chi}{192\pi} \frac{m_\chi^2 - 2 m_{\text{med}}^2 }{m_{\text{med}}^2 \left(m_{\text{med}}^2 -m_\chi^2 \right)^2} \, ,  \nonumber \\
 g_G^{(1)} &\approx \frac{\alpha_s \Gamma_{s/b}^2 }{96\pi m_\chi^3 (m_{\text{med}}^2 -m_\chi^2)} \left[-2m_\chi^4 \log{\left( \frac{m_q^2}{m_{\text{med}}^2} \right)} - m_\chi^2 (m_{\text{med}}^2 + 3m_\chi^2 ) \right. \\ 
 & \left. + (m_{\text{med}}^2 - 3m_\chi^2)(m_{\text{med}}^2 + m_\chi^2) \log{\left( \frac{m_{\text{med}}^2}{m_{\text{med}}^2 - m_\chi^2} \right)} \right] \, ,  \nonumber \\
 g_G^{(2)} &\approx \alpha_s \Gamma_{s/b}^2 \frac{-2 m_{\text{med}}^2 m_\chi^2 + 2(m_{\text{med}}^2 - m_\chi^2)^2 \log{\left(\frac{m_{\text{med}}^2}{m_{\text{med}}^2 - m_\chi^2} \right) +3m_\chi^4 }}{48\pi m_\chi^3 (m_{\text{med}}^2 - m_\chi^2)^2} \, . \nonumber
\end{align}

We will refer to these analytic expressions in Section \ref{sec:Results}, where we present the numerical results, as they play an important role in the SI direct detection procedure in quarkphilic DM Majorana models. Note that in the expressions above, $m_\text{med}$ corresponds to the mass of the colored particle directly coupling to DM.

\subsection{Indirect Detection} \label{sec:IndirectDetection}
In this section we shortly discuss indirect detection limits for the model classes presented in this work. 
Majorana type models are typically unconstrained by standard indirect searches due to the kinematic suppression of the pair annihilation cross section $\sigma(\bar{\chi} \chi \to \bar{f}f) \sim v^2$. However, this suppression is lifted for dark matter annihilations into a three-body final state containing a fermion pair and a photon $\bar{\chi} \chi \to \bar{f}f \gamma$ \cite{Bringmann:2007nk}. Such processes, commonly called Virtual Internal Bremsstrahlung (VIB) lead to a special spectral feature that is distinct from astrophysical backgrounds. Analyses searching for these features by FermiLAT \cite{Bringmann:2012vr} and HESS \cite{Abramowski:2013ax} can be used constrain models that allow a DM coupling to light fermions \cite{Garny:2013ama,Belanger:2015nma}, such as the model classes discussed in this article.

We calculate the VIB contribution to the annihilation cross section using \textsc{micrOMEGAs} 5.0 and compare the results to the the 95\% C.L. upper limits on the annihilation cross section $\langle \sigma v \rangle_{\bar{f}f\gamma} + 2\langle \sigma v \rangle_{\gamma \gamma}$ for $\kappa=1.1,1.01$ obtained by \cite{Garny:2013ama}. As in the case of DD, we rescale the experimental constraints with a fraction of the observed relic density so that
 \begin{align}
 \label{eqn:BoundRescaling2}
 \sigma_{\text{ID}}  \leq \left(\frac{\Omega_{\text{DM}}}{\Omega_\chi}\right)^2 \text{Bound}(m_\chi) \, .
\end{align}
Note that for $\kappa < 2$, the contribution of the two-photon final state is negligible. Moreover, for large $\kappa$ VIB is suppressed by $\kappa^{-4}$ and thus we do not present limits for these cases.    

Our analysis shows that even for coannihilating scenarios the parameter space of the models analyzed in this work \textit{cannot} be constrained by VIB. Principally, for large couplings to leptons (in the case of leptophilic/amphiphilic DM) or quarks (quarkphilic/amphiphilic) the DM annihilation cross section can be relatively large. Nevertheless, the corresponding points in the parameter space typically feature a small relic density, which results in relaxed ID constraints because of the quadratic rescaling given in Eq. \eqref{eqn:BoundRescaling2}. 

Since our numerical results indicate that the relevant parameter space for Dirac models is excluded by DD already, we do not apply the standard ID searches constraining the DM annihilation into a pair of SM fermions for these models.

\section{Results of the Numerical Analysis}  \label{sec:Results}

In this section we present the results of our numerical analysis. The section itself is further divided into separate discussion of the individual models. Since the discussed models all contain a singlet fermion DM candidate, both Dirac and Majorana versions of the model are presented. Since the parameter space of this model is vast and might vary from one model to the other, several assumptions are made throughout this analysis:
\begin{itemize}
	\item[•] We present all plots in the $M_{DM}$-$\Gamma_\mu$ plane. We scan logarithmically over $10^4$ parameter sets in the $[100\, \text{GeV},40\, \text{TeV}] \times [10^{-5},4 \pi]$ intervals. 
	\item[•] We restrict ourselves to parameter sets respecting the DM stabilizing symmetry and satisfying the vacuum stability conditions as well as perturbativity bounds (see Sections \ref{sec:YukConstraints} and \ref{sec:ScalarPotConstraints}). The bounds on $R_K$ and $(g-2)_\mu$ , however, are not regarded as stringent bounds on our parameter sets as we perform a scan over $\Gamma_\mu$. 
	\item[•] We keep the product $\Gamma_s \Gamma_b^*$ at the upper bound \footnote{If the bound on $\Gamma_\mu$ posed by $R_K$ is surpassed, $\Gamma_s \Gamma_b^*$ is scaled down  accordingly to still be able to solve the $R_K$ anomaly.}. This generally facilitates the realization of a solution to the $R_K$ anomaly, as smaller leptonic couplings $\Gamma_\mu$ suffice in this case. As only the product is fixed, we differentiate between two choices of coupling structures for the new quark Yukawa couplings: democratic and hierarchical. In the democratic scenario we choose the second and third generation couplings $\Gamma_s$ and $\Gamma_b$ to be almost identical \footnote{By almost identical we mean as close by as the bound on $\Gamma_s$ from $D$-$\bar{D}$ mixing allows. Typically, the difference is within $\mathcal{O}(1\%)$.}, meaning $\Gamma_s \approx \Gamma_b \approx \sqrt{\mathcal{B}^\text{model}_{bs}(\kappa) \cdot \nicefrac{M_{DM}}{\text{GeV}}}$.
	
In the hierarchical setup we impose a strong flavor structure. As $\Gamma_b$ is unconstrained, it is possible to set its value up to the perturbative limit. In the upcoming plots, we present the most extreme hierarchical case of $\Gamma_b = 4\pi \, \wedge \, \Gamma_s = \mathcal{B}^\text{model}_{bs}(\kappa) \cdot \nicefrac{M_{DM}}{4\pi \text{GeV}}$ to illustrate the possible impact of a non-democratic flavor structure.  
	\item[•] We fix all non-vanishing Higgs portal couplings to $0.1$, since they do not alter the DM phenomenology in a substantial manner. 
	\item[•] We present scenarios where strong coannihilation effects are present ($\kappa=1.01 , 1.1$) as well as scenarios where those are absent ($\kappa=5,15$) in each model.
\item[•] We assume mass degeneracy between the new non-dark matter particles $\phi_Q$ and $\phi_L$ in a-type models and $\psi_Q/\psi_L$ and $\phi$ in b-type models respectively. \\ In addition to the scenarios presented in this article, we have studied an examplatory case for each model class (lepto/quark/amphiphilic) allowing for a non-degenerate, non-DM dark sector. For quarkphilic and leptophilic scenarios, only coannihilating scenarios are affected by a non-degenerate, non-DM dark sector, while amphiphilic models can be affected in both setups. Adopting the point of view that the mass degenerate setup refers to the case of the smallest possible dark sector masses involved for a given $\kappa$, all regions of viable parameter space are shifted to smaller DM masses when considering a non-mass-degenerate, non-DM dark sector. While an analysis of the cases specified above could be interesting, a detailed study of this scenario is beyond the scope of this paper. 	
\end{itemize}

In all plots, the area allowed by direct detection (DD), $R_K$, and muon $g-2$ ($\Delta a_\mu$) are presented in colors blue, gray, and green respectively. 
 The lines belonging to DM relic density (RD) correspond to the observed relic density $\Omega_{DM}h^2=0.120\pm0.001$ in the universe. On the one hand, the parameter space to the left of these lines features underproduction of relic density, while on the other hand the region to the right of the line features overproduction of DM.


\FloatBarrier
\subsection{bIIA}
\FloatBarrier
Following Table \ref{tbl:SUclassification}, the representations of the dark sector particles in bIIA are
\begin{align}
 \psi_L = (\textbf{1},\textbf{1})_0, \, \psi_Q= (\textbf{3},\textbf{1})_{\nicefrac{2}{3}}, \, \phi=(\textbf{1},\textbf{2})_{-\nicefrac{1}{2}} \nonumber \\
 \stackrel{\text{EWSB}}{\Rightarrow} \psi_L \to \psi_L^0 , \, \, \psi_Q \to \psi_Q^{+\nicefrac{2}{3}} , \, \, \phi \to  \begin{pmatrix} \frac{1}{\sqrt{2}}\left(\phi^0 + \phi^{0'} \right) \\ \phi^{-} \end{pmatrix}
 \,.
\end{align}
As $\psi_L$ qualifies as the DM candidate, bIIA belongs to the class of \textit{leptophilic DM} models. 

\subsubsection{Dirac DM} \label{sec:bIIADirac}
\FloatBarrier
\begin{figure}
\centering
	\includegraphics[width=0.5\textwidth]{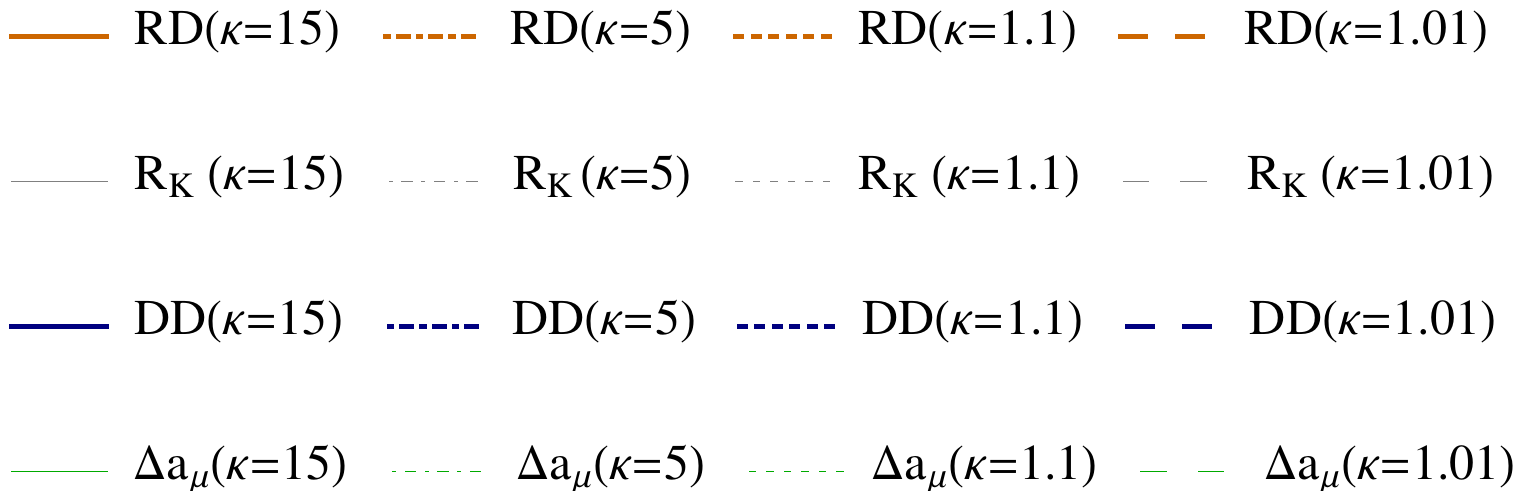} \\
	\subfigure[democratic]{
        \includegraphics[width=0.5\textwidth]{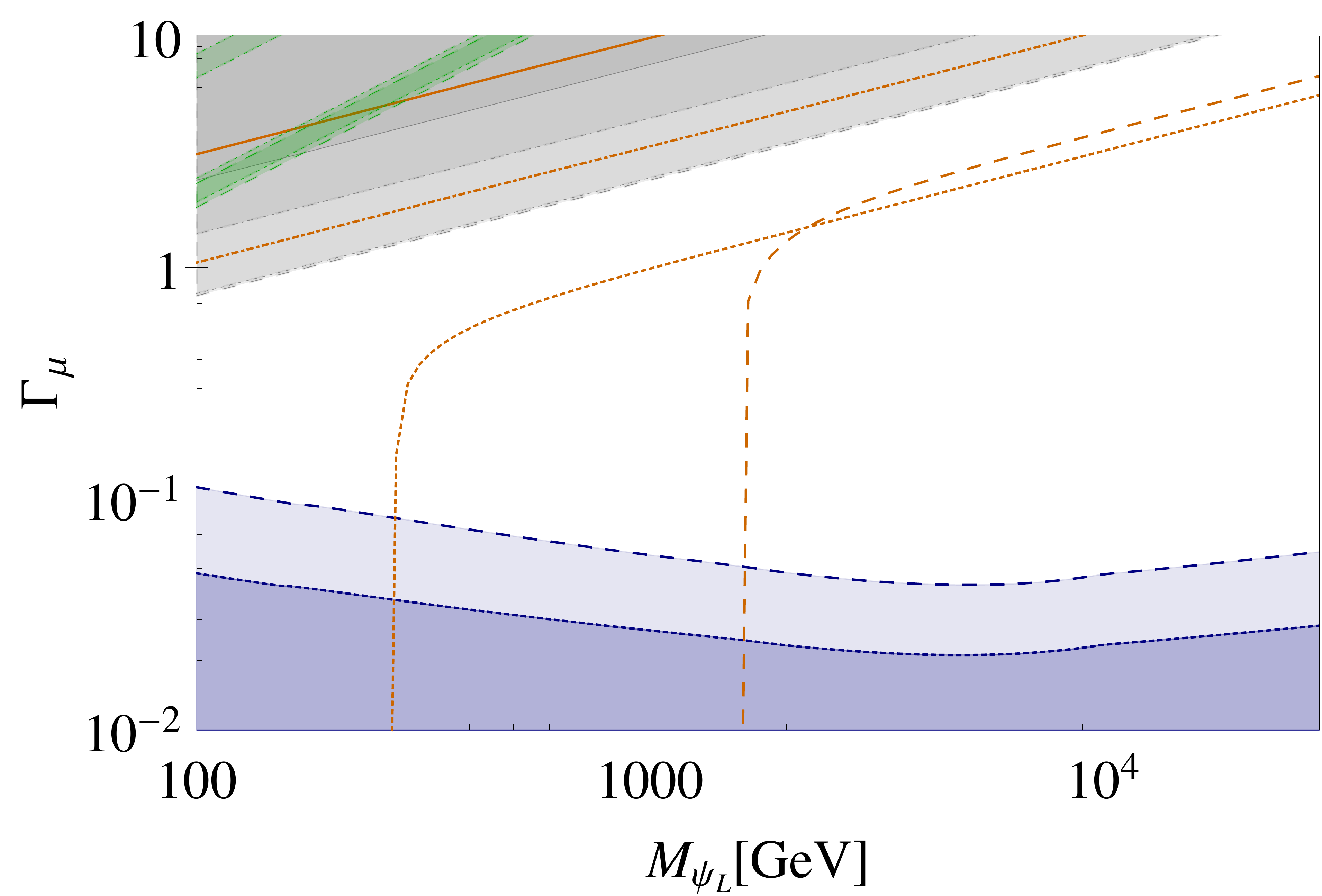}}%
	\subfigure[hierarchical]{
        \includegraphics[width=0.5\textwidth]{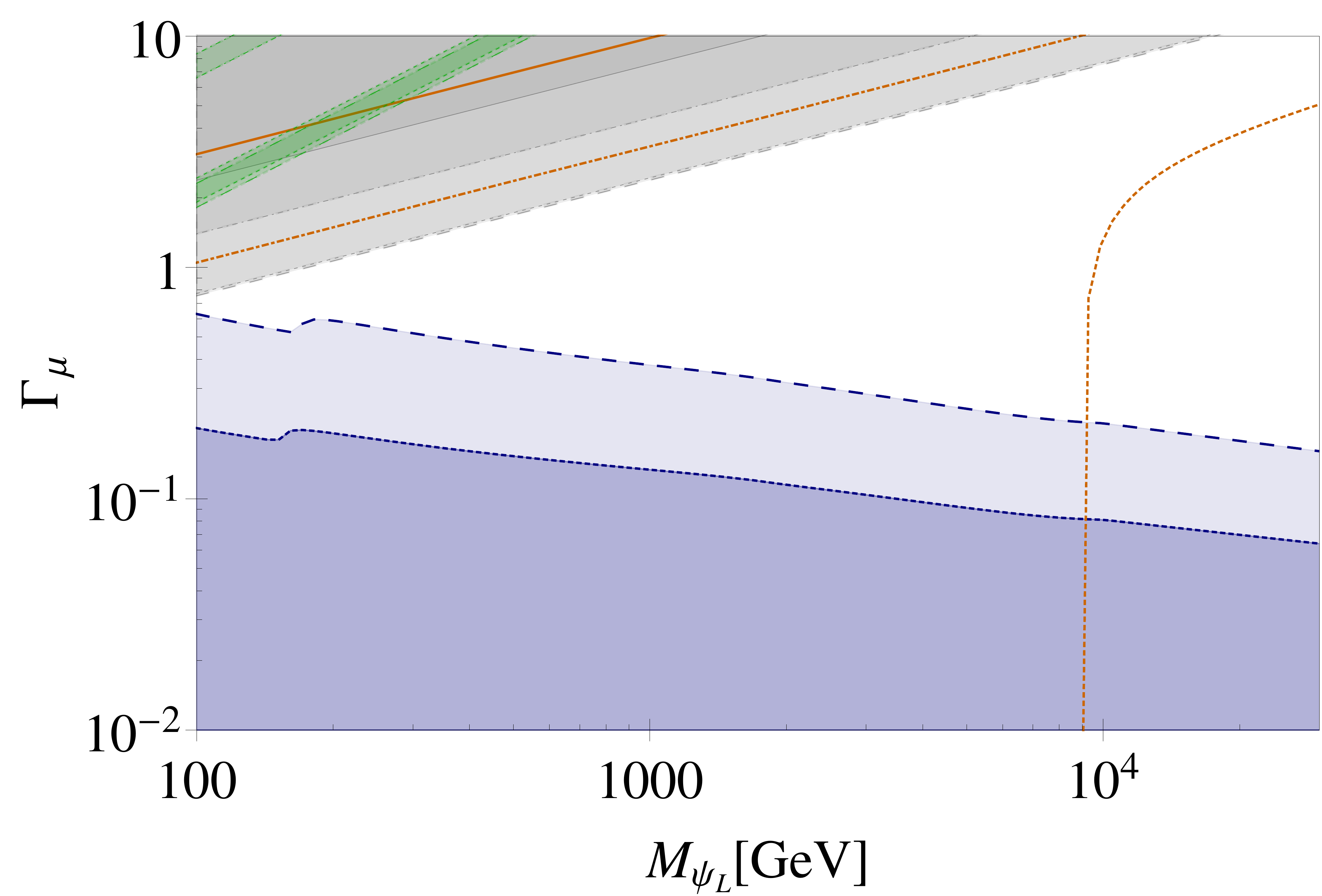}} 
    \caption{Summary plot for Dirac DM in bIIA. The plot provides the implications on the parameter space of the relic density (RD), the $R_K$ observable ($R_K$), direct detection (DD) and the anomalous magnetic moment of the muon ($\Delta a_\mu$) for $\kappa=1.01$ (dashed line), $\kappa=1.1$ (dotted line), $\kappa=5$ (dot-dashed line) and $\kappa=15$ (solid line). The orange lines give the parameters that generate the observed relic density. Parameter points to the left of those lines underproduce DM, while DM is overproduced to the right. The gray region displays the region where the $R_K$ anomaly can be addressed at $2 \sigma$. The blue region indicates the parameter space \emph{allowed} by direct detection experiments including both SI and SD constraints. Lastly, the green regions show areas where $\Delta a_\mu$ can be reproduced at $1 \sigma$. The area above the green bands lead to overly large contributions to $\Delta a_\mu$ and is therefore excluded.}
    \label{fig:SummarybIIADirac}
\end{figure}

The numerical results obtained for bIIA Dirac are summarized in Figure \ref{fig:SummarybIIADirac}.
Figure \ref{fig:SummarybIIADirac} indicates that of the depicted mass configurations only $\kappa=15$ principally allows for a solution of $R_K$, since the corresponding the RD line lies within the allowed $R_K$ region. Typically, the correct relic density is not just a line, but a small allowed band. However, as the relative experimental error of $\Omega_\text{DM} h^2$ measured by Planck is $\approx 0.8\%$ the allowed band is considerably tiny. Since it is a purely non-coannihilating scenario, the DM freeze-out is dominated by $\bar{\psi}_L \psi_L \to \bar{L} L$ annihilations and we can estimate the slope of the RD graph using Eq. \eqref{eqn:correctRD}, thus ultimately indicating a $\Gamma_\mu \sim \sqrt{\nicefrac{M_\text{DM}}{\text{GeV}}}$ dependence. The '$\Gamma_\mu$-intercept' or the 'height' of the RD line is thus determined by the individual $\mathcal{B}_{RD}^m(\kappa)$ of the model, $\mathcal{B}_{RD}^\text{bIIA}(\kappa)$ is this case.  
As shown in Eq. \eqref{eqn:RDleptophilic}, a common solution to the correct relic density and the $R_K$ anomaly can only exist if $\kappa \gtrsim 11.8$, while $\kappa \lesssim 11.8$ configurations on the other hand are factually excluded because of DM overproduction. 

Coannihilating scenarios are not always dominated by the aforementioned annihilations and therefore do not behave as plainly as their non-coannihilating counterparts in this model. As can be seen in Fig. \ref{fig:SummarybIIADirac}, there is a lower mass threshold on successful RD reconstruction stemming from annihilations of $\psi_Q$ via the strong gauge coupling or via the new quark Yukawa and also $\phi$ via the Higgs portal (see Figure \ref{fig:annihilation}). The threshold exists also in the case of vanishing Higgs portal coupling and the new quark Yukawa couplings $\Gamma_s$ and $\Gamma_b$, since the contribution from the strong gauge coupling $g_3$ always exists. Typically, Higgs portal interactions are dominated by the other two types of interaction and therefore we choose to fix the corresponding coupling $\lambda_{\phi, H, 1}$ during the entire scan.
The visible $\kappa$-dependence of the threshold is explained by the typical coannihilation suppression factor of $\sim \exp{(-2x_f \kappa)}$  (see Eqs. \eqref{eqn:equilibrium}-\eqref{eqn:sigmaveff}): As $\kappa$ increases, coannihilation channels become more and more suppressed and thus the thermally averaged total cross section shrinks, leading to an increased DM relic density. This way, the threshold shifts towards lower masses for increasing $\kappa$. 

Furthermore, the annihilation cross section into quarks scales like $\sim \Gamma_b^4 + \Gamma_b^2 \Gamma_s^2 +\Gamma_s^4$ at any given $\kappa$, which explains the shift of the threshold to higher masses in the hierarchical scenario, since this factor is maximized for large values of $\Gamma_b$ realized in the hierarchical coupling structure.
As the masses grow larger, coannihilating scenarios with the correct relic density become dominated by direct annihilations into leptons, as a larger $\Gamma_\mu$ is needed to achieve the correct thermally averaged cross section, leading to the same scaling behavior as in non-coannihilating scenarios.  

In the democratic setup, another effect can be observed in the coannihilating scenarios: The intersection of the RD lines of $\kappa = 1.1$ and $\kappa = 1.01$. This effect cannot be understood from annihilations described in Eq. \eqref{eqn:CrossSectionAnnihi} alone, but by taking into account effective conversions of the dark sector particles. We provide a general estimate of the thermally averaged cross section in \hyperref[sec:AppendixA]{Appendix A}. In there, we also shortly the discuss the $\kappa$ interval, where such intersections occur.

\begin{figure}
    \centering
	\includegraphics[width=0.7\textwidth]{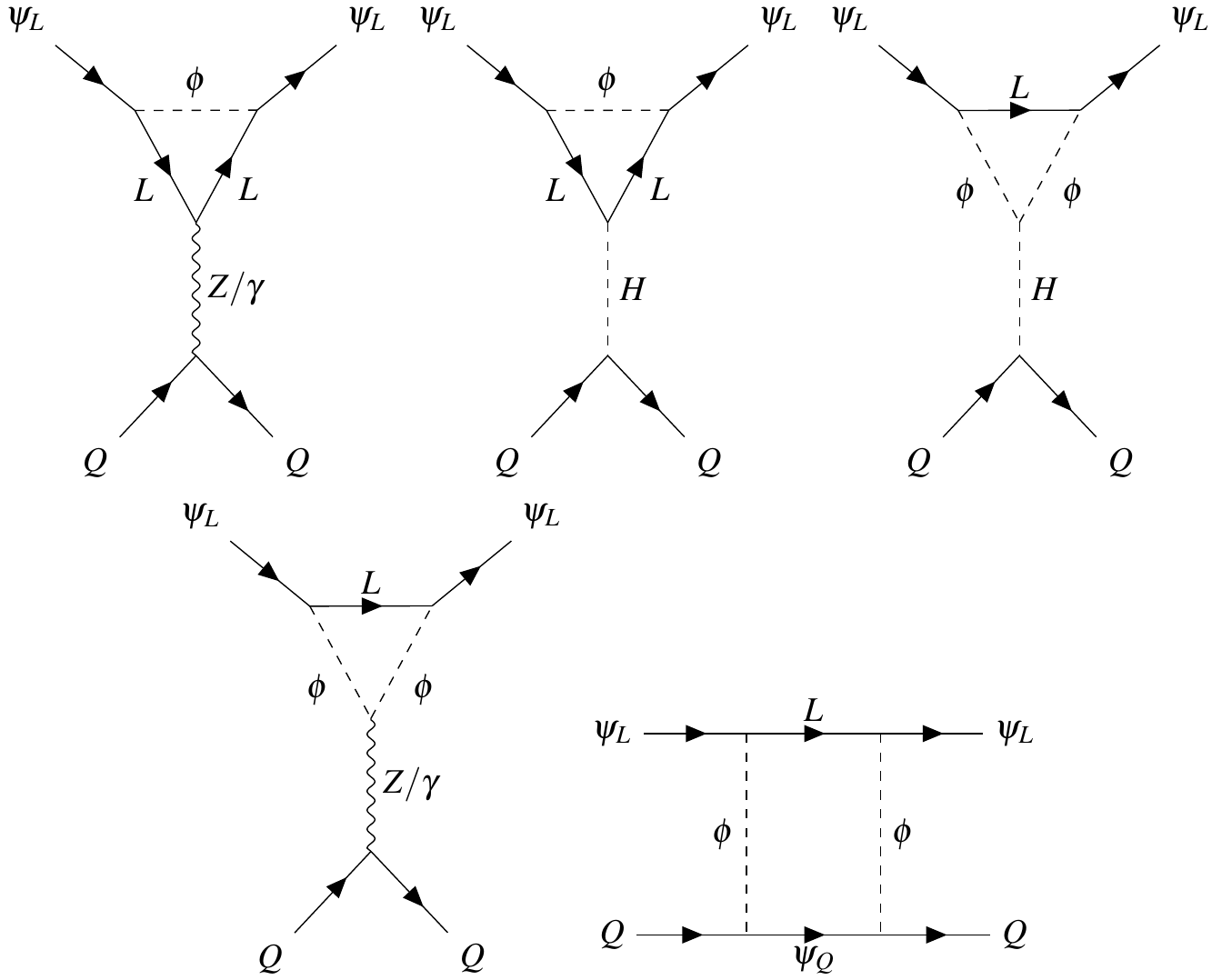}
    \caption{Leading order DM-quark diagrams contributing to the DD cross section of $\psi_L$ DM.}
    \label{fig:DiagramsDDPsiL}
\end{figure}

Direct detection is mediated at leading order at one-loop level in this model, as there are no tree-level vertices with quarks for leptophilic DM. The corresponding leading order diagrams are depicted in Figure \ref{fig:DiagramsDDPsiL}. It is important to note at this point that the box diagram exhibits one additional heavy dark sector particle propagator. Different from the penguin diagrams also, the final state quarks are of second or third generation, as $\Gamma_d=0$, which suppresses the diagram even further due to the reduced fractions of higher generation quark parton distribution functions (PDFs).  \\ \indent
Taking DD results into account, we can state that a solution of the $R_K$ anomaly is impossible in this setup, since the allowed regions for DD and $R_K$ are completely disjoint in both hierarchical and democratic implementation. It is important to stress that DD even excludes configurations where DM is underproduced. For this model to not be excluded, coannihilations are essential, since only those configurations feature non-overproduced RD within the allowed DD regions.
The tightness of the constraints present in this model is due to the strong effective vector coupling of $\psi_L$ to the $Z$-boson, which in turn lead to a strong constraint from the SI DM-nucleon cross section. This constraint is weaker in coannihilating scenarios by $1-2$ orders of magnitude, which stresses the tension between a solution to $R_K$ and DD even more.

In general, two competing effects are at work in this model: On the one hand, the effective $\bar{\psi}_L \psi_L Z$-coupling becomes smaller with increasing $\kappa$, since the masses of the dark sector particles within the loop increase accordingly, thus alleviating the bounds. An increasing $\kappa$ on the other hand generally increases $\Omega_\text{DM} h^2$ by lowering the thermally averaged cross section $\langle \sigma v \rangle$. As the bound posed by XENON is dependent on the fraction of $\psi_L$-DM compared in the observed DM density according to Eq. \eqref{eqn:BoundRescaling}, bounds are tightened by this rescaling. This interplay explains the intuitively unexpected flipping behavior of the DD bounds of $\kappa=15$ and $\kappa=5$, since the competing effects scale differently and thus a turning point of hierarchies is now expected. \\
\indent
A comparison between Fig. \ref{fig:SummarybIIADirac} (a) and (b) leads to the conclusion that the flavor structure of the new quark Yukawas only influences the DD bounds of coannihilating scenarios. This is due to the fact that in leptophilic DM models only the suppressed box diagram exhibits a $\Gamma_{s/b}$ dependence, while the relic density in turn is indeed affected by these couplings in coannihilating scenarios, as mentioned above. As a hierarchical flavor structure tends to deplete DM more severely, the fraction of $\psi_L$-DM shrinks and DD bounds are softened as a consequence. 
\\ \indent
As a last point, an explanation of the $g-2$ of the muon cannot be constructed in all versions of the bIIA Dirac model, as the allowed region and the DD allowed region are also disjoint. 

\subsubsection{Majorana DM}
\FloatBarrier
\begin{figure}
\centering
	\subfigure[democratic]{
        \includegraphics[width=0.5\textwidth]{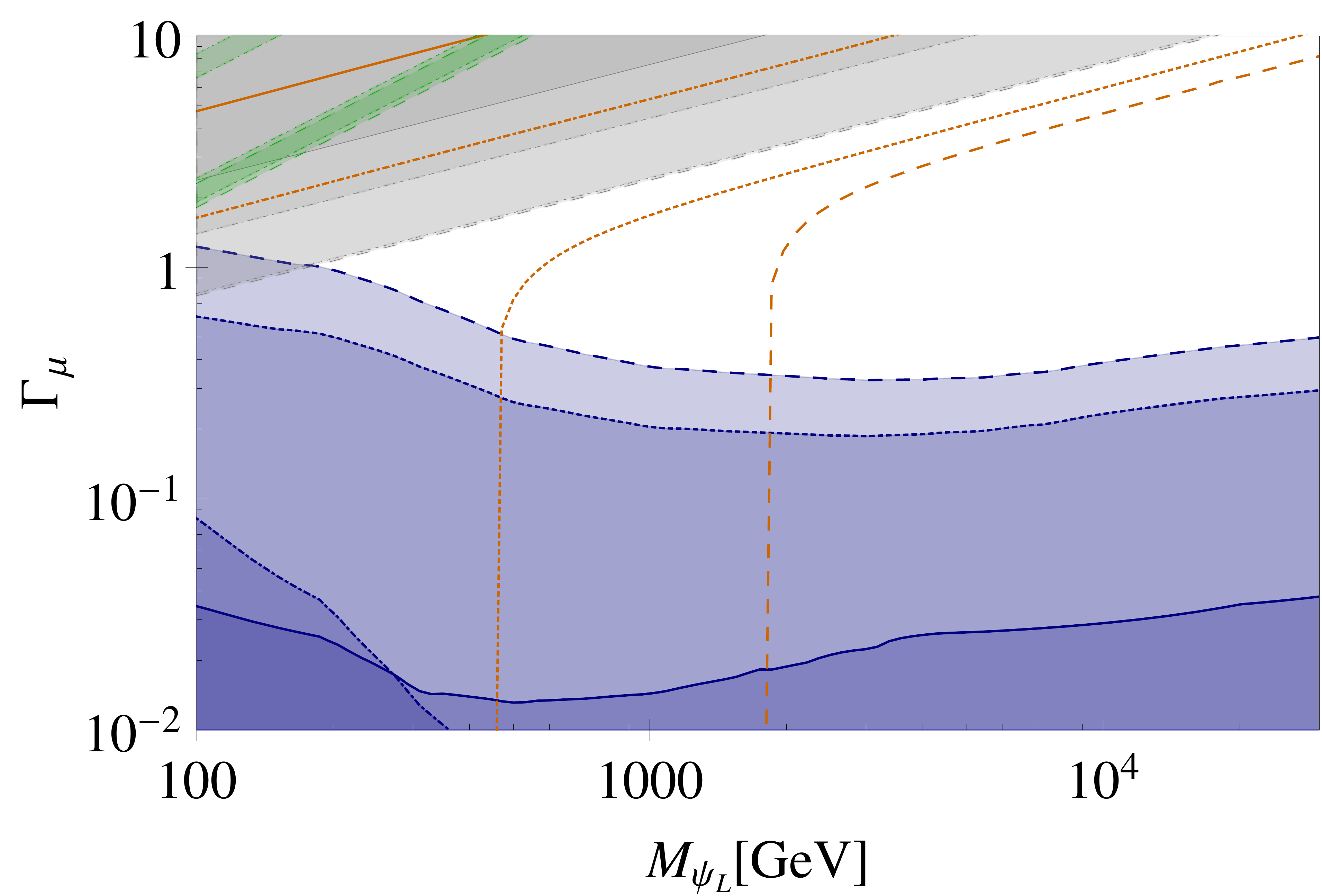}}%
	\subfigure[hierarchical]{
        \includegraphics[width=0.5\textwidth]{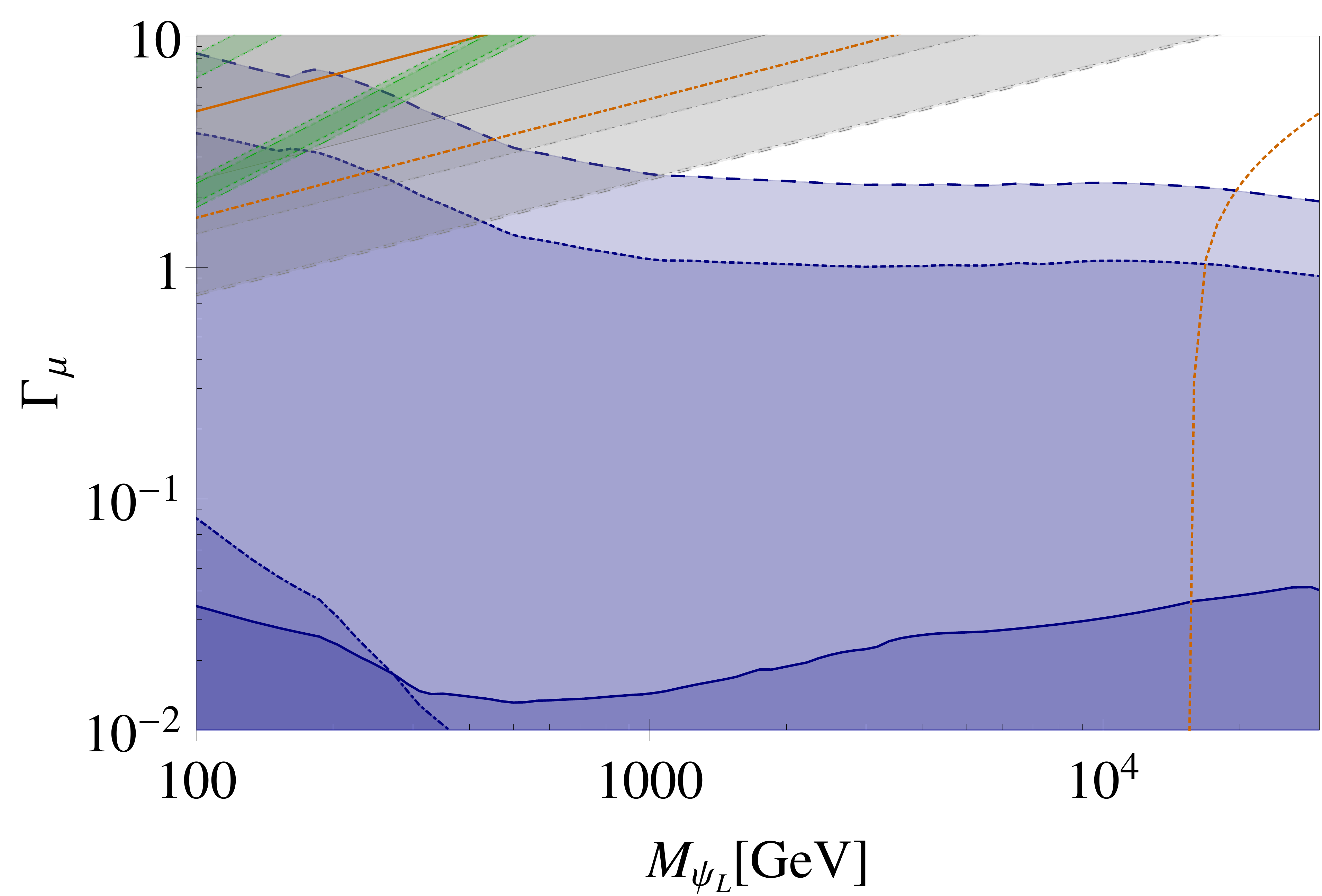}}%
    \caption{Summary plot for Majorana DM in bIIA. The legend and an explanation of the color scheme are given in Fig. \ref{fig:SummarybIIADirac}.}
    \label{fig:SummarybIIAMajorana}
\end{figure}

The Majorana version's results are presented in Figure \ref{fig:SummarybIIAMajorana}.
In comparison to the Dirac version, the most striking differences of the Majorana version are the DD bounds, which are weakened by about one order of magnitude in $\Gamma_\mu$. This is mainly due to the vanishing of vector current contribution to the $\bar{\psi}_L \psi_L Z$ vertex, which is forbidden in the Majorana case. In the absence of this contribution, the most stringent bounds do not come from the SI DD cross section, but from the SD DM-nucleon cross section generated by the axial vector current contribution to $\bar{\psi}_L \psi_L Z$ vertex, although the SD limits are $\sim 6$ orders of magnitude weaker. The bounds on SD DM-neutron cross section also come from XENON \cite{Aprile:2019dbj}, while IceCube puts the more stringent bounds on the DM-proton cross section \cite{Aartsen:2016zhm} in this mass range. The DM-neutron cross section tends to constrain the lower mass region up to $\sim 300\,$GeV, while higher masses are typically constrained by the DM-proton cross section bounds from IceCube. We always apply the strongest bound at any given mass\footnote{In the discussion of the upcoming models, the mass regions where the aforementioned experiments are most constraining vary but are not explicitly stated}. 

As the blue regions allowed by DD now overlap with the gray ($R_K$) and green ($(g-2)_\mu$) regions, valid solutions to $R_K$ or $(g-2)_\mu$ (though not simultaneous) exist. We state here, however, that these solutions do not solve the DM problem, as DM is strongly underproduced in this part of the parameter space. Such constellations thus require a multi-component DM solution \footnote{Note that the other DM component(s) must stem from a separate dark sector to the one presented in this work in this case.} that goes beyond this model. 

In the democratic setup, the overlap with the $R_K$ region exists in the mass region $\lesssim 180\,$GeV, while a slightly bigger window arises in the hierarchical setup with DM masses ranging up to $\sim 420\,$GeV for $\kappa=1.1$ and $\sim 1000\,$GeV for $\kappa=1.01$. For a valid $(g-2)_\mu$ solution a flavor hierarchy is beneficial, as the allowed mass range for a democratic case is up to less than $100\,$GeV, whereas in the hierarchical case we observe viable masses up to $\sim 170\,$GeV for $\kappa=1.1$ and even $\sim 290\,$GeV for $\kappa=1.01$. The mass regions mentioned above appear to be accessible at the LHC, since they feature a colored particle of a mass less than a TeV. We briefly review results of searches of similar setups in Section \ref{sec:collider}, which suggest such low-massive regions are excluded. A dedicated collider study of this model beyond the contents of Section \ref{sec:collider} could be interesting. 

Another feature of the Majorana variant is that $R_K$ and RD can be reconciled with lower $\kappa$ values, as Majorana DM annihilations are p-wave at leading order and thus suffer from velocity suppression. This in turn requires a larger $\Gamma_\mu$ to achieve to the observed relic density, ultimately pushing relic density lines of lower $\kappa$ into the gray area. Mathematically quantified, this effect leads to a new lower bound on $\kappa$ for a reconciliation of $R_K$ and RD of $\kappa \gtrsim 4.7$ as indicated in Eq.\eqref{eqn:RDleptophilic}. 

Furthermore, we note at this point that the relic density lines of the coannihilating scenarios do not intersect in this version of the model. The reason for this effect is an alteration of the degrees of freedom of DM because of the Majorana nature (see \hyperref[sec:AppendixA]{Appendix A} for more details).


\subsection{bVA}
\FloatBarrier
Following Table \ref{tbl:SUclassification}, the representations of the dark sector particles in bVA are
\begin{align}
 \psi_L = (\textbf{1},\textbf{1})_0, \, \psi_Q= (\textbf{3},\textbf{3})_{\nicefrac{2}{3}}, \, \phi=(\textbf{1},\textbf{2})_{-\nicefrac{1}{2}} \nonumber \\
 \stackrel{\text{EWSB}}{\Rightarrow} \psi_L \to \psi_L^0 , \, \, \psi_Q \to \begin{pmatrix} \psi_Q^{+\nicefrac{5}{3}} \\ \psi_Q^{+\nicefrac{2}{3}} \\ \psi_Q^{-\nicefrac{1}{3}} \end{pmatrix} , \, \, \phi \to  \begin{pmatrix} \frac{1}{\sqrt{2}}\left(\phi^0 + \phi^{0'} \right) \\ \phi^{-} \end{pmatrix}
 \, .
\end{align}
This model can be called the 'sibling' of bIIA, since the fields are almost in the same representations. The only difference is that $\psi_Q$ is an $SU(2)_L$-triplet rather than a singlet. 

\subsubsection{Dirac DM}\label{sec:bVADirac}
\FloatBarrier
\begin{figure}[H]
\centering
	\subfigure[democratic]{
        \includegraphics[width=0.5\textwidth]{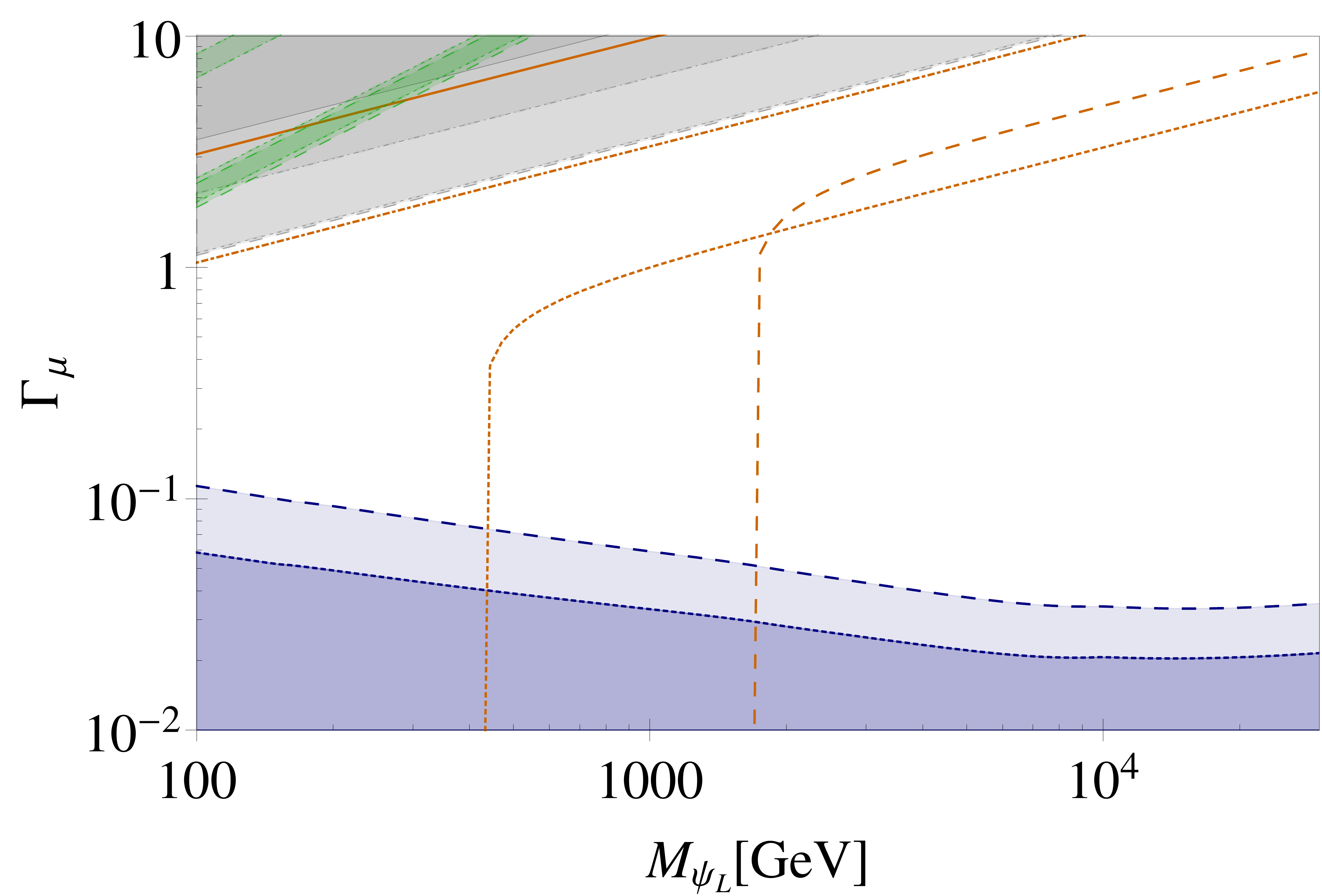}}%
	\subfigure[hierarchical]{
        \includegraphics[width=0.5\textwidth]{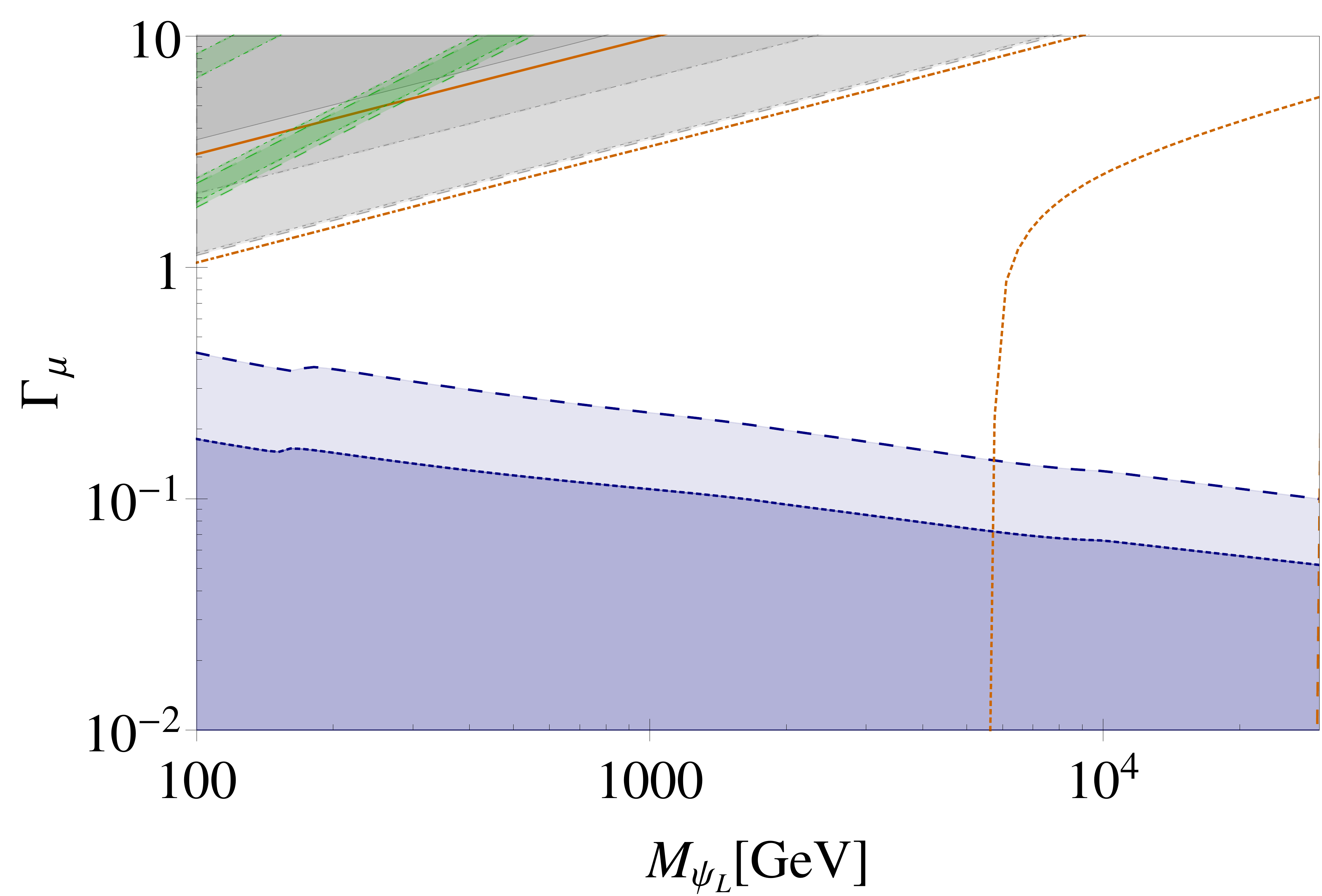}}%
    \caption{Summary plot for Dirac DM in bVA. The legend and an explanation of the color scheme are given in Fig. \ref{fig:SummarybIIADirac}.}
    \label{fig:SummarybVADirac}
\end{figure}

The numerical results of Dirac version of bVA are summarized in Figure \ref{fig:SummarybVADirac}. 
As the singlet $\to$ triplet shift of $\psi_Q$ is the only distinction, no difference in the relic densities of non-coannihilating scenarios occur, since they are dominated by direct annihilation of DM into leptons. This fact is well confirmed by a comparison of Figs. \ref{fig:SummarybIIADirac} and \ref{fig:SummarybVADirac}.
 
Compared to bIIA, bVA has a greater number of coannihilation channels and therefore features an increased $\langle \sigma v \rangle$ in coannihilation scenarios, leading to a higher mass threshold. As Eq. \eqref{eqn:RDleptophilic} suggests, the model dependent coefficient $\mathcal{B}^\text{bVA}_{\mu}(\kappa)$ is larger than the one belonging to bIIA. This stems from the fact that there are now also more contributions to $B$-$\bar{B}$ mixing and $b \to s l^+ l^-$ transitions. This results in a comparably higher $\kappa_0^\text{bVA}$.  

Concerning the direct detection results, the main statement from bIIA also applies to bVA. There are no viable solutions to either the $R_K$ or $(g-2)_\mu$ anomalies in this model. The DD results are virtually the same as in bIIA. This is due to two competing effects. The RD is slightly more depleted for coannihilating scenarios, which weakens the bounds by a factor of $\nicefrac{\Omega}{\Omega_{DM}}$
, thus rendering the DD bounds weaker than in the bIIA model by $\mathcal{O}(10\%)$.

\subsubsection{Majorana DM}
\FloatBarrier
\begin{figure}
\centering
	\subfigure[democratic]{
        \includegraphics[width=0.5\textwidth]{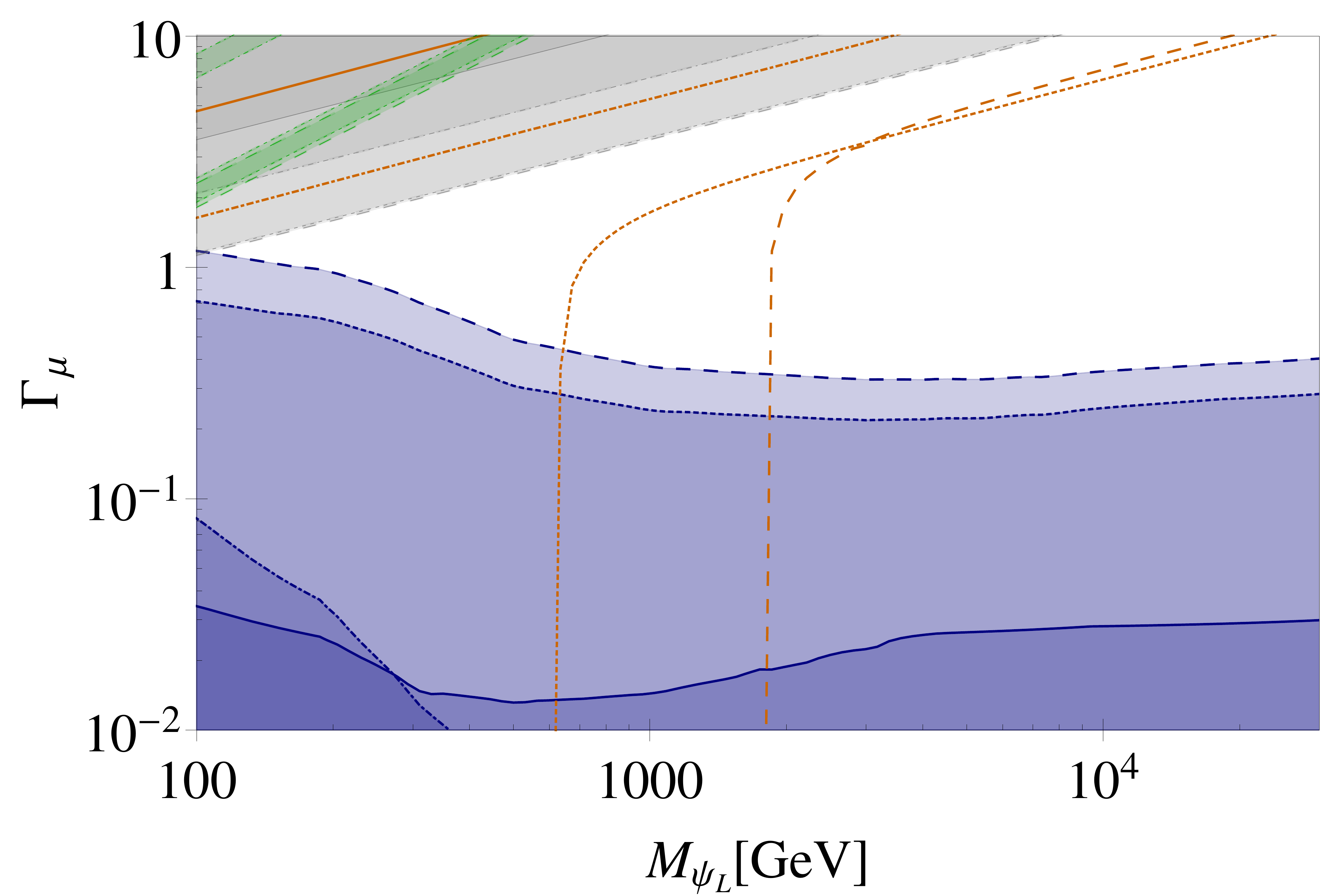}}%
	\subfigure[hierarchical]{
        \includegraphics[width=0.5\textwidth]{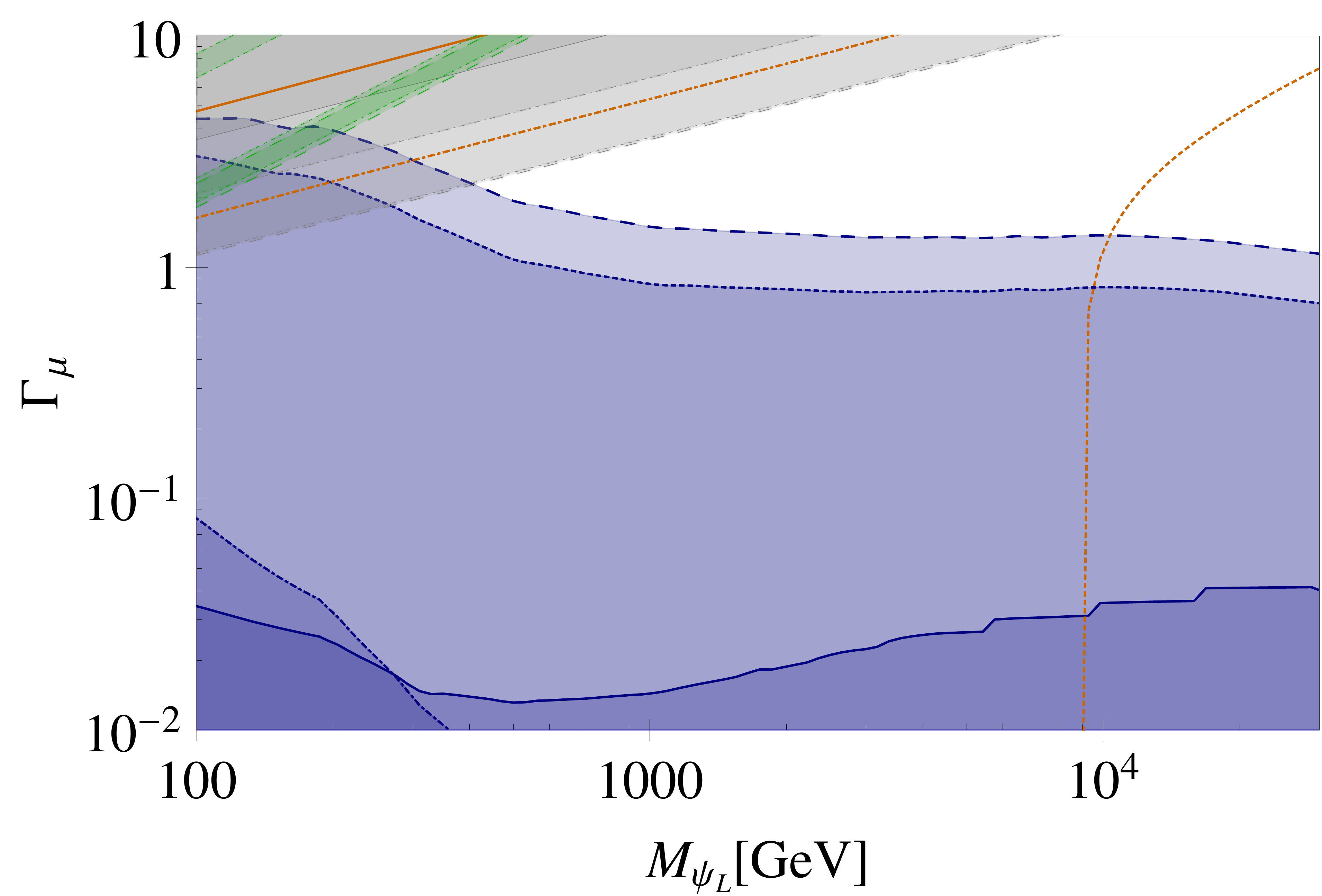}}%
    \caption{Summary plot for Majorana DM in bVA. The legend and an explanation of the color scheme are given in Fig. \ref{fig:SummarybIIADirac}.}
    \label{fig:SummarybVAMajorana}
\end{figure}

Figure \ref{fig:SummarybVAMajorana} presents the results for the bVA Majorana version. 
As indicated in Section \ref{sec:bVADirac}, there is a strong resemblance between bIIA and bVA.
The most interesting difference is the parameter space viable for an $R_K$ solution, which shrinks substantially due to the more severe $R_K$ bounds. The range is diminished to an upper bound on the DM mass of $\sim 260\,$GeV for $\kappa=1.1$ and $\sim 410\,$GeV for $\kappa=1.01$ in the hierarchic version of this model. 
 
Another minor difference is the intersection of the coannihilating RD lines compared to the non-intersecting behavior in bIIA Majorana. The reason for this is the change in degrees of freedom of the color charged fermion and the subsequent change in number density of the non-DM dark sector particles (see \hyperref[sec:AppendixA]{Appendix A}).

\FloatBarrier
\subsection{bIIB} \label{sec:bIIB}
\FloatBarrier
Following Table \ref{tbl:SUclassification}, the representations of the dark sector particles in bIIB are
\begin{align}
 \psi_L = (\bar{\textbf{3}},\textbf{1})_{-\nicefrac{2}{3}}, \, \psi_Q= (\textbf{1},\textbf{1})_0 , \, \phi=(\textbf{3},\textbf{2})_{\nicefrac{1}{6}} \nonumber \\
 \stackrel{\text{EWSB}}{\Rightarrow} \psi_Q \to \psi_Q^0 , \, \, \psi_L \to \psi_L^{-\nicefrac{2}{3}} , \, \, \phi \to  \begin{pmatrix} \phi^{+\nicefrac{2}{3}}  \\ \phi^{-\nicefrac{1}{3}} \end{pmatrix}
 \, .
\end{align}
This model's DM candidate is $\psi_Q$, which qualifies bIIB as a \textit{quarkphilic} DM model.
A notable feature of quarkphilic DM is a tree-level contribution to the direct detection cross section via the new Yukawa couplings $\Gamma_{s/b}$. However, since the first generation down Yukawa $\Gamma_d$ is set to zero, these contributions are suppressed via PDF. 

\subsubsection{Dirac DM}
\FloatBarrier
\begin{figure}
\centering
	\subfigure[democratic]{
        \includegraphics[width=0.5\textwidth]{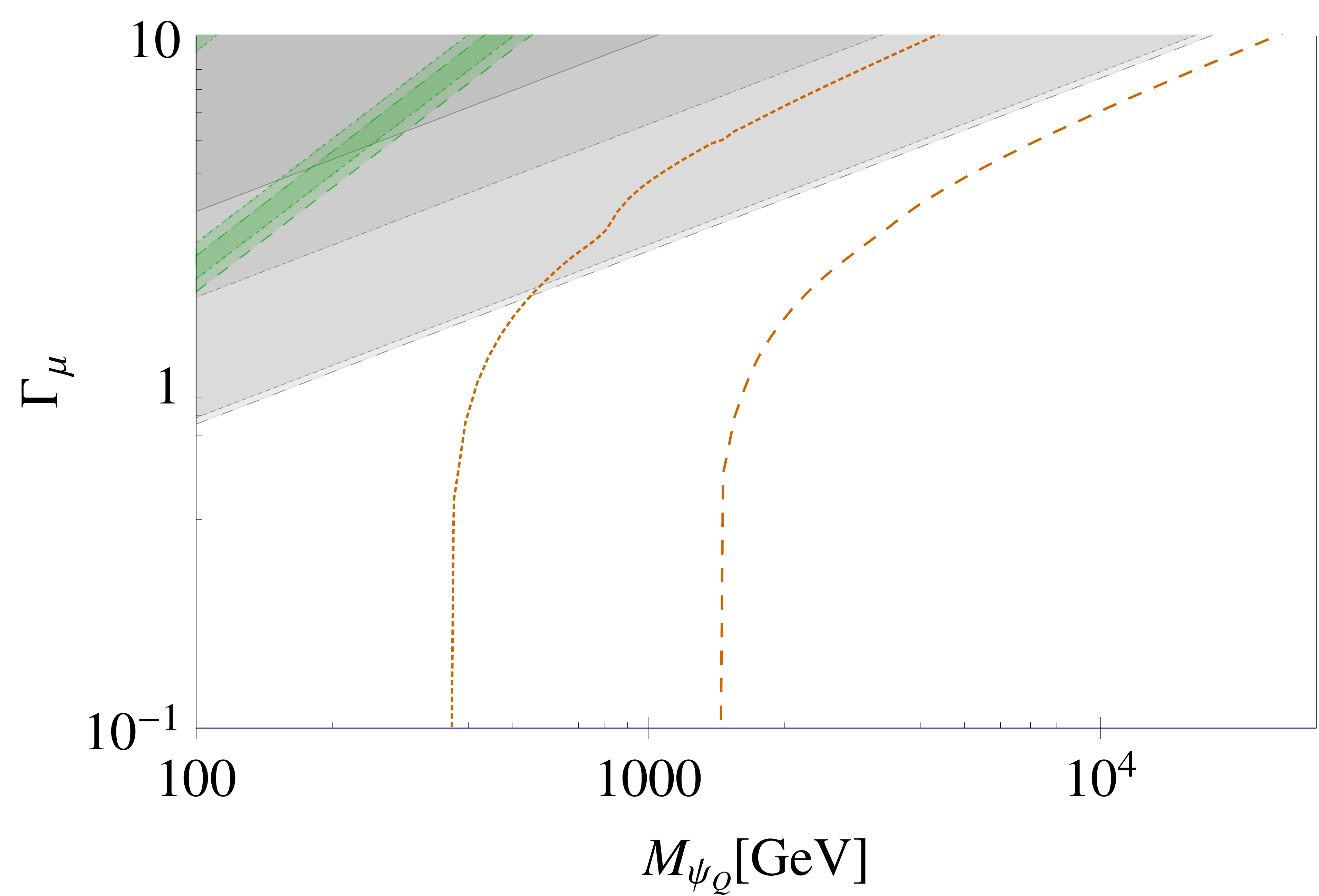}}%
	\subfigure[hierarchical]{
        \includegraphics[width=0.5\textwidth]{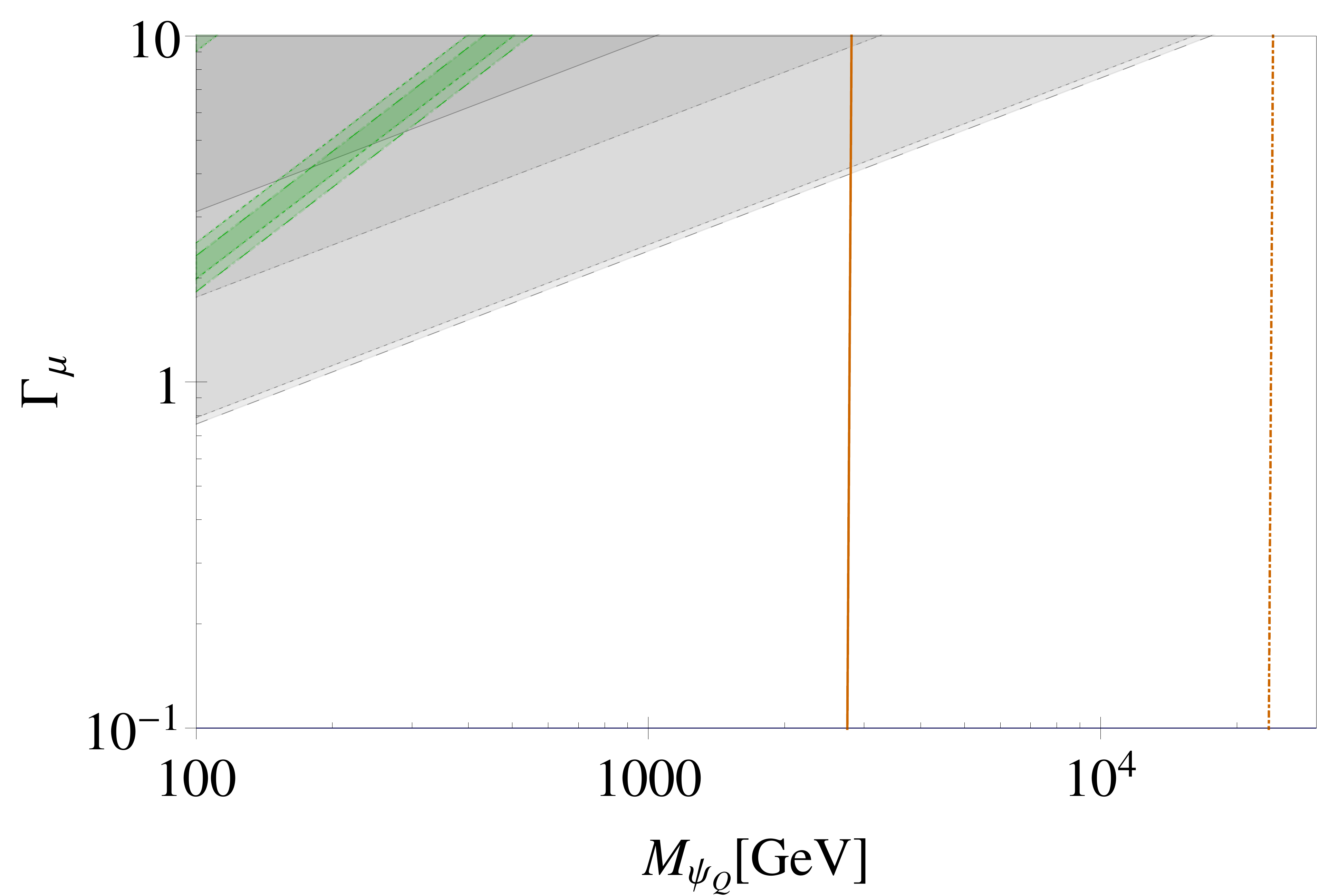}}%
    \caption{Summary plot for Dirac DM in bIIB. The legend and an explanation of the color scheme are given in Fig. \ref{fig:SummarybIIADirac}.}
    \label{fig:SummarybIIBDirac}
\end{figure}

The results of the numerical scans for bIIB Dirac are summarized in Figure \ref{fig:SummarybIIBDirac}. 
As $\psi_L$ is the coannihilation partner in quarkphilic DM models like bIIB, $\Gamma_\mu$, the Yukawa coupling directly contributing to $R_K$ and $\Delta a_\mu$ is unimportant in scenarios where $\kappa \gtrsim 1.2$. As can be seen in Fig. \ref{fig:SummarybIIBDirac}, in both democratic and hierarchical versions, $\kappa=5$ and $\kappa=15$ cannot reproduce the observed DM relic density for DM masses $> 100\,$GeV. This is due to the fact that $\Gamma_{s/b}$ are not strong enough to deplete the RD by themselves. Further, a $\Gamma_\mu$ within the perturbative bounds is not sufficient to overcome the mass suppression by increasing $\langle \sigma v \rangle$ in a way that DM is not overproduced. However, successful relic density reproduction is possible within coannihilation scenarios, where this mass suppression is lowered and thus contributions from $\psi_L$ annihilation are sizable enough to offer some viable parameter space. Furthermore, a hierarchical flavor structure between second and third generation quark couplings lowers the influence of $\Gamma_\mu$ on the RD, because the thermal freeze-out is dominated by direct annihilation into third generation quarks. This behavior is well illustrated by a comparison between Fig. \ref{fig:SummarybIIBDirac} (a) and (b), as no change in the $\Gamma_\mu$-direction is apparent. We also observe the shift of the RD mass threshold to higher masses already described in Sec. \ref{sec:bIIADirac}.

Presumably the most obvious feature of all in this model is the complete absence of parameter space allowed by direct detection. The reasons for this include the aforementioned tree-level diagrams ($s$ and $t$ channel) inducing a sizable contribution to $\sigma^\text{SI}_\text{DD}$ and a vector current contribution to the $\bar{\psi}_Q \psi_Q Z$-vertex.

\begin{figure}
    \centering
    \subfigure[tree-level]{\centering
	\includegraphics[width=0.25\textwidth]{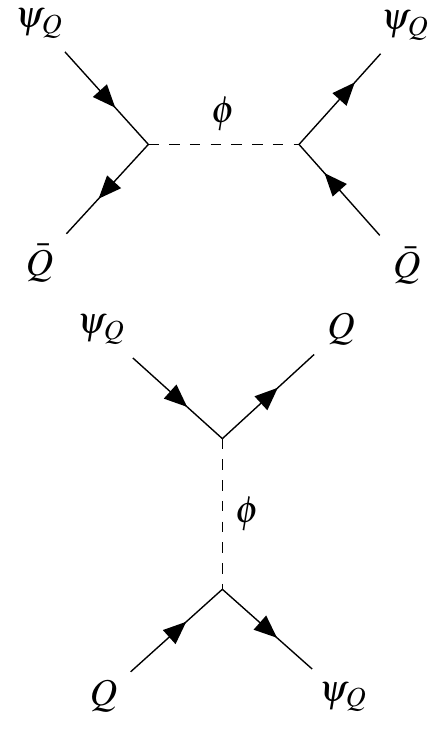}}
    \subfigure[one-loop]{\centering
	\includegraphics[width=0.7\textwidth]{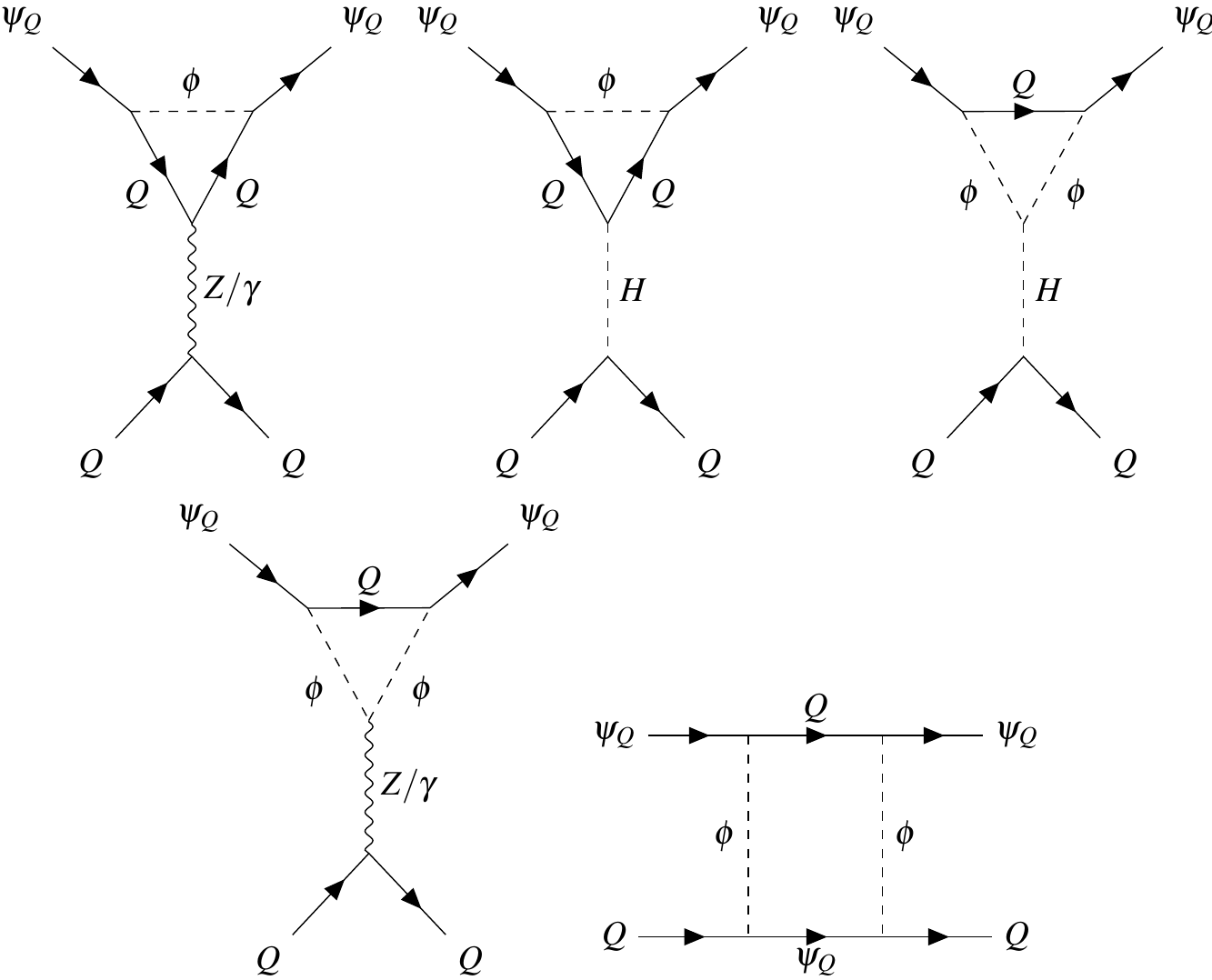}}
    \caption{Tree-level (a) and one-loop (b) DM-quark diagrams contributing to the DD cross section of $\psi_Q$ DM. The legend and an explanation of the color scheme are given in Fig. \ref{fig:SummarybIIADirac}.}
    \label{fig:DiagramsDDPsiQ}
\end{figure}

All contributions to DD up to one-loop are presented in Figure \ref{fig:DiagramsDDPsiQ}. Note here, that contribution of the effective $\bar{\psi}_Q \psi_Q H$ vertex is small compared to the ones from the $\bar{\psi}_Q \psi_Q Z$-vertex. This is the case, although the top-quark Yukawa $y_t$ enters in these processes and thus the Higgs exchange is not suppressed by a small Yukawa coupling, which is the case in leptophilic DM models. Moreover, the Higgs portal diagram suffers suppression due to an additional heavy scalar propagator and is thus less important.

\newpage
\subsubsection{Majorana DM}
\FloatBarrier

\begin{figure}
\centering
	\subfigure[democratic]{
        \includegraphics[width=0.5\textwidth]{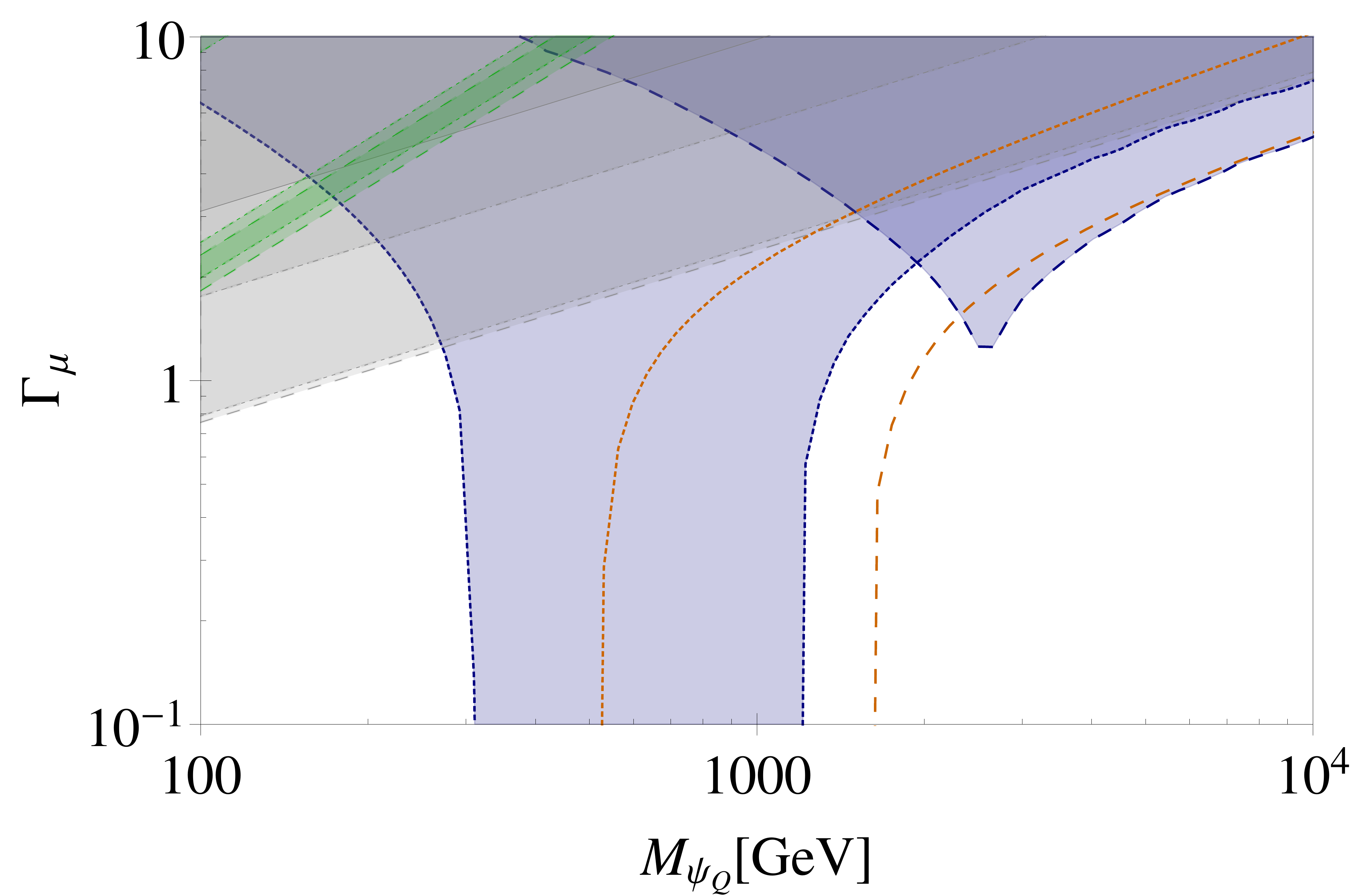}}%
	\subfigure[hierarchical]{
        \includegraphics[width=0.5\textwidth]{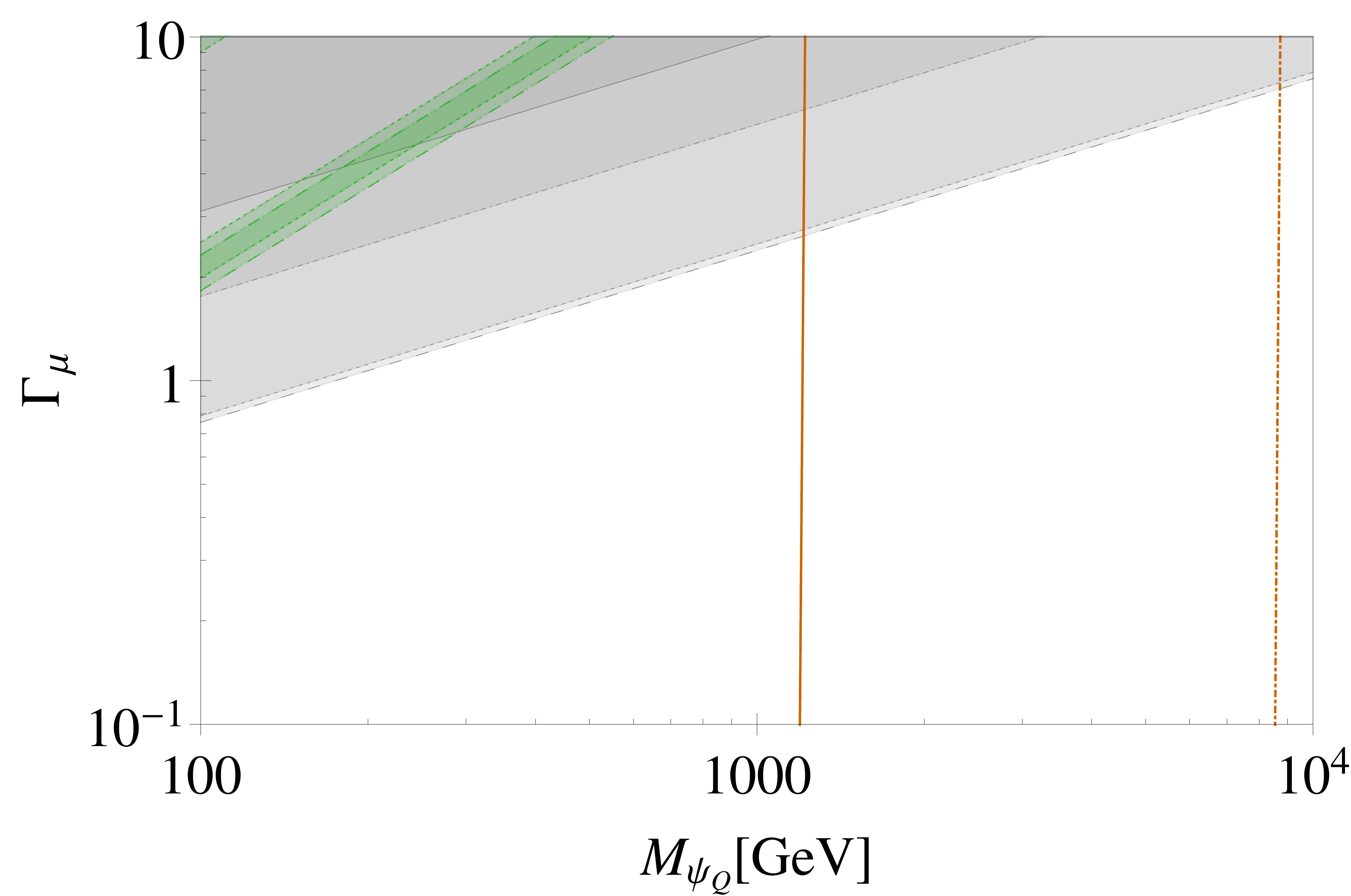}}%
    \caption{Summary plot for Majorana DM in bIIB. The legend and an explanation of the color scheme are given in Fig. \ref{fig:SummarybIIADirac}.}
    \label{fig:SummarybIIBMajorana}
\end{figure}

Figure \ref{fig:SummarybIIBMajorana} presents the numerical results of the Majorana DM version of bIIB. The behavior of the relic density is basically the same for Majorana as for Dirac with a slight difference in the mass threshold, which is due to the annihilation being $p$-wave rather than $s$-wave. 

With regards to direct detection, the Majorana version exhibits parameter space in agreement with DD limits. As mentioned in earlier sections, the vector current contribution to the effective $\bar{\psi}_Q \psi_Q Z$ vertex vanishes in Majorana DM models, so that the main contribution to the SI DM-nucleon cross section on one-loop level is absent.

The tree-level diagrams (see Fig. \ref{fig:DiagramsDDPsiQ}) are of the scalar mediated $s$-channel type\footnote{Note that tree-level t-channel diagram has an operator structure of $(\bar{\psi_Q} Q) (\bar{Q} \psi_Q)$, which resembles the structure of an $s$-channel diagram.} and can be related to $t$-channel type DD via Fierz transformations. According to the discussion in Section \ref{sec:Fierz}, this leads to $(v,v)$, $(v,a)$, $(a,v)$  and $(a,a)$ contributions. Since the $(v,v)$ structure does not exist in Majorana models, there is no unsuppressed contribution to the SI DM-nucleon cross section in this case and the twist-$2$ operators discussed in Section \ref{sec:twist2} become important, which depend on two powers of $\Gamma_{s/b}$ (see Fig. \ref{fig:DMGluonScattering}). The mass scaling of $\sigma^{\text{SI}}$ can be derived from Eq. \eqref{eqn:sigmaTwist2}.
As $\sigma^{\text{SI}}$ effectively diminishes with increasing DM mass and the bound posed by XENON1T softens as $\sim m_\text{DM}$, twist-$2$ contributions generally constrain lower masses stronger than higher masses. 
An analogous behavior can be observed for $\kappa$. The cross section $\sigma^{\text{SI}}$ is lowered with increasing $\kappa$, pushing the threshold of DD to lower masses effectively. 

In the democratic scenario, where $\Gamma_{s/b}$ take moderate values, a considerable amount of parameter space is left open in coannihilation scenarios. We observe a complementary behavior of SD and SI DD, as SI limits rule out small masses, whereas SD limits rule out larger masses.
In the area of small $M_{\psi_Q}$ and large $\Gamma_\mu$, RD is significantly depleted by coannihilation channels involving $\psi_L$ and thus the bound is softened, while the contributions to the SI cross section do not depend on $\Gamma_\mu$ so that high couplings to muons become viable ultimately. \\ \indent
The shape of the SD exclusion can be explained by the DD bound rescaling due to the relic density. Areas in the parameter space where RD is strongly overproduced feature tighter bounds compared to the underproduced part of the parameter space \footnote{Note that this parameter space is ruled out by RD anyways.}. \\ \indent
Non-coannihilating scenarios are ruled out by SD DD as the complete parameter space overproduces DM, although they are completely unconstrained by SI DD.

In this model, there is viable parameter space for a solution to $R_K$ and DM in the $\kappa=1.1$ scenario for DM masses $M_{\psi_Q} \gtrsim 1.5\,$TeV and couplings $\Gamma_\mu \gtrsim 3$. In this window, the correct RD and $R_K$ can be solved while the sets of parameters are still allowed by DD. This window, however, is narrow and requires a finely tuned mass gap $\kappa$, as both larger and smaller values of $\kappa$ studied in this work either exclude the parameter space via direct detection ($\kappa=5$) or underproduce DM ($\kappa=1.01$).
This model can in principle also explain $R_K$ and $(g-2)_\mu$ simultaneously within the $\kappa=1.1$ scenario with the caveat of DM being heavily underproduced. This solution, however, is also unstable in $\kappa$, as both lower and higher values of $\kappa$ lead to exclusion via DD. 

The hierarchical scenario is completely excluded by DD due to the large new Yukawa coupling $\Gamma_b$. The magnitude of this coupling drives the SI threshold to larger masses and causes the SI and SD exclusion areas to overlap completely. Additionally, the RD rescaling in the large $\Gamma_\mu$ does not soften the bound sufficiently to open up the parameter space in this scenario as opposed to the democratic one. 


\FloatBarrier
\subsection{bVIB} \label{sec:bVIB}
\FloatBarrier
\begin{align}
 \psi_L = (\bar{\textbf{3}},\textbf{3})_{-\nicefrac{2}{3}}, \, \psi_Q= (\textbf{1},\textbf{1})_0 , \, \phi=(\textbf{3},\textbf{2})_{\nicefrac{1}{6}} \nonumber \\
 \stackrel{\text{EWSB}}{\Rightarrow} \psi_Q \to \psi_Q^0 , \, \, \psi_L \to \begin{pmatrix} \psi_L^{+\nicefrac{1}{3}} \\ \psi_L^{-\nicefrac{2}{3}} \\  \psi_L^{-\nicefrac{5}{3}} \end{pmatrix} , \, \, \phi \to  \begin{pmatrix} \phi^{+\nicefrac{2}{3}}  \\ \phi^{-\nicefrac{1}{3}} \end{pmatrix}
 \, .
\end{align}
For the same reason bVA is considered the 'sibling' model to bIIA, the singlet to triplet shift, bVIB can be considered the 'sibling' of bIIB. 
\FloatBarrier
\subsubsection{Dirac DM}
\FloatBarrier
\begin{figure}
\centering
	\subfigure[democratic]{
        \includegraphics[width=0.5\textwidth]{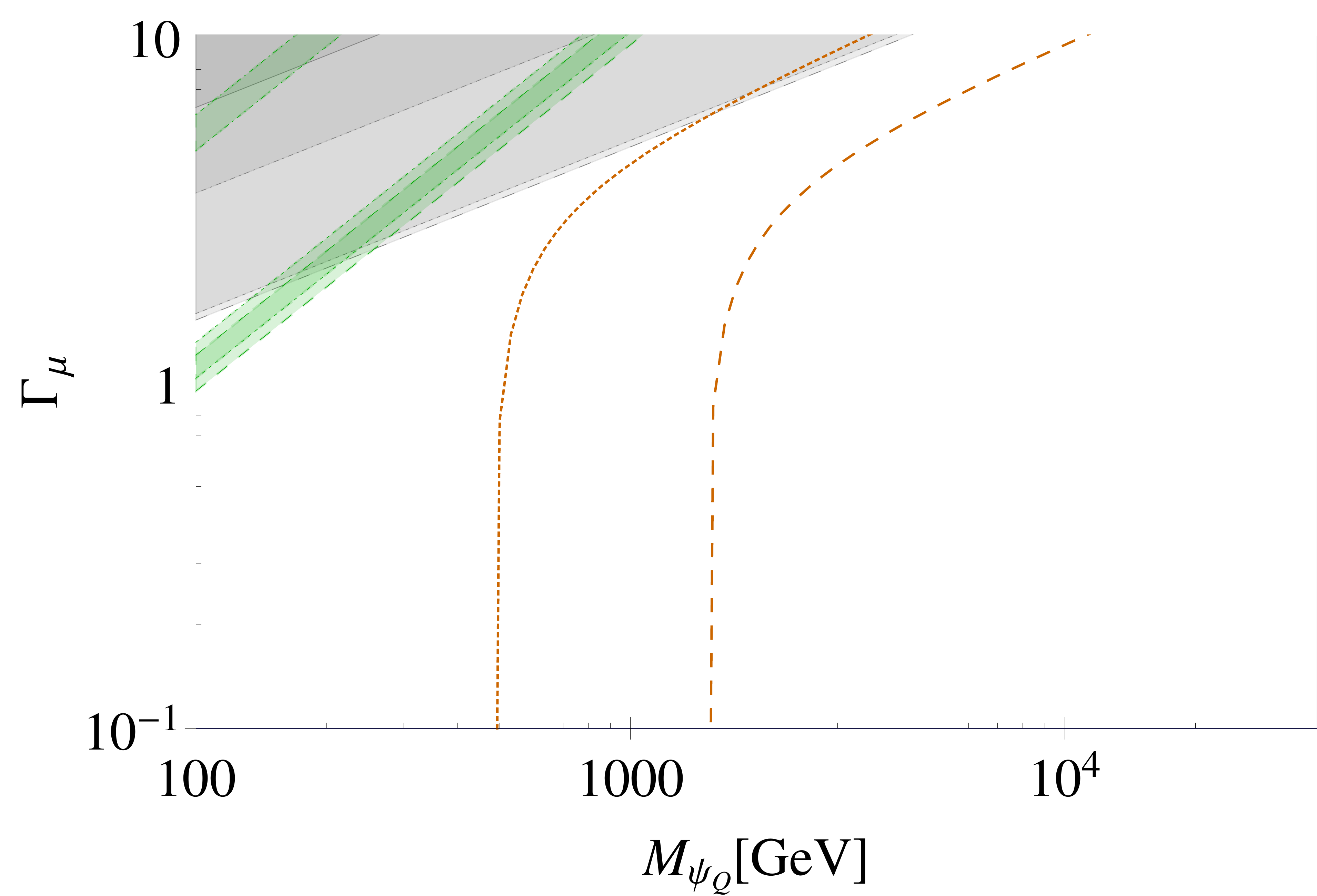}}%
	\subfigure[hierarchical]{
        \includegraphics[width=0.5\textwidth]{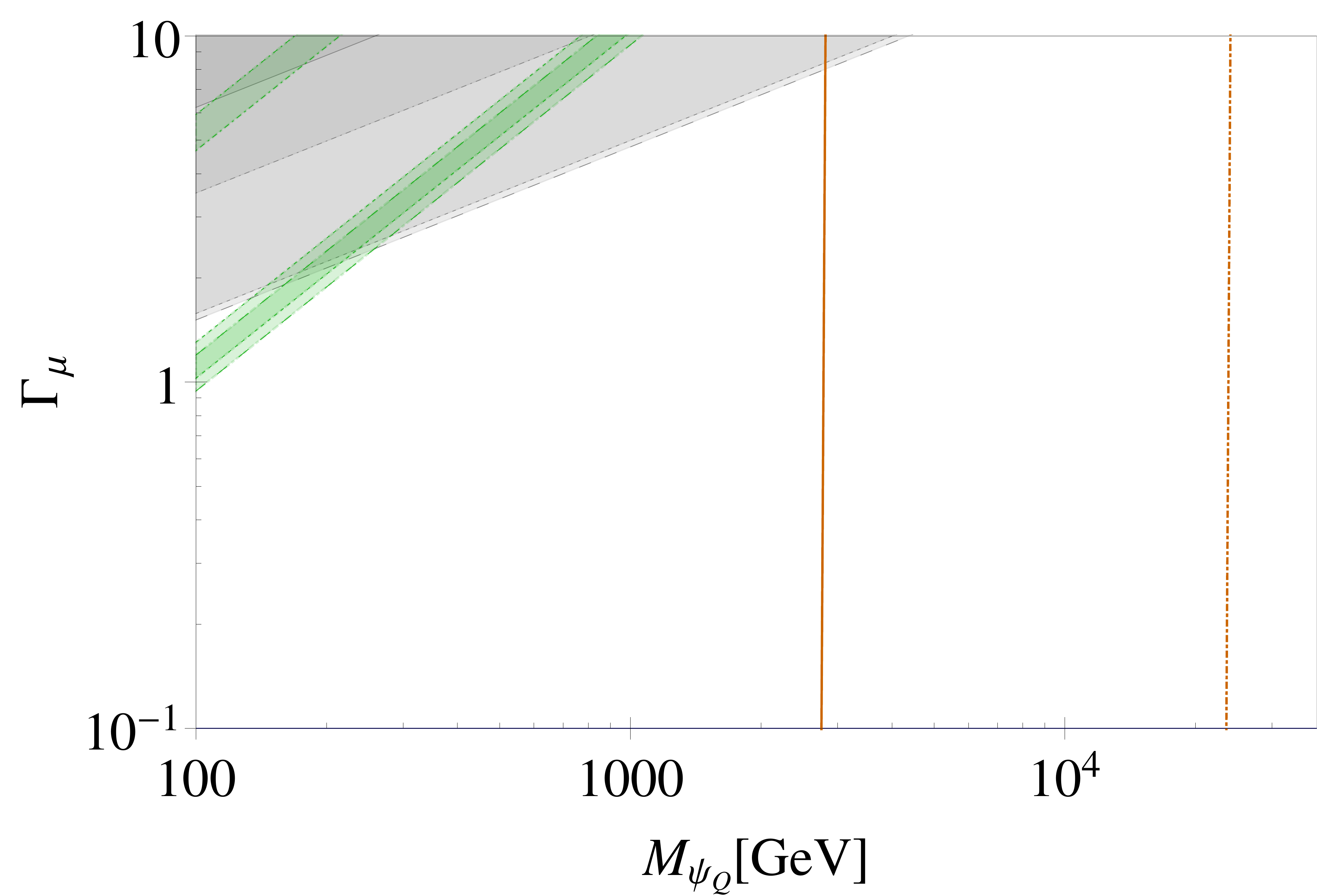}}%
    \caption{Summary plot for Dirac DM in bVIB. The legend and an explanation of the color scheme are given in Fig. \ref{fig:SummarybIIADirac}.}
    \label{fig:SummarybVIBDirac}
\end{figure}


The results for Dirac DM in bVIB are summarized in Figure \ref{fig:SummarybVIBDirac}. A comparison between Figs. \ref{fig:SummarybIIBDirac} and \ref{fig:SummarybVIBDirac} shows that the two sibling models bIIB and bVIB expectedly lead to similar results. In bVIB we can observe a more stringent bound on $\Gamma_\mu$ for an $R_K$ solution than in bIIB. This feature can also be observed in the bIIA vs. bVA comparison, where the triplet-model bVA exhibits more stringent $R_K$ bounds. The reason for this is again the increased number of diagrams contributing to the $b \to s l^+ l^-$ transitions in a triplet model.   
The relic densities of the non-coannihilating scenarios $\kappa=5,15$ are completely unaltered in comparison to bIIB, since only the $\psi_L$ gauge representation differ. 

Regarding direct detection, it is left to state that the whole parameter space studied in this work is excluded for this model, as is the case in bIIB. 
The two competing effects, namely rescaling of the DD bounds from lower RD for coannihilation scenarios and more contributions to DD cross section
from additional diagrams involving triplet particles can have an $\mathcal{O}(10\%)$-effect on the DD bounds (see Section \ref{sec:bVADirac}). This, however, is not enough to render any of the parameter space viable in the quarkphilic Dirac DM model. 
\subsubsection{Majorana DM}
\FloatBarrier
\begin{figure}
\centering
	\subfigure[democratic]{
        \includegraphics[width=0.5\textwidth]{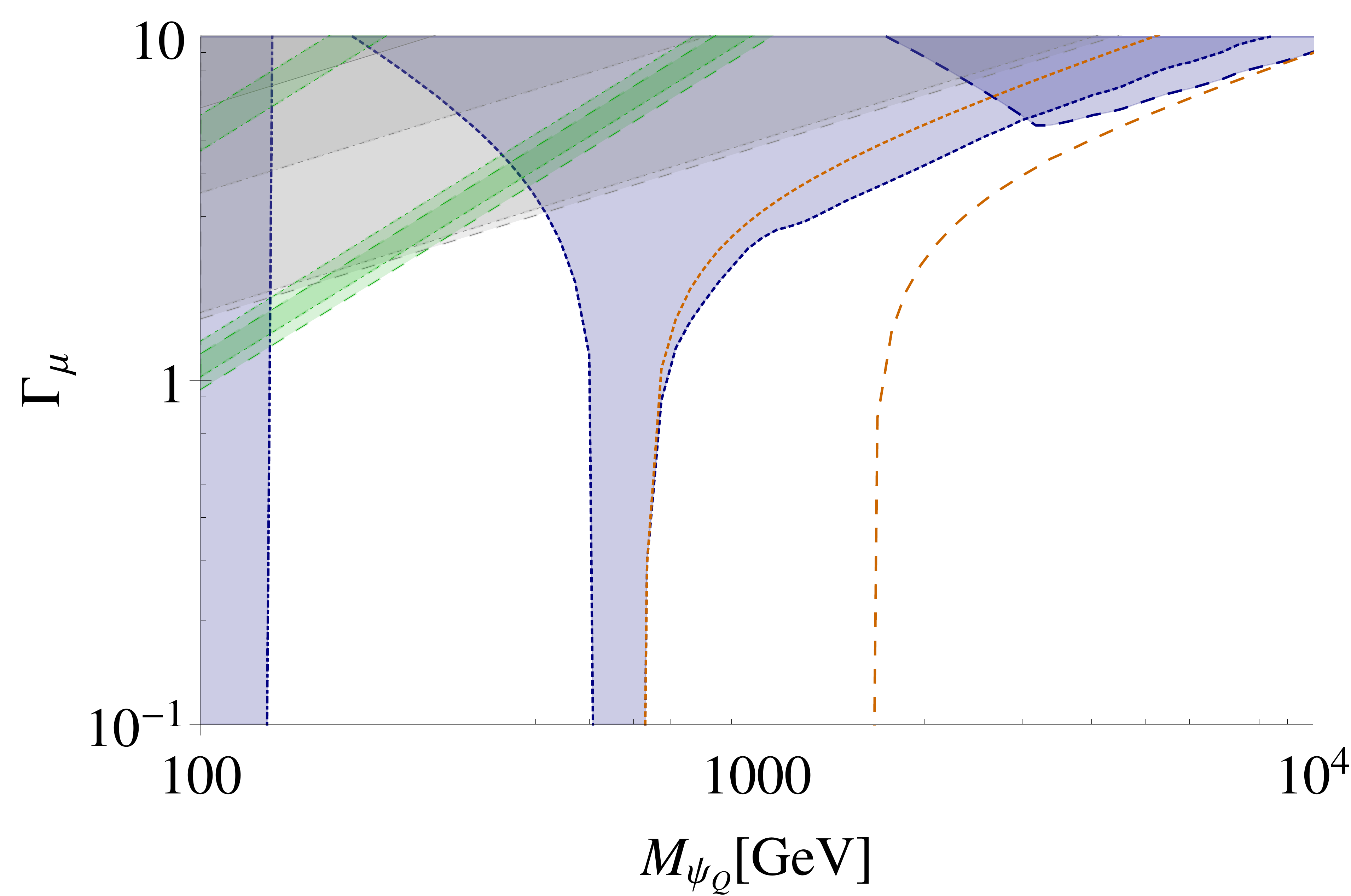}}%
	\subfigure[hierarchical]{
        \includegraphics[width=0.5\textwidth]{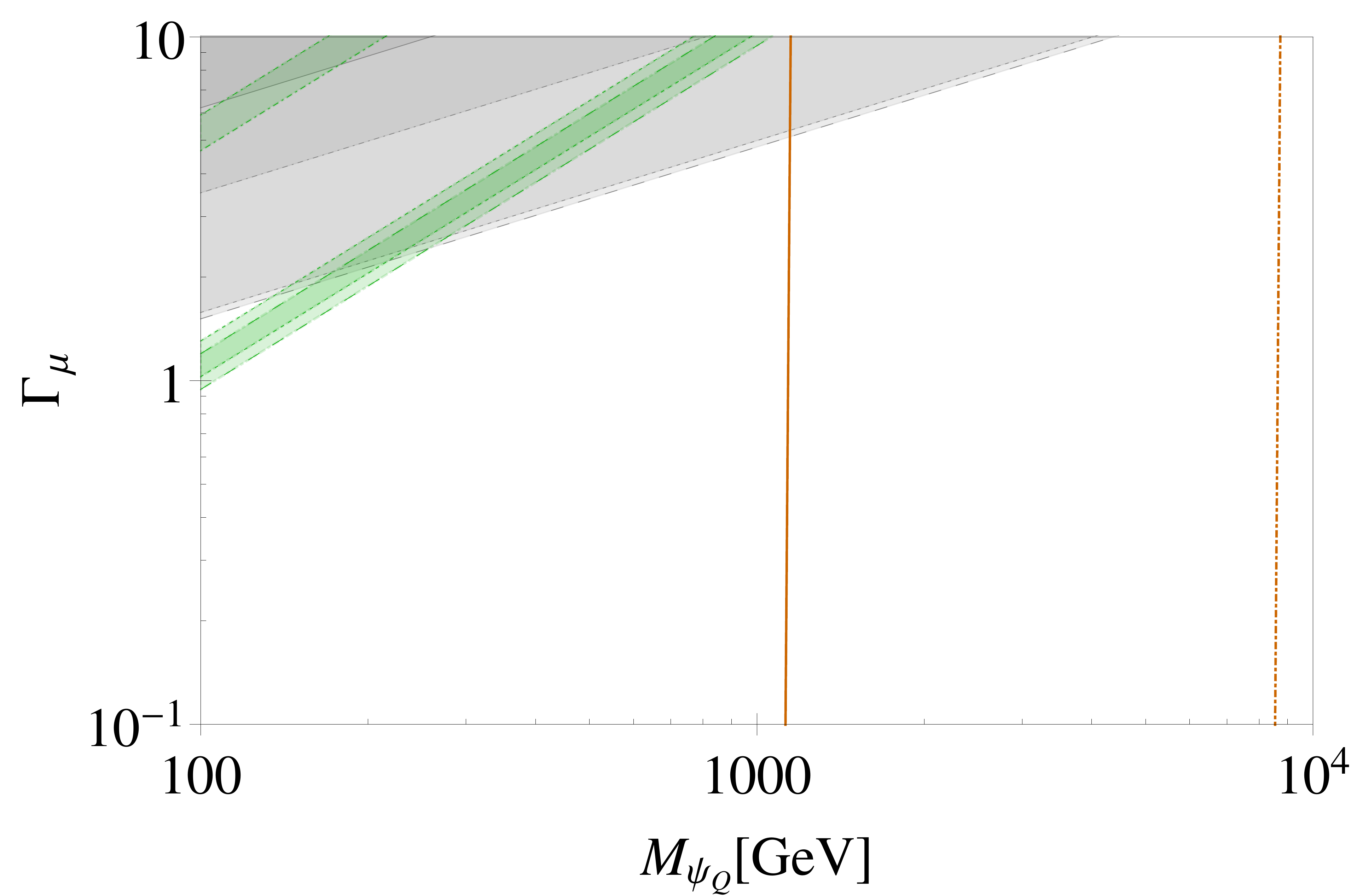}}%
    \caption{Summary plot for Majorana DM in bVIB. The legend and an explanation of the color scheme are given in Fig. \ref{fig:SummarybIIADirac}.}
    \label{fig:SummarybVIBMajorana}
\end{figure}
The results of the Majorana version of bVIB are summarized in Figure \ref{fig:SummarybVIBMajorana}. 
This model's results mostly resemble the results obtained for bIIB Majorana, due to the minor differences between the models.
The hierarchical scenario is still completely ruled out by direct detection and does therefore not show any significant difference to the hierarchical bIIB Majorana model.

The democratic scenario on the other hand exhibits some notable differences. The most important difference is that there is no parameter space for a simultaneous solution of $R_K$ and DM, as the $R_K$ region shrinks significantly due to the shift in contributions to both $b\to s l^+ l^-$ transitions and $B$-$\bar{B}$ mixing. Therefore, $\psi_Q$-DM is a possibility in this model only if the original motivation of the model is abandoned. 

Another minor difference is that parameter space allowed by DD in the $\kappa=5$-scenario reaches mass values $>100\,$GeV, as the mass thresholds of SI DD are generally shifted to higher masses. This is due to stronger underproduction of DM and subsequent relaxation of the DD bounds compared to bIIB Majorana.

Furthermore, the conclusions regarding a solution of $(g-2)_\mu$ remain almost the same as in bIIB. There is a possibility for a correct $\Delta a_\mu$ with the downside of underproduced $\psi_Q$-DM. The main difference between the solutions of bIIB and bVIB is the mass range where such a realization is possible. This model shows also that in principle there is a possibility of a solution to $(g-2)_\mu$ involving a relatively light $\psi_Q$, as demonstrated in the case of $\kappa=5$. In this case, however, we expect collider searches to become more and more restrictive, as we outline in Section \ref{sec:collider}.


\FloatBarrier
\subsection{aIA} \label{sec:aIA}
\FloatBarrier
\begin{align}
 \psi= (\textbf{1},\textbf{1})_0 , \, \phi_L = (\textbf{1},\textbf{2})_{-\nicefrac{1}{2}} , \, \phi_Q=(\textbf{3},\textbf{2})_{\nicefrac{1}{6}} \nonumber \\
 \stackrel{\text{EWSB}}{\Rightarrow} \psi \to \psi^0 , \, \, \phi_L \to \begin{pmatrix} \frac{1}{\sqrt{2}} \left( \eta^0 + \eta^{0'} \right) \\ \eta^- \end{pmatrix} , \, \, \phi_Q \to  \begin{pmatrix} \sigma^{+\nicefrac{2}{3}}  \\ \sigma^{-\nicefrac{1}{3}} \end{pmatrix}
 \, .
\end{align}
As aIA is the only a-type model with a fermionic singlet DM candidate, it makes up it's own category of \textit{amphiphilic} DM, meaning that DM couples to both quarks and leptons. Thus, aIA possesses with properties of both quark- and leptophilic models. Moreover, this model is the only model that features different $R_K$ bounds in Dirac and Majorana versions, as the $SU(2)_L$ structure allows for additional diagrams to contribute to the $B$-$\bar{B}$ mixing and $b \to s l^+l^-$ transitions (see \cite{Arnan:2016cpy} for a more detailed discussion).

\subsubsection{Dirac DM}
\FloatBarrier

\begin{figure}
\centering
	\subfigure[democratic]{
        \includegraphics[width=0.5\textwidth]{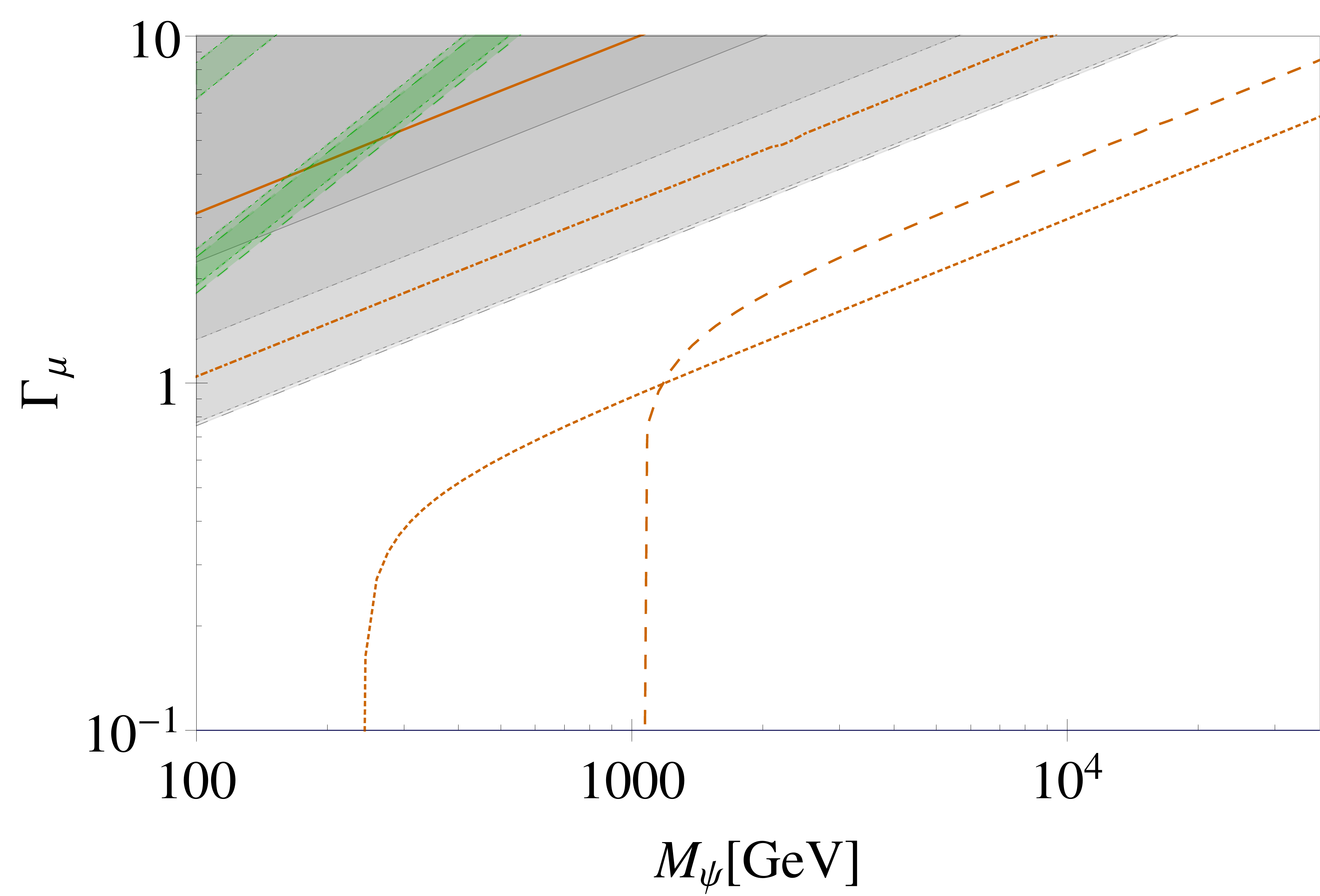}}%
	\subfigure[hierarchical]{
        \includegraphics[width=0.5\textwidth]{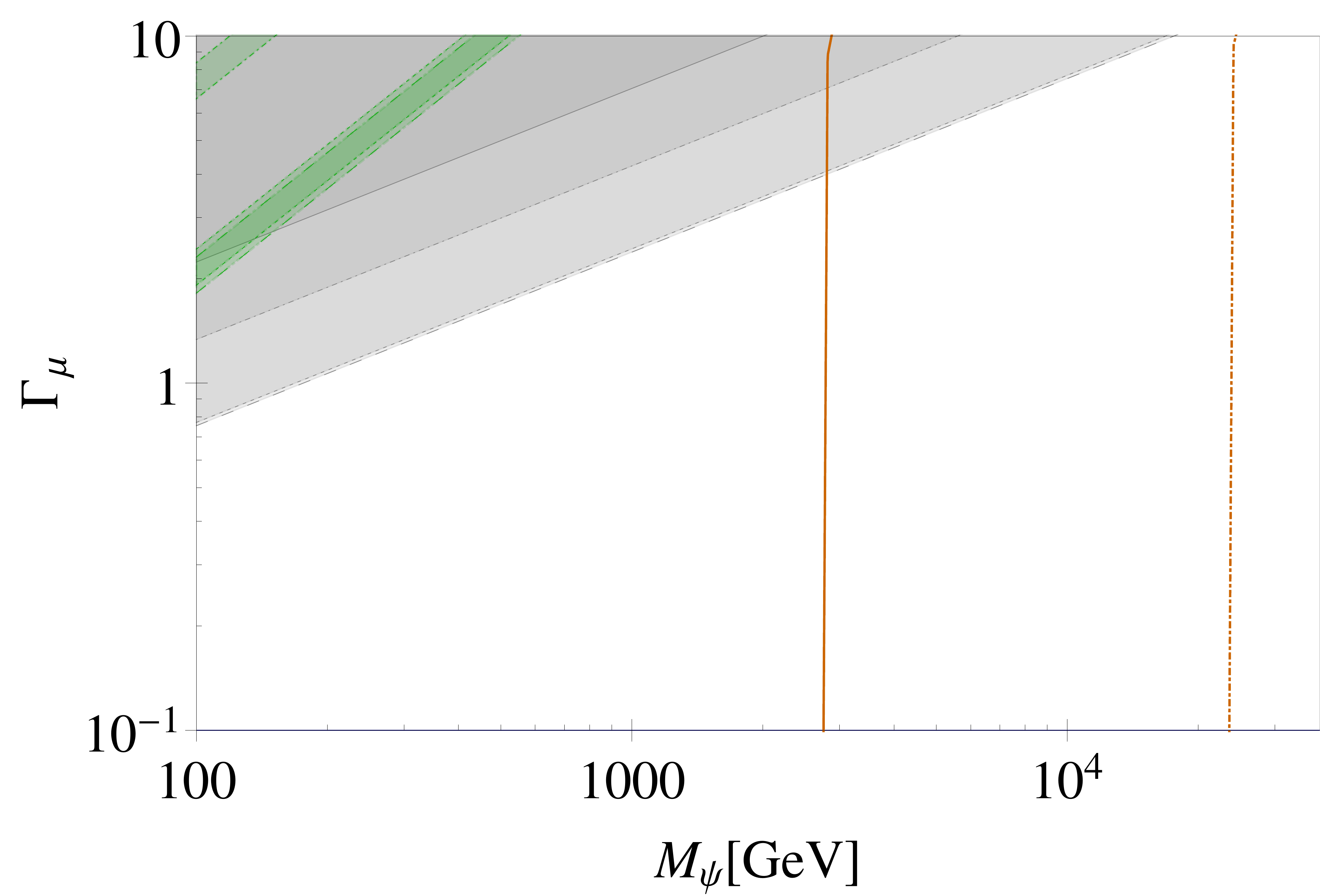}}%
    \caption{Summary plot for Dirac DM in aIA. The legend and an explanation of the color scheme are given in Fig. \ref{fig:SummarybIIADirac}.}
    \label{fig:SummaryaIADirac}
\end{figure}

We present the results for aIA in Figure \ref{fig:SummaryaIADirac}.
As the parameter space principally allowing for an $R_K$ solution is significantly enlarged in comparison to all other models in this study ($\sim$ 1-2 order(s) of magnitude), all $\kappa$-configurations have the potential to provide a simultaneous solution DM for relic density and $R_K$ in the democratic scenario. We can observe the leptophilic DM characteristics especially in the scaling of the non-coannihilating configurations. The hierarchical scenario on the other hand shows very high mass thresholds. This is caused by direct annihilations into quarks, as the annihilation cross section is compeletely dominated by annihilation into third generation quarks. This feature can be also observed in purely quarkphilic DM models (see Figs. \ref{fig:SummarybIIBDirac} and \ref{fig:SummarybVIBDirac}). 

\begin{figure}
    \centering
    \subfigure[tree-level]{\centering
	\includegraphics[width=0.25\textwidth]{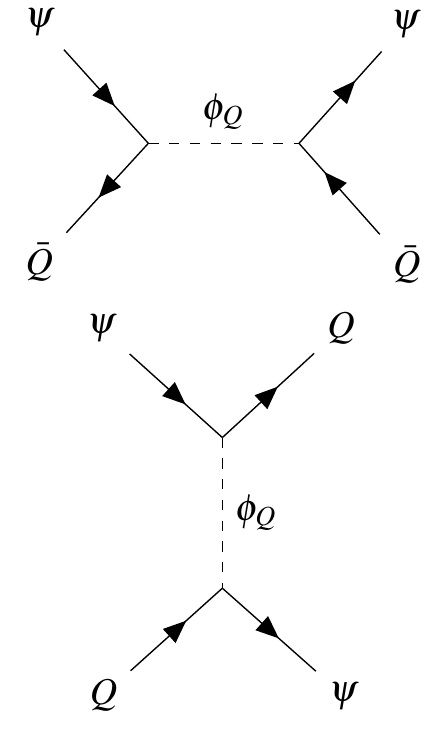}}
    \subfigure[one-loop]{\centering
	\includegraphics[width=0.7\textwidth]{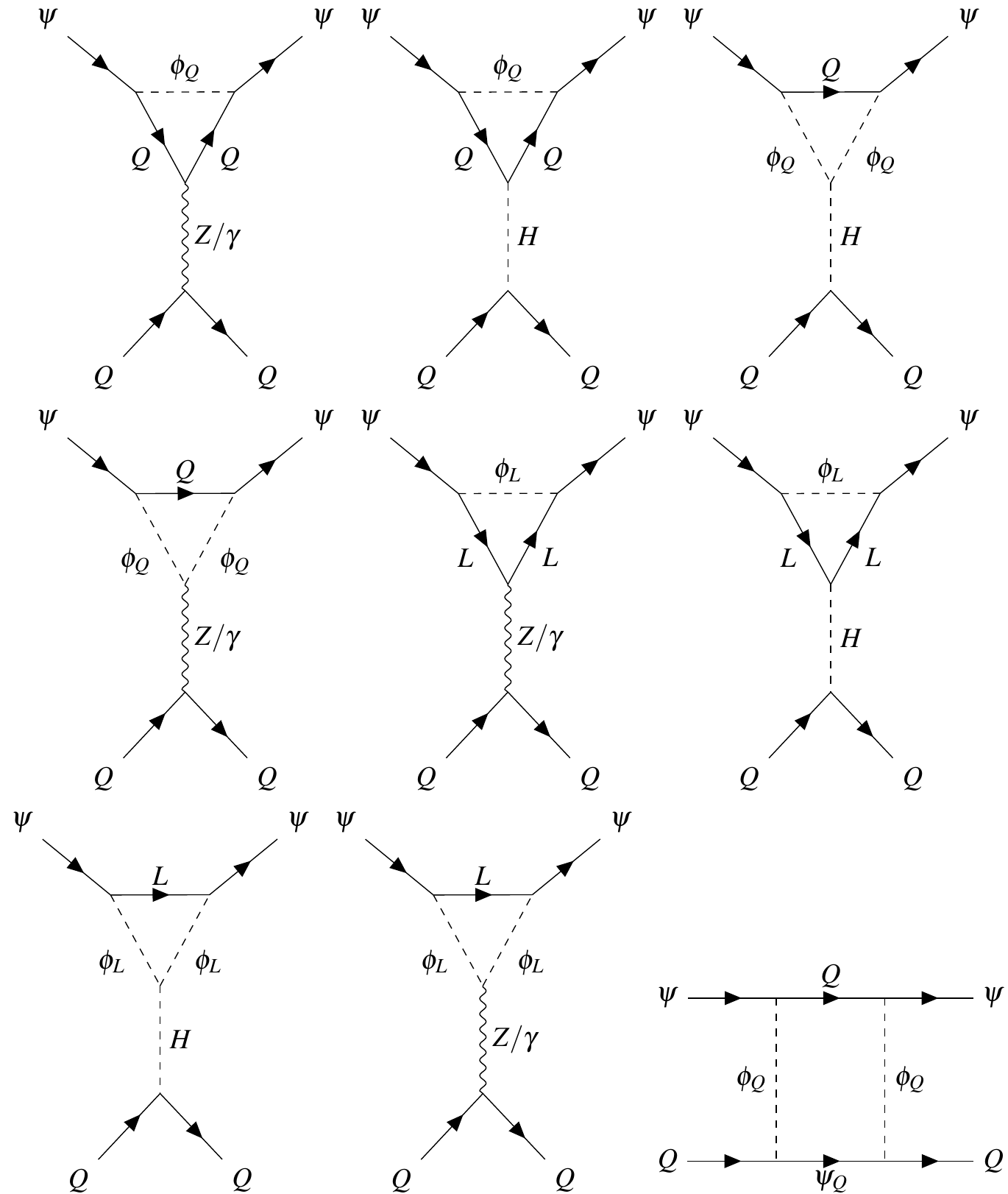}}
    \caption{Tree-level (a) and one-loop (b) DM-quark diagrams contributing to the DD cross section of $\psi_Q$ DM.}
    \label{fig:DiagramsDDPsi}
\end{figure}

Direct detection rules out all of the interesting parameter space in this model. For the diagrams contributing to DD, we refer to Fig. \ref{fig:DiagramsDDPsi}. First of all, we obtain the tree-level contributions to the SI DM-nucleon cross section typical for quarkphilic DM. Furthermore, aIA Dirac exhibits $t$-channel $Z$-diagrams depending on both new leptonic and quark Yukawa couplings. This leads to even more enhanced vector contributions to the $\bar{\psi} \psi Z$-vertex compared to leptophilic DM or quarkphilic DM models.

\newpage
\subsubsection{Majorana DM}
\FloatBarrier
\begin{figure}
\centering
	\subfigure[democratic]{
        \includegraphics[width=0.5\textwidth]{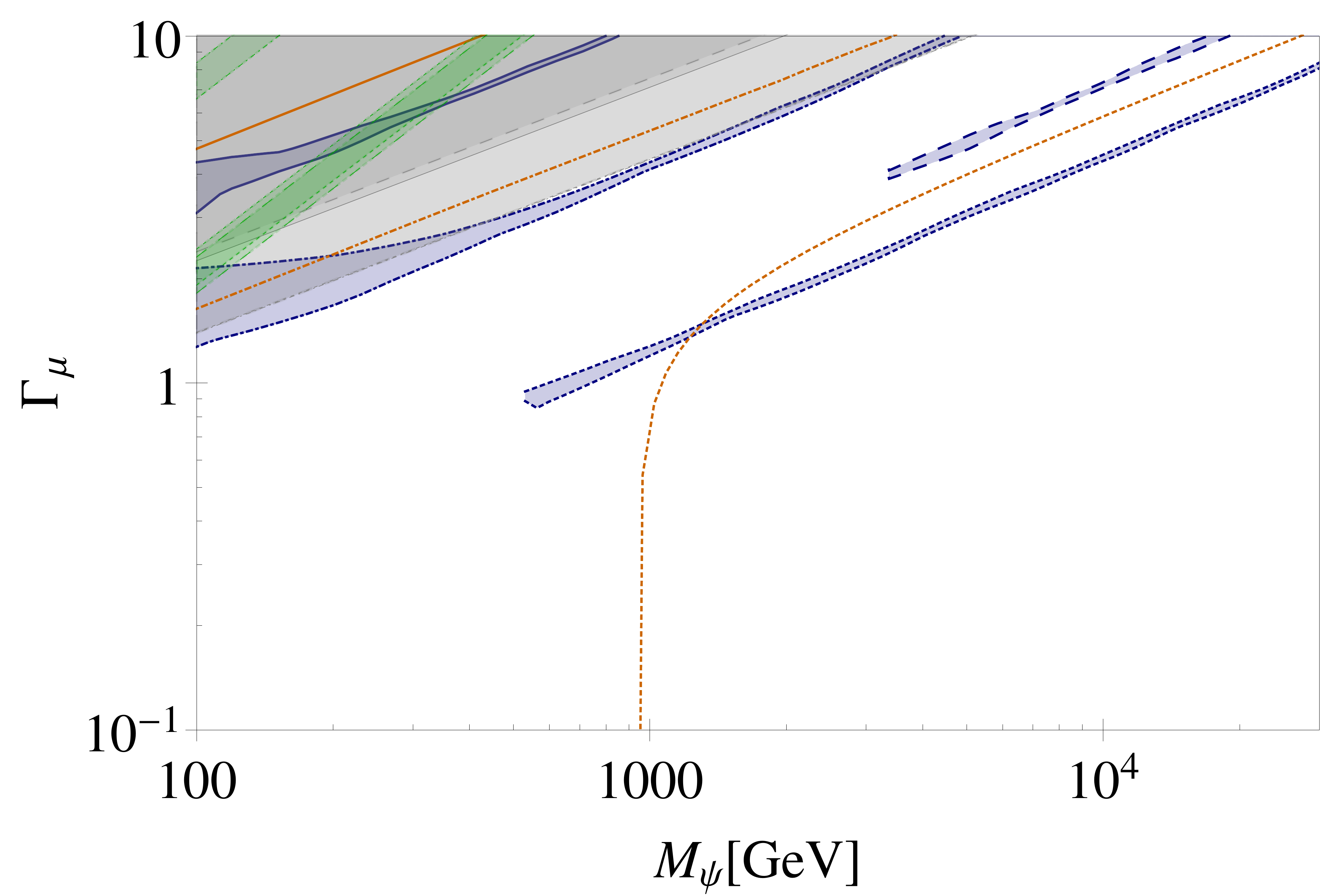}}%
	\subfigure[hierarchical]{
        \includegraphics[width=0.5\textwidth]{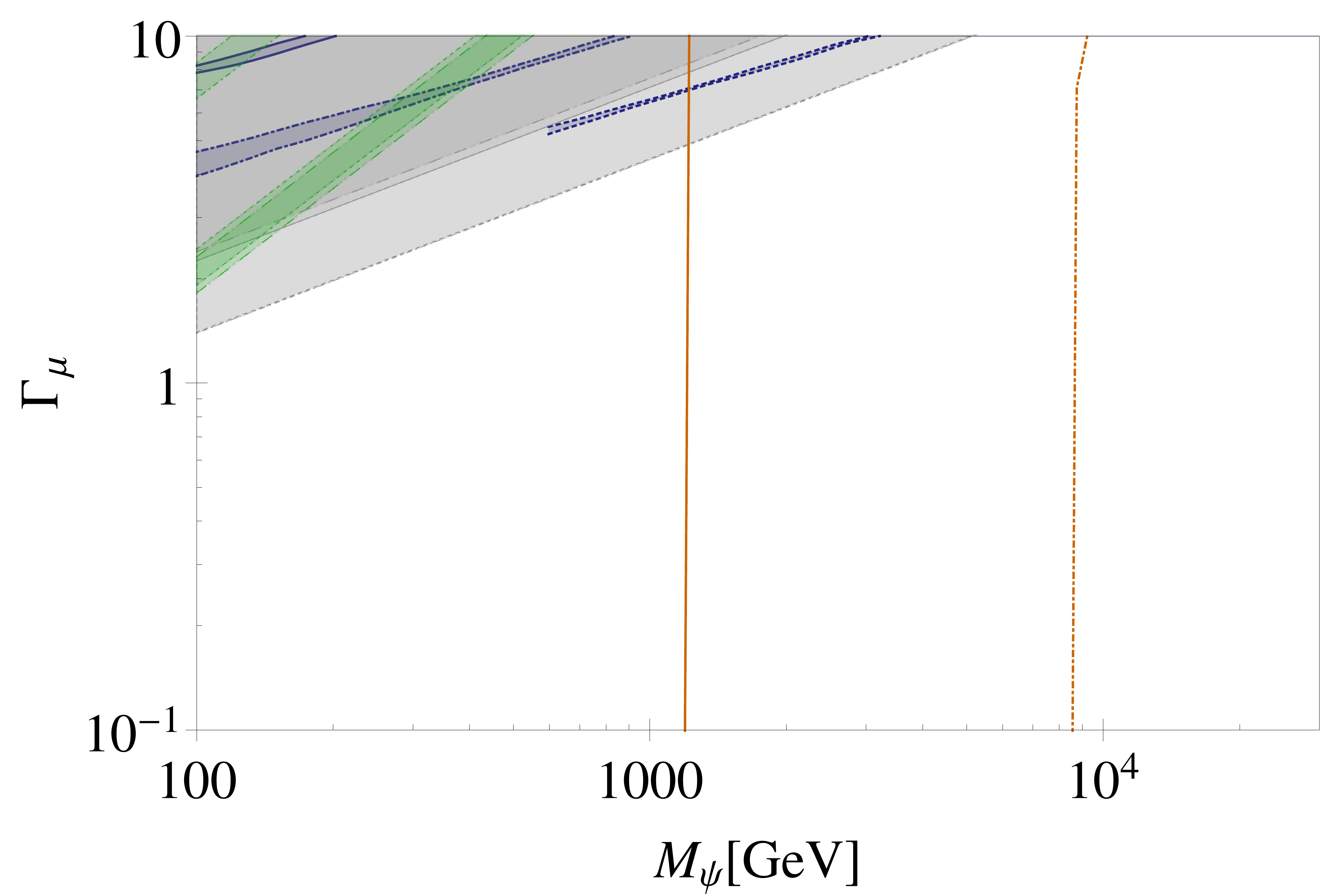}}%
    \caption{Summary plot for Majorana DM in aIA. The legend and an explanation of the color scheme are given in Fig. \ref{fig:SummarybIIADirac}.}
    \label{fig:SummaryaIAMajorana}
\end{figure}

Figure \ref{fig:SummaryaIAMajorana} summarizes the results for the Majorana version of aIA. 
Compared to the Dirac version, the Majorana version of aIA features shifted relic density lines, which is due to the p-wave annihilation. The coannihilating scenarios exhibit higher mass thresholds, which is a feature in all models. 

As indicated at the beginning of Section \ref{sec:aIA}, the $R_K$ regions follow an altered hierarchy regarding the coannihilation parameter $\kappa$ in the Majorana version compared to the Dirac version. This is a unique feature of aIA among all models studied in this work, as it is the only model where additional diagrams enter in the Majorana case. The $R_K$ bound of $\kappa=1.01$ is more restrictive than the $\kappa=15$ bound, while $\kappa=5$ and $\kappa=1.1$ differ only at the $\mathcal{O}(1\%)$-level\footnote{The $R_K$ bound as a function of $\kappa$ has a global minimum at $\kappa \approx 1.78$ and is divergent at $\kappa=1,\infty$.}. 

Generally speaking, the direct detection results are the most interesting in this model. Since aIA features traits of both quarkphilic and leptophilic models, areas allowed by DD stem from a dynamic interplay between these model characteristics. \\ \indent
As is characteristic for leptophilic models, the DM-nucleon cross section is constrained for large $\Gamma_\mu$, because of the axial vector part of the one-loop diagrams involving leptons.

The shape of the exclusion line in this model, however, differs from leptophilic models. There are two major reasons for this: One the hand, the relic density rescaling is not only influenced by direct annihilation into leptons but also by direct annihilation into quarks in this model. This is the reason for the much more parallel alignment of DD exclusion curves compared to leptophilic models. On the other hand, SD loop-contributions differ because of additional quark loops, which come with an opposite sign compared to leptonic contributions because of their hypercharge structure\footnote{Note that the additional quark contributions do not necessarily enhance the SD DM-nucleon cross section because of potential cancellations with the leptonic contributions.}. \\ \indent
The exclusion from below is a typical characteristic of one-loop contributions involving quarks in the loop, as they are dependent on $\Gamma_{s/b}$ rather than $\Gamma_\mu$ but still sensitive to the masses of the particles running in the loop (and therefore also $\kappa$). \\ \indent
The allowed regions visible in Fig. \ref{fig:SummaryaIAMajorana} therefore arise due to the interplay of the leptonic and hadronic loop contributions and relic density rescaling of the DD bound. 

The above mentioned contributions are all SD but as discussed in the Sections \ref{sec:bIIB} and \ref{sec:bVIB}, SI DD induced by DM-gluon scattering can also occur in quarkphilic models. This effect is visible in coannihilation scenarios, where the allowed regions feature a cut-off at a certain DM mass, such that smaller DM masses are excluded. In the hierarchical scenario, the allowed area for $\kappa=1.01$ disappears completely. 

In coannihilation scenarios, we find a 'sweet spot' for a valid DM production in the region $M_\psi \in[1.22,\, 1.32]\,$TeV. This, however, does not offer a solution to either the $R_K$ or the $(g-2)_\mu$ anomalies.
In the non-coannihilation scenario $\kappa=5$, the mass region $M_\psi \lesssim 200\,$GeV offers a solution to both $R_K$ and DM in this model.   

Simultaneous solutions for $R_K$, DM and $(g-2)_\mu$ do not exist in this model. Moreover, even individual solutions to $(g-2)_\mu$ are excluded by DD in this setup.

\FloatBarrier
\subsection{Collider constraints}
 \label{sec:collider}
In this section, we review existing searches of setups similar to the scenarios discussed in the previous sections. While the results cannot be explicitly applied to the setups studied in this work, they can provide an indication of the excluded parameter regions. Note that we do not perform a detailed collider study in this article. 

For the leptophilic DM scenarios, {\it i.e.} for models bIIA and bVA, $\psi_Q$ can be pair-produced at tree-level, subsequently decaying through $\psi_Q \rightarrow Q \phi^{\dagger}$~($\phi^{\dagger} \rightarrow \psi_L \bar{L}$). Such channels of $\psi_L$ production can be constrained by dilepton+jets+$\slashed{E}_T$ searches at the LHC~\cite{ATLAS:2016ljb,Sirunyan:2020tyy}: for example, dilepton searches can rule out $M_{\psi_L} \lesssim 600$~GeV for $M_{\psi_Q} \lesssim 800$~GeV~\cite{Cline:2017qqu}. 

For the quarkphilic DM scenarios, such as bIIB and bVIB, $\psi_Q$ can be produced in $t$-channel interactions mediated by the colored scalar  $\phi$~\cite{Racco:2015dxa,Aaboud:2017phn}. ATLAS constrained such scenarios through monojet+$\slashed{E}_T$ searches with an integrated luminosity of 36~$fb^{-1}$ at $\sqrt{s} = 13$~TeV~\cite{Aaboud:2017phn}. In our context, for $\Gamma_{Q} = 1$, this rules out $M_{\psi_Q} \lesssim 600$~GeV for $M_{\phi} \sim 700$~GeV. For lower values of $M_{\psi_Q}$ it can rule out even higher values of $M_{\phi}$ up to $\sim 1.6$~TeV (see Fig.~8 of Ref.~\cite{Aaboud:2017phn}).
In the model aIA, the DM candidate $\psi$ has tree-level couplings to the SM quarks, leading to the same production channels at the LHC as the bIIB and bVIB models. Thus, one can expect similar constraints on model aIA as well.  

In the b-type models, for small enough values of $\Gamma_Q$ or $\Gamma_L$, colored fermions can become sufficiently long-lived at collider scales and may decay outside the detector. Displaced vertex+$\slashed{E}_T$ searches at the LHC can be recast into constraints on the mass and lifetime of such a long-lived particle~\cite{Belanger:2018sti}. This can rule out the new physics Yukawa couplings in the range $\Gamma \sim 10^{-2} - 10^{-5}$ for fermion masses $\lesssim 1.8$~TeV. 

\FloatBarrier

\section{Summary and Conclusions}\label{sec:conclusion}
In this article, we investigated the DM phenomenology of SM extensions addressing the $R_K$ and $(g-2)_\mu$ anomalies at one-loop level, which also contain a fermionic singlet DM candidate.
In order to stabilize the fermionic singlet against a decay, we assume the dark sector particles to be charged under an unbroken stabilizing symmetry.
We found five Dirac and five Majorana models divided into three different classes (labeled quarkphilic, leptophilic and amphiphilic) sharing similar phenomenological features.
For each model, we studied DM relic density, spin-dependent and spin-independent direct detection in a hierarchical coupling set-up ($\Gamma_b \gg \Gamma_s$), as well as in democratic coupling set-up ($\Gamma_b \sim \Gamma_s$). Furthermore, we studied different mass-hierarchies between the DM and the other dark-sector particles within each model, ranging from coannihilation scenarios ($\kappa=1.01, 1.1$) to non-coannihilation scenarios ($\kappa=5,15$). 

Since these models typically offer a vast parameter space, we focused this study on the models' ability to solve the aforementioned anomalies in the major part of the parameter space under study. This is achieved by keeping the product of the second and third generation quark couplings $\Gamma_s^* \Gamma_b$ at the maximal value allowed by $B$-$\bar{B}$-mixing, while $\Gamma_\mu$ is left unconstrained.

Tables \ref{tbl:ModelAchievementsSimul} and \ref{tbl:ModelAchievementsSingle} summarize the models' achievements in each set-up regarding their potential to provide simultaneous (Tbl. \ref{tbl:ModelAchievementsSimul}) or individual solutions (Tbl. \ref{tbl:ModelAchievementsSingle}) to the $R_K$ and $(g-2)_\mu$ anomalies and the observed relic density under the premise of accordance with direct detection bounds. 

\begin{table}
 \centering
 \begin{tabular}{ll|c|c|c|c|c|c}
  & & RD + $R_K$ + $\Delta a_\mu$ & RD + $R_K$  & RD + $\Delta a_\mu$  \\ \hline \hline 
  \multirow{4}{*}{\rotatebox{90}{leptophilic}}& bIIA & \xmark & \xmark  & \xmark   \\[5pt] 
  & $\text{bIIA}^\text{Maj}$ & \xmark & \xmark & \xmark \\[5pt] 
  &bVA & \xmark & \xmark & \xmark  \\[5pt] 
  &$\text{bVA}^\text{Maj}$ & \xmark & \xmark & \xmark  \\[5pt] \hline 
  \multirow{4}{*}{\rotatebox{90}{quarkphilic}}& bIIB & \xmark & \xmark & \xmark  \\[5pt]
  &$\text{bIIB}^\text{Maj}$ & \xmark & dem.: $M_{\psi_Q}\gtrsim 1.5\,\text{TeV}\left.\right|_{\kappa=1.1}$ & \xmark \\[5pt]
  &bVIB & \xmark & \xmark & \xmark  \\[5pt] 
  &$\text{bVIB}^\text{Maj}$ & \xmark & \xmark & \xmark  \\ [5pt]  \hline
  \multirow{2}{*}{\rotatebox{90}{amphi-} \rotatebox{90}{philic}}& aIA & \xmark & \xmark & \xmark  \\[5pt] 
  &$\text{aIA}^\text{Maj}$ & \xmark &  \thead{dem.: $M_{\psi}\lesssim 200\,\text{GeV}\left.\right|_{\kappa=5}$} & \xmark\\[5pt] \hline 
 \end{tabular}
 \caption{We list the viable mass regions for various simultaneous solutions to relic density, $R_K$ and $\Delta a_\mu$ for each model under the premise of accordance with direct detection limits.}
 \label{tbl:ModelAchievementsSimul}
\end{table}

\begin{sidewaystable}[!p]
 \begin{tabular}{ll|c|c|c}
  & & $R_K$ only & RD only & $\Delta a_\mu$ only \\[5pt] \hline \hline 
  \multirow{4}{*}{\rotatebox{90}{leptophilic}}& bIIA & \xmark & \thead{dem.: $M_{\psi_L}\approx 270\, \text{GeV}\left.\right|_{\kappa=1.1}$, $M_{\psi_L}\approx 1.62\,\text{TeV}\left.\right|_{\kappa=1.01}$ \\ hier.:$M_{\psi_L} \approx 9.1\, \text{TeV}\left.\right|_{\kappa=1.1}$}  & \xmark   \\[5pt] 
  & $\text{bIIA}^\text{Maj}$ & \thead{dem.: $M_{\psi_L}\lesssim 180\,\text{GeV}\left.\right|_{\kappa=1.01}$ \\ hier.:$M_{\psi_L}\lesssim 420\,\text{GeV}\left.\right|_{\kappa=1.1}$,\\ $M_{\psi_L}\lesssim 1000\,\text{GeV} \left.\right|_{\kappa=1.01}$  } & \thead{dem.:$M_{\psi_L}\approx 470\,\text{GeV} \left.\right|_{\kappa=1.1}$, $M_{\psi_L}\approx 1.85\,\text{TeV} \left.\right|_{\kappa=1.01}$ \\
  hier.: $M_{\psi_L}\lesssim 16.8\,\text{TeV} \left.\right|_{\kappa=1.1}$ } & \thead{hier.: $M_{\psi_L}\lesssim 170\,\text{GeV} \left.\right|_{\kappa=1.1}$,\\ $M_{\psi_L}\lesssim 290\,\text{GeV} \left.\right|_{\kappa=1.01}$} \\[5pt] 
  &bVA &  \xmark & \thead{dem.:$M_{\psi_L}\approx 440\,\text{GeV}\left.\right|_{\kappa=1.1}$,$M_{\psi_L}\approx 1.7\,\text{TeV}\left.\right|_{\kappa=1.01}$ \\hier.: $M_{\psi_L}\approx 5.7\,\text{TeV}\left.\right|_{\kappa=1.1}$,$M_{\psi_L}\approx 30\,\text{TeV}\left.\right|_{\kappa=1.01}$}  & \xmark \\[5pt] 
  &$\text{bVA}^\text{Maj}$ & \thead{hier.: $M_{\psi_L}\lesssim 410\,\text{GeV}\left.\right|_{\kappa=1.01}$,\\ $M_{\psi_L}\lesssim 260\,\text{GeV}\left.\right|_{\kappa=1.1}$} & \thead{dem.: $M_{\psi_L}\lesssim 630\,\text{GeV} \left.\right|_{\kappa=1.1}$, $M_{\psi_L}\lesssim 1.85\,\text{TeV} \left.\right|_{\kappa=1.01}$ \\ hier.: $M_{\psi_L}\lesssim 9\,\text{TeV} \left.\right|_{\kappa=1.1}$, $M_{\psi_L}\lesssim 32.4\,\text{TeV} \left.\right|_{\kappa=1.01}$ 
  } & \thead{hier.: $M_{\psi_L}\lesssim 140\,\text{GeV}\left.\right|_{\kappa=1.1}$,\\ $M_{\psi_L} \lesssim 210\,\text{GeV}\left.\right|_{\kappa=1.01}$} \\[5pt] \hline 
  \multirow{4}{*}{\rotatebox{90}{quarkphilic}}& bIIB & \xmark & \xmark  & \xmark \\[5pt]
  &$\text{bIIB}^\text{Maj}$ & \thead{dem.:$M_{\psi_Q}\gtrsim 270\,\text{GeV}\left.\right|_{\kappa=1.1}$,\\ $M_{\psi_Q}\gtrsim 1.5\,\text{TeV}\left.\right|_{\kappa=1.01}$} & dem.: $M_{\psi_Q}\gtrsim 520\,\text{GeV}\left.\right|_{\kappa=1.1}$ $M_{\psi_Q}\gtrsim 2.32\,\text{TeV}\left.\right|_{\kappa=1.01}$ & \thead{dem.: $M_{\psi_Q} \in [150,500]\,\text{GeV}\left.\right|_{\kappa=1.1}$,\\ $M_{\psi_Q} \in [400, 550] \,\text{GeV}\left.\right|_{\kappa=1.01}$}  \\[5pt]
  &bVIB & \xmark & \xmark  & \xmark \\[5pt] 
  &$\text{bVIB}^\text{Maj}$ & \thead{dem.: $M_{\psi_Q}\gtrsim 410\,\text{GeV}\left.\right|_{\kappa=1.1}$,\\ $M_{\psi_Q}\gtrsim 2.4\,\text{TeV}\left.\right|_{\kappa=1.01}$} & dem.: $M_{\psi_Q}\gtrsim 620\,\text{GeV}\left.\right|_{\kappa=1.1}$ &  \thead{dem.: $M_{\psi_Q} \in [350,970]\,\text{GeV}\left.\right|_{\kappa=1.1}$\\  $M_{\psi_Q} \lesssim 130 \,\text{GeV}\left.\right|_{\kappa=5}$} \\[5pt] \hline
  \multirow{2}{*}{\rotatebox{90}{amphi-} \rotatebox{90}{philic}}& aIA & \xmark & \xmark  & \xmark \\[5pt] 
  &$\text{aIA}^\text{Maj}$ & \thead{dem.: $M_{\psi}\lesssim 860\,\text{GeV}\left.\right|_{\kappa=15}$,\\ $M_{\psi}\lesssim 420\,\text{GeV}\vee M_\psi\gtrsim 1.5\,\text{TeV}\left.\right|_{\kappa=5}$ \\
  hier.:$M_{\psi}\lesssim 200\,\text{GeV}\left.\right|_{\kappa=15}$, $M_{\psi}\lesssim 900\,\text{GeV}\left.\right|_{\kappa=5}$\\
  $M_{\psi} \in [600\,\text{GeV},\,3.2\,\text{TeV} ]\left.\right|_{\kappa=1.1}$} & dem:  $M_{\psi}\in [1.22,1.32]\,\text{TeV}\left.\right|_{\kappa=1.1}$  & \xmark \\[5pt] \hline 
 \end{tabular}

 \caption{We list the viable mass regions for various individual solutions to relic density, $R_K$ and $\Delta a_\mu$ for each model under the premise of accordance with direct detection limits.}
    \label{tbl:ModelAchievementsSingle}
\end{sidewaystable}

We found that all Dirac DM models are excluded as viable solutions to the $R_K$ and $(g-2)_\mu$ anomalies by direct detection, as the DM-nucleon cross section  
receives a large contribution from the kinematically unsuppressed ($v,v$) currents in $Z$-boson exchange. This is a particularly strong statement, when we remind ourselves that the one-loop-type models addressing the $R_K$ anomaly usually require a large coupling to muons, which diminishes the relic density substantially, leading to significant relaxation of the direct detection bounds. 
Majorana DM models, on the other hand, can offer open parameter space due to the vanishing vector currents in these type of models.

Leptophilic Majorana models in principle allow for a solution to $R_K$ with underproduced DM for coannihilation scenarios in the region $M_\text{DM} \lesssim 1000\,$GeV.
The $(g-2)_\mu$-anomaly on the other hand can only be solved for DM masses $\lesssim 290\,$GeV. 
An individual solution to DM only is generally possible in the low-$\Gamma_\mu$-regions of coannihilation scenarios. Note that this statement is even true for Dirac DM leptophilic models while still $R_K$ and $(g-2)_\mu$ cannot be explained \footnote{This is in contrast to quarkphilic and amphiphilic Dirac DM models, where observed relic density cannot be reproduced in agreement with direct detection in the parameter space studied in this work.}. 
We also have shown that a hierarchical flavor structure generally allows for larger DM masses and relaxes direct detection bounds in this context. 

In contrast, quarkphilic and amphiphilic Majorana models conceptually allow for a simultaneous solution of both $R_K$ and DM. While our study only found a window for large DM masses $M_\text{DM} \gtrsim 1.5\,$TeV in a coannihilation scenario for the quarkphilic model $\text{bIIB}^\text{Maj}$, the amphiphilic $\text{aIA}^\text{Maj}$ offers a low-$M_\text{DM}$ solution for a non-coannihilation set-up. Additionally, we found a small window in the TeV range for a coannihilation scenario, where $\psi$ is a viable DM candidate without solving the $R_K$ anomaly \footnote{Besides the low-massive solution, which we expect to be strongly constrained by a collider study, all remaining viable scenarios feature colored coannihilation. Those setups might receive sizable corrections to the relic density from non-perturbative effects such as the Sommerfeld enhancement and bound state formation arising in the annihilations of the colored particles, as discussed for instance \cite{Harz:2018csl} for a simplified colored coannihilations scenario. Those effects are not considered in \textsc{micrOMEGAs} and an inclusion lies beyond the scope of this work. }. 
Quarkphilic models also offer individual solutions to $R_K$, $(g-2)_\mu$ and DM in democratic coannihilation scenarios. In the case of a hierarchical flavor structure, direct detection rules out all viable parameter space due to large, $\Gamma_b$-driven contributions from quark loops to spin-independent DM-gluon scattering and spin-dependent DM-quark scattering.
Amphiphilic models are not able to pose even an individual solution for $(g-2)_\mu$ but provide a number of solutions to the $R_K$ anomaly, including coannihilation and non-coannihilation scenarios. 
\\ \indent
As Tbl. \ref{tbl:ModelAchievementsSimul} suggests, the most interesting models studied in this work contain colored particles in the TeV-range, which raises the question of detectability at colliders. Collider searches in similar set-ups can exclude dark sector particle masses up to the TeV scale. These results, however, assume fixed values of DM couplings, which motivates a customized collider study. 

\newpage

\textbf{Note added:} During the finishing stages of this work, the article \cite{Arcadi:2021glq} appeared that studies the problem discussed from a slightly different perspective. Both works discuss one-loop extension addressing \textit{B}-anomalies in the context of \textit{t}-channel DM. Out of the ten models discussed in this paper four are covered in the analysis of \cite{Arcadi:2021glq}, namely aIA and bIIB for both Majorana and Dirac fermions, albeit with a more elaborate collider phenomenology. 

Moreover, the DM analysis in both works is complementary, since the works study different parts of the parameter space and the fixed mass ratios of DM to the other dark sector particles employed here are particularly useful to study coannihilating scenarios while the approach in \cite{Arcadi:2021glq} is more suitable for collider phenomenology. 

Finally, we point out here that the spin-dependent limits are especially relevant for the case of Majorana DM where they can be more constraining than the spin-independent searches.

\section*{Acknowledgements} 
MB and DD thank Subhendu Rakshit for the kind hospitality at IIT Indore. SK thanks TU Dortmund for the kind hospitality.
MB and DD thank Alexander Pukhov for useful discussion about \textsc{calcHEP} and \textsc{micrOMEGAs}, as well as Susanne Westhoff and C{\'e}dric Delaunay for their comments on VIB in such flavor models. MB acknowledges support from the DFG Emmy Noether Grant No. HA 8555/1-1.
This work was supported by the German Academic Exchange Service (DAAD) and the Department of Science and Technology (DST), India, entitled "Flavour Physics and Dark Matter: Reconnecting the Dots" under Projekt-ID 57389505 and grant number INT/FRG/DAAD/P-22/2018.
\\
\\
The Feynman diagrams in this work are produced with \textsc{FeynTikZ} \cite{Ellis:2016jkw}.

\appendix
\section*{Appendix} \label{sec:appendix}
\FloatBarrier
\subsection*{Appendix A} \label{sec:AppendixA}
In Section \ref{sec:RD} we provide a simple estimate of the relic density assuming that the annihilation cross section is dominated by direct annihilations of DM via the new Yukawa couplings $\Gamma_i$. The relic density is inversely proportional to the effective annihilation cross section given in Eq.\@ \eqref{eqn:sigmaveff}. In this appendix, we illustrate that even in the scenario where only the direct annihilation of DM is relevant for the effective annihilation cross section, efficient conversions of DM, encoded in the ratio of equilibrium densities in Eq.\@ \eqref{eqn:sigmaveff}, might alter the predictions of the relic density significantly in the case of small mass splittings. More precisely, we address the interplay of the Yukawa coupling $\Gamma_i$ required to reproduce the observed relic density for a given mass ratio $\kappa$. \\
If we only consider the thermally averaged cross section of the direct DM annihilation in Eq.\@ \eqref{eqn:CrossSectionAnnihi}, we find that the cross section decreases for an increasing $\kappa$ for both Majorana and Dirac DM. This corresponds to a decreasing cross section for an increasing mediator mass. \\
If we additionally consider the contribution of the ratio of equilibrium densities at the time of thermal freeze-out, $x_f \approx 25$, the $\kappa$ dependence of the effective thermally averaged annihilation cross section results in
\begin{align}
\left \langle \sigma_\text{eff} v \right \rangle &\sim \frac{1}{\left( 1+\kappa^2 \right)^2} \left[ 1+\frac{g_\text{non-DM}}{g_\text{DM}} \kappa^\frac{3}{2} \exp \left( -\kappa x_f + x_f \right) \right]^{-2}  \times \nonumber \\
&\times \left\{\begin{array}{ll} \frac{1}{8}, & \text{for Dirac DM} \\
         \left( 1 + \kappa^4 \right) \left( 1+ \kappa^2 \right)^{-2}, & \text{for Majorana DM}\end{array}\right. \, . \tag{A.1}
\end{align}
Evidently, the effective annihilation cross section develops a maximum at e.g. $\kappa \sim 1.15$ for Dirac DM in the bIIA model. Consequently, we expect to match the observed  relic density with the smallest $\Gamma$ for $\kappa=1.15$ in this scenario. This effect can also be seen for the models bIIA, bVA and aIA in Figure \ref{fig:SummarybVADirac}, \ref{fig:SummarybIIADirac} and \ref{fig:SummaryaIADirac} where the line for the correct relic density with $\kappa=1.1$ lies below the line for $\kappa=1.01$. Note that such an effect cannot be seen for the quarkphilic scenarios (bIIB and bVIB), as the annihilation cross section is dominated by $\Gamma_s$ and $\Gamma_b$ and the results are presented in the $M_\text{DM}-\Gamma_\mu$ plane.  \\
In case of Majorana DM the situation is more complicated. As the direct DM annihilation is p-wave suppressed due to the Majorana nature of DM, s-wave coannihilations of the other dark sector particles can become dominant for smaller mass splittings. In the case of leptophilic DM (bIIA and bVA), the annihilation is dominated by annihilations of the scalar doublet into SM leptons for $\kappa=1.01$, which are as well mediated by $\Gamma_\mu$. For $\kappa=1.1$ however, direct DM annihilations dominate the effective annihilation cross section. \\ 
To identify the $\kappa$ dependence of the effective cross section in this case we need to add up the contributions from scalar and Majorana fermion annihilations according to Eq.\@ \eqref{eqn:sigmaveff}. The only difference between the two models is the increased number of colored fermions in the dark sector for the bVA model (their number increases from $6$ to $18$). This change is sufficient to alter the $\kappa$ dependence significantly via the efficient conversions in the dark sector. While the effective annihilation cross section for the bIIA Majorana scenario is decreasing monotonically, the effective annihilation cross section of the bVA scenario develops a maximum between $\kappa=1.01$ and $\kappa=1.1$. This explains the different ordering of the relic density lines of $\kappa=1.01$ and $\kappa=1.1$ in these two scenarios. \\
      
\FloatBarrier
\subsection*{Appendix B} \label{sec:AppendixB}
In this section, we present the constraints on the Yukawa couplings $\Gamma_s \Gamma_b^*$ and $\Gamma_\mu$ from $B$-$\bar{B}$, $R_K$ and $g-2$ of the muon in Tables \ref{tbl:couplingConstraintsYuk} and \ref{tbl:g-2}.
\begin{sidewaystable}[!p]
    \centering
        \begin{tabular}{c|c|c} 
        & $\mathcal{B}_{bs}^\text{model}(\kappa)$ & $\mathcal{B}_\mu^\text{model}(\kappa)$  \\ \hline \hline
        aIA & $2.45\cdot 10^{-6}  \left( \sqrt{( \frac{-0.000792 + 0.000792 \kappa^4 - 
        0.001583 \kappa^2 \log{(\kappa^2)}}{(\kappa^2 -1)^3}} \right)^{-1}$ &
        $0.00958 \left(\sqrt[4]{\frac{-0.000792 + 0.000792 \kappa^4 - 0.001583 \kappa^2 \log{(\kappa^2)}}{(
        (\kappa^2 -1)^3}}\right)^{-1}$  \\ \hline
        aI$\text{A}_{\text{M}}$ & $2.45\cdot 10^{-6} \left(\sqrt{\frac{-0.003958 + 0.003166 \kappa^2 + 
        0.0007916 \kappa^4 + (-0.001583 - 
            0.003166 \kappa^2) \log{(\kappa^2})}{(\kappa^2 -1)^3}} \right)^{-1}$  & 
            $0.00958 \left( \sqrt[4]{\frac{-0.003958 + 0.003166 \kappa^2 + 
        0.0007916 \kappa^4 + (-0.001583 - 
            0.003166 \kappa^2) \log{(\kappa^2)}}{( \kappa^2 -1)^3}} \right)^{-1}$  \\ \hline
        bIIA & $0.000151  \kappa$ & $0.538516 \left(\sqrt{ \frac{\kappa (-230.909 + 307.878 \kappa^2 - 
        76.970 \kappa^4 - 153.939 \log{(\kappa^2}))}{(\kappa^2 -1 )^3} } \right)^{-1}$  \\ \hline
        bVA & $0.000270 \kappa$ &  $0.538516 \left( \sqrt{\frac{\kappa (-103.266 + 137.687 \kappa^2 - 
        34.4218 \kappa^4 - 68.8437 \log{(\kappa^2)})}{(\kappa^2 -1)^3}} \right)^{-1}$  \\ \hline
        bIIB & $0.000087 \left( \sqrt{ \frac{-1 +  \kappa^4 - 2 \kappa^2 \log{(\kappa]^2)}}{(\kappa^2 -1)^3}} \right)^{-1}$ & $0.080783    \sqrt[4]{(\kappa^2 -1)^3} \left| \sqrt{\frac{-3 + 4 \kappa^2 -  \kappa^4 - 4 \log{(\kappa)}}{
        \sqrt{ -1 + \kappa^4 - 4 \kappa^2 \log{(\kappa)}}}} \right|^{-1}$  \\ \hline
        bVIB & $2.45\cdot 10^{-6} \left( \sqrt{\frac{-0.000791572 + 0.000791572 \kappa^4 - 
        0.00158314 \kappa^2 \log{(\kappa^2)}}{( \kappa^2 -1)^3}} \right)^{-1}$ & 
        $0.538516 \left( \sqrt{ \frac{-33.3288 + 44.4384 \kappa^2 - 
        11.1096 \kappa^4 - 22.2192 \log{(\kappa^2)}}{( \kappa^2 -1)^3 \sqrt{\frac{-1 + \kappa^4 - 2 \kappa^2 \log{(\kappa^2)}}{ (\kappa^2 -1)^3}}}} \right)^{-1}$
        \end{tabular}
    \caption{Presented here are the constraints on new Yukawa coupling $\Gamma_\mu$ and the product $\Gamma_s \Gamma_b^*$. The entries of this table represent the upper bound on $\Gamma_s \Gamma_b^*$ from $B$-$\bar{B}$-mixing and thus a lower bound on $\Gamma_\mu$ from $R_K$.}
    \label{tbl:couplingConstraintsYuk}
\end{sidewaystable}

\begin{sidewaystable}
    \centering
        \begin{tabular}{c|c} 
             & $\Gamma_\mu \in $ \\ \hline \hline
        bIIA & $\frac{[0.0000438, 0.0000557] (1-\kappa^2)^2 \nicefrac{M_{\psi_L}}{\text{GeV}} }{\sqrt{\left| -0.0000235648 - 0.0000353472 \kappa^2 + 0.0000706944 \kappa^4 - 
        0.0000117824 \kappa^6 - 0.0000706944 \kappa^2 \log{(\kappa^2)} \right|}} $ \\ \hline
        bVA  & $\frac{[0.0000438, 0.0000557] (1-\kappa^2)^2 \nicefrac{M_{\psi_L}}{\text{GeV}} }{\sqrt{\left| -0.0000235648 - 0.0000353472 \kappa^2 + 0.0000706944 \kappa^4 - 
        0.0000117824 \kappa^6 - 0.0000706944 \kappa^2 \log{(\kappa^2)} \right|}} $ \\ \hline
        bIIB & $[0.01805,0.02294] \kappa \nicefrac{M_{\psi_Q}}{\text{GeV}}$ \\ \hline
        bVIB & $[0.00932,0.01185] \kappa \nicefrac{M_{\psi_Q}}{\text{GeV}}$ \\ \hline
        aIA  & $\frac{[0.0000438, 0.0000557]  (1-\kappa^2)^2 \nicefrac{M_{\psi}}{\text{GeV}}}{\sqrt{\left| -0.0000235648 - 0.0000353472 \kappa^2 + 0.0000706944 \kappa^4 - 
        0.0000117824 \kappa^6 - 0.0000706944 \kappa^2 \log{(\kappa^2)} \right|}} $
        \end{tabular}
        \caption{Presented here are the constraints on new Yukawa coupling $\Gamma_\mu$ for a solution of $(g-2)_\mu$. These constraints contain upper and lower bound on $\Gamma_\mu$ formulated as an interval.}
        \label{tbl:g-2}
\end{sidewaystable}

\end{document}